%% file: MainFinal.tex
\documentclass[journal ,12]{IEEEtran}
\usepackage{colortbl}
\usepackage[table]{xcolor}

\usepackage{tikz}
\usepackage{url}
\usepackage{float} 
\usepackage{amsmath,amsfonts,amssymb,mathrsfs,amsthm,amsbsy,graphicx,color,bm,balance,mathtools,xfrac,nccmath,siunitx,cite,soul}
\usepackage{multirow}

\newtheorem{exmp}{Example}
\newtheorem{rem}{Remark}

\usepackage[font=scriptsize,labelfont=bf]{caption}
\DeclareMathOperator{\Tr}{\rm{Tr}}

\usepackage{pgfplots}
\pgfplotsset{compat=newest}
\usetikzlibrary{plotmarks}
\usetikzlibrary{arrows.meta}
\usepgfplotslibrary{patchplots}
\usepackage{grffile}

\usepackage[font=scriptsize,labelfont=bf]{subcaption}

\usepackage{algpseudocode}
\usepackage{algorithm}
\newcommand{\bc}{\text{BackCom}\xspace}
\newcommand{\mbc}{\text{MoBC}\xspace} 
\newcommand{\bbc}{\text{BiBC}\xspace}
\newcommand{\abc}{\text{AmBC}\xspace} 
 \usepackage[utf8]{inputenc}
\usepackage{booktabs, caption, makecell,xspace}

\usepackage{threeparttable}

\usepackage[autostyle]{csquotes}

\usepackage{pifont}% http://ctan.org/pkg/pifont
\newcommand{\cmark}{\ding{51}}%
\newcommand{\xmark}{\ding{55}}%
\newcommand{\lmark}{\textbf{--}}

 \newtheorem{theorem}{Theorem}

 \theoremstyle{definition}
 \makeatletter
\newcommand{\printfnsymbol}[1]{%
  \textsuperscript{\@fnsymbol{#1}}%
}
\makeatother
\begin{document}
\bstctlcite{IEEEexample:BSTcontrol}

\title{Time-Spread Pilot-Based Channel Estimation for Backscatter Networks}
\author{Fatemeh Rezaei\textsuperscript{\textasteriskcentered}\thanks{\printfnsymbol{1}F. Rezaei and D.  Galappaththige contributed equally to this work.}, \IEEEmembership{Member, IEEE}, Diluka  Galappaththige\printfnsymbol{1}, \IEEEmembership{Member, IEEE},   Chintha Tellambura, \IEEEmembership{Fellow, IEEE,}   Amine Maaref, \IEEEmembership{Senior Member, IEEE}
\thanks{F. Rezaei, D.  Galappaththige, and C. Tellambura are with the Department of Electrical and Computer Engineering, University of Alberta, Edmonton, AB, T6G 1H9, Canada (e-mail: \{rezaeidi, diluka.lg, ct4\}@ualberta.ca).  \\
\indent A. Maaref is with Huawei Canada, 303 Terry Fox Drive, Suite 400, Ottawa, Ontario K2K 3J1 (e-mail: amine.maaref@huawei.com).}  \vspace{-0mm}}

\maketitle
\begin{abstract} Current backscatter channel estimators employ an inefficient silent pilot transmission protocol, where tags alternate between silent and active states. To enhance performance, we propose a novel approach where tags remain active simultaneously throughout the entire training phase. This enables a one-shot estimation of both the direct and cascaded channels and accommodates various backscatter network configurations. We derive the conditions for optimal pilot sequences and also establish that the minimum variance unbiased (MVU) estimator attains the Cram\'{e}r-Rao lower bound. Next, we propose new pilot designs to avoid pilot contamination.  We then present several linear estimation methods, including least square (LS), scaled LS, and linear minimum mean square error (MMSE), to evaluate the performance of our proposed scheme. We also derive the analytical MMSE  estimator using our proposed pilot designs.
Furthermore, we adapt our method for cellular-based passive Internet-of-Things (IoT) networks with multiple tags and cellular users. {Extensive numerical and simulation results} are provided to validate the effectiveness of our approach. Notably, at least \qty{10}{\dB m} and \qty{12}{\dB m}  power savings compared to the prior art are achieved when estimating the direct and cascaded channels.  These findings underscore the practical benefits and superiority of our proposed approach. 
\end{abstract}


\IEEEpeerreviewmaketitle
\section{Introduction}

Backscatter communication (\bc) networks use electronic tags to reflect external radio frequency (RF) signals to transmit data.  They enable passive or ambient Internet of Things (IoT) networks \cite{Van2018, Toro2022, Diluka2022, Rezaei2023, rezaei2020large, Song2022}. They comprise multiple interconnected devices that can sense, collect, and exchange information about their environment without explicit human intervention \cite{Van2018, Toro2022, Diluka2022 }. They have a wide range of applications, such as smart homes, smart cities, industrial IoT, and healthcare \cite{Van2018, Toro2022, Diluka2022 }. They are under intense consideration by the third-generation  (3GPP) standards for passive  IoT networks \cite{Huawei, Huawei_ambient}.

\begin{figure*}[!t]\centering \vspace{-0mm}
 	\def\svgwidth{450pt} 
 	\fontsize{8}{8}\selectfont 
 	\graphicspath{{Figures/}}
 	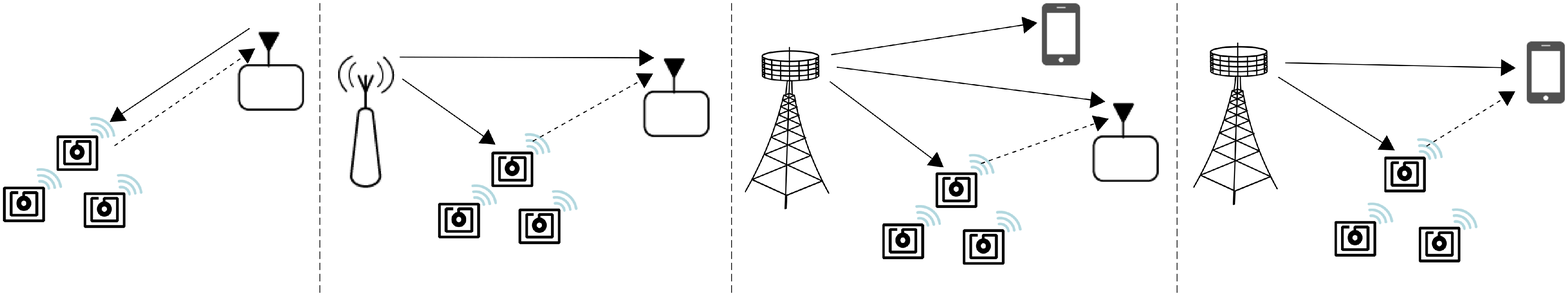 %\vspace{-0mm}
 	\caption{Different  \bc configurations.}\vspace{-0mm} \label{fig:config}
 \end{figure*}

A typical \bc network comprises an RF  emitter, a reader, and one or multiple tags \cite{Diluka2022}, where the basic  configuration can be monostatic, bistatic, or ambient (Fig.~\ref{fig:config}). In monostatic \bc (\mbc) (Fig.~\ref{fig:config}a), the reader emits the RF signal, the tag reflects it, and the reader decodes the reflected signal. In contrast,  the bistatic \bc (\bbc)  employs dedicated carrier emitters (Fig.~\ref{fig:config}b). On the other hand,  ambient \bc  (\abc) (Fig.~\ref{fig:config}c) utilizes  ambient
RF emitters, such as TV towers, cellular
base stations (BS), Wi-Fi access points (AP), and others, to communicate with the reader. \abc can also be a form of symbiotic radio (SR) (Fig.~\ref{fig:config}d), where the reader acts as a cooperative receiver, e.g., a smartphone, which decodes the signals of both the RF source and tags \cite{Long2020}.

\subsection{Channel Estimation for \bc  Networks}\label{sec:introduction} 
Channel estimation is an essential, ubiquitous task in any wireless network, which is necessary for ensuring performance, reliability, and security \cite{Lu2022}. For example, the reader in a \bc network needs accurate channel state information (CSI)   for decoding, beamforming, signal detection, interference mitigation, security and privacy enhancements, and other tasks \cite{Lu2022}.  This need can be met by pilot-based,  blind, or semi-blind techniques CSI estimators \cite{kim2015wireless, Ozdemir2007}. Pilots, which are known symbols, improve the  estimation accuracy,  while blind techniques exploit received signal statistics without pilot symbols.  However, their accuracy levels (in terms of mean square error (MSE)) may be low  due to the absence of pilot symbols  \cite{Ozdemir2007}. 

Although pilot-based techniques are widely used, they reduce spectral efficiency by limiting the amount of time available for data transmission and also increase energy consumption \cite{Hassibi2003, Emil2016}. Therefore, optimizing the trade-off between channel estimation accuracy and spectral and energy efficiencies is critical for the overall performance \cite{Hassibi2003, Emil2016}.

% ================================== %
% ===================================== %
However, the problem of 
\bc channel estimation is distinct from that for conventional wireless networks and presents challenges due to the specific characteristics of the channel and the limitations of the tags. The  \bc channel is a cascaded (or dyadic) one when the signal propagates from the RF  emitter to the tag and then from the tag to the reader, and a direct channel describes signal propagation from the emitter to the reader. The cascaded channel has fading characteristics that are significantly different from those of direct channels and are subject to a double path loss. These specific channel characteristics are described in \cite{Diluka2022} and references therein.

The other challenges emanate from the limitations of the tags.  Firstly, tags do  not generate pilot signals by themselves as they lack active RF circuits but  simply reflect incident  RF signals \cite{Diluka2022}.   Secondly, passive tags harvest (i.e., absorb)  a part of the received RF power to operate their internal circuits \cite{Diluka2022}, resulting in low reflected power levels that may be insufficient for accurate channel estimation. Furthermore, because passive tags must harvest energy,   they often have low-cost, power-integrated circuits with limited processing capabilities \cite{Qualcomm2022, Huawei}. Consequently, they may not be able to transmit complex, high-power pilot sequences.  These limitations indicate that traditional pilot-based channel estimation methods must be adapted carefully in this context. 

Due to the vast application potential of \bc networks, several useful channel estimation techniques have already been developed. In the following, we will briefly describe these works, emphasizing their strengths and weaknesses in order to identify gaps in the existing literature. By doing so, we set the background and context for our work.

\subsection{Previous Contributions}

We have listed the ones for \abc     \cite{Ma2018, Abdallah2021, Zhao2018, Zhao2019, Liu2021, Zhu2018, Zheng2019} in Table~\ref{Table_prior_art}. 
As can be seen, most works focus on single-tag ($K=1$) networks.  For a network of $K \geq 1 $ tags,  these works rely on what may be referred to as the silent protocol. It has the essential idea of just one node transmitting pilots at a time while every other node remains silent.  Thus, it divides the channel estimation interval into $K+1$ equal slots. All the tags remain silent on the first slot except for the RF emitter.  During each one of the remaining $K$ slots, one tag reflects pilots while others remain silent.  Consequently, the direct (RF source-reader) channel and the cascaded channels are estimated one by one in a round-robin manner.

However, the silent protocol   (Fig.~\ref{fig_TagTrans}) has several drawbacks. Firstly, as the number of tags increases, the number of slots required also increases, resulting in longer pilot sequence lengths, and reduced energy and spectral efficiencies. Secondly, cascaded channel estimates are obtained by subtracting the direct channel estimate, resulting in error propagation and increased MSE of the estimates. To reduce the MSE, longer pilot sequences are necessary, further degrading the energy and spectral efficiencies. These critical drawbacks make the silent protocol inefficient for handling a large number of tags.  We further discuss these problems in  Section \ref{Prior_art}.

Based on the silent protocol, these works employ classical and machine learning methods, including expectation maximization \cite{Ma2018, Shuo2018}, eigenvalue decomposition of the received signal's covariance matrix \cite{Zhao2018} for blind estimation, discrete Fourier transformation (DFT) \cite{Zhao2019}, deep neural networks \cite{Liu2021}, and iterative estimation \cite{Zhu2018} for pilot-based estimation. Moreover, pilot-based channel estimation has also been investigated in large intelligent surface (LIS)-aided \abc networks  \cite{Zheng2019}, building on the basic idea explored in the previous works.

In  \cite{Abdallah2023},  the direct channel and cascaded channel are estimated simultaneously for the full-duplex ambient RF source and the legacy user.  This work develops a  maximum-likelihood (ML) estimator, but the extension to multiple tags is challenging. Reference \cite{Wang2022} also leverages a deep learning approach for the joint pilot design and channel estimation in a multi-tag SR network. It first uses a deep residual network with three blocks to eliminate the noise and then exploits successive interference cancellations to estimate the direct channel and the cascaded links sequentially. 

The works \cite{Mishra2019, Yerzhanova2021} focus on \mbc channel estimation, addressing single and multiple tag scenarios.    Reference \cite{Abdallah2021} develops direct and cascaded channel estimators of a single tag network while accounting for carrier-frequency offsets (CFO) and in-phase/quadrature imbalance.   The study \cite{Darsena2018} develops joint channel estimation, interference suppression, and data detection for an orthogonal frequency division multiplexing-based \abc system with a single tag. The method employs the space alternating generalized expectation maximization algorithm.  
{Reference \cite{Dai2023} presents a two-phase uplink-training-based channel estimation approach for estimating direct and cascaded channels in a cell-free SR network with a single user and single tag. The study \cite{Wei2021} proposes two iterative channel estimate approaches for reconfigurable intelligent surfaces (RIS)-enabled multi-user communication networks.  }

{\linespread{1.0}
\begin{table*}[!t]\vspace{-0mm}
\caption{Summary of related works. } \vspace{-0mm}
\label{Table_prior_art}
\begin{center}
\begin{threeparttable}
% \scalebox{0.9}{
\begin{tabular}{|l|l|l|lll|ll|l|}
\hline
\multirow{2}{*}{Config.}       & \multirow{2}{*}{Reference}                            & \multirow{2}{*}{Pilot-based} & \multicolumn{3}{c|}{Setup}                                                                          & \multicolumn{2}{c|}{Channel}                                   & \multirow{2}{*}{Key  idea} \\ \cline{4-8}
                                     &                                                       &                              & \multicolumn{1}{l|}{Tags}              & \multicolumn{1}{l|}{Reader}            & Emitter           & \multicolumn{1}{l|}{Fading}            & Path loss             &                                      \\ \hline \hline
                  & \cite{Ma2018}                        & \xmark\,(Blind)                       & \multicolumn{1}{l|}{$K=1$}              & \multicolumn{1}{l|}{SA}                & SA                & \multicolumn{1}{l|}{Rayleigh}          & \xmark &  Silent                                    \\ \cline{2-9}
 & \multirow{2}{*}{\cite{Abdallah2021}} & \cmark                  & \multicolumn{1}{l|}{\multirow{2}{*}{}} & \multicolumn{1}{l|}{\multirow{2}{*}{}} & \multirow{2}{*}{} & \multicolumn{1}{l|}{\multirow{2}{*}{}} & \multirow{2}{*}{}     &                                      \\ \cline{3-3} %\cline{9-9} 
                                     &                                                       & \xmark \,(Semi-blind)                    & \multicolumn{1}{l|}{\multirow{-2}{*}{$K=1$}}                  & \multicolumn{1}{l|}{\multirow{-2}{*}{SA}}                  &    \multirow{-2}{*}{SA }              & \multicolumn{1}{l|}{\multirow{-2}{*}{Rayleigh}}                  &  \multirow{-2}{*}{\xmark }                   &     \multirow{-2}{*}{Silent  }                                  \\ \cline{2-9}
                                     & \cite{Zhao2018}                      &   \xmark \,(Blind)                          & \multicolumn{1}{l|}{$K=1$}                  & \multicolumn{1}{l|}{MA}                  &  SA                 & \multicolumn{1}{l|}{Rayleigh}                  &      \xmark                 &    Silent                                  \\ \cline{2-9}
                                     & \cite{Zhao2019}                      &   \cmark                         & \multicolumn{1}{l|}{$K=1$}                  & \multicolumn{1}{l|}{MA}                  &   SA                & \multicolumn{1}{l|}{Rayleigh}                  &     \xmark                  &    Silent                                  \\ \cline{2-9}
                                     & \cite{Liu2021}                       &  \cmark                            & \multicolumn{1}{l|}{$K=1$}                  & \multicolumn{1}{l|}{MA}                  &      SA             & \multicolumn{1}{l|}{Rayleigh}                  &   \xmark                    &    Silent                                  \\ \cline{2-9}
                                     & \cite{Zhu2018}                       &  \cmark                            & \multicolumn{1}{l|}{$K=1$}                  & \multicolumn{1}{l|}{SA}                  &  SA                 & \multicolumn{1}{l|}{Rayleigh}                  &                \xmark       &     Silent                                 \\ \cline{2-9}
                                     & \cite{Zheng2019}$^{*}$                     &  \cmark                            & \multicolumn{1}{l|}{LIS}                  & \multicolumn{1}{l|}{SA}                  &  SA                 & \multicolumn{1}{l|}{Rayleigh}                  &    \xmark                   &  Silent                                    \\ \cline{2-9}
  & \multirow{2}{*}{\cite{Abdallah2023}$^\dagger$ } & \cmark                  & \multicolumn{1}{l|}{\multirow{2}{*}{}} & \multicolumn{1}{l|}{\multirow{2}{*}{}} & \multirow{2}{*}{} & \multicolumn{1}{l|}{\multirow{2}{*}{}} & \multirow{2}{*}{}     &                                      \\ \cline{3-3} %\cline{9-9} 
                                     &                                                       & \xmark   \,(Semi-blind)                & \multicolumn{1}{l|}{\multirow{-2}{*}{$K=1$}}                  & \multicolumn{1}{l|}{\multirow{-2}{*}{MA}}                  &    \multirow{-2}{*}{SA }              & \multicolumn{1}{l|}{\multirow{-2}{*}{Nakagami}}                  &  \multirow{-2}{*}{\cmark }                   &   \multirow{-2}{*}{One-shot}                                    \\ \cline{2-9}      
\multirow{-10}{*}{\abc}                                    & \cite{Darsena2018}                   &    \xmark  \,(Semi-blind)                         & \multicolumn{1}{l|}{$K=1$}                  & \multicolumn{1}{l|}{SA}                  & SA                  & \multicolumn{1}{l|}{Rayleigh}                  & \cmark                      &   Iterative                                   \\ \hline
 SR                                    & \cite{Wang2022}                      &  \cmark                            & \multicolumn{1}{l|}{$K\geq 1$}                  & \multicolumn{1}{l|}{SA}                  & MA                  & \multicolumn{1}{l|}{Rayleigh}                  & \xmark                      &    Silent                                  \\ \hline
                                     & \cite{Mishra2019}                    &    \cmark                           & \multicolumn{1}{l|}{$K=1$}                  & \multicolumn{1}{l|}{MA}                  & \lmark                  & \multicolumn{1}{l|}{Rayleigh}                  & \cmark                      &     \lmark                                 \\ \cline{2-9}
 \multirow{-2}{*}{\mbc}                                    & \cite{Yerzhanova2021}                &     \cmark                          & \multicolumn{1}{l|}{$K\geq 1$}                  & \multicolumn{1}{l|}{MA}                  &   \lmark                & \multicolumn{1}{l|}{Rician}                  & \cmark                      &  Silent                                    \\ \hline
Any                                     & \textbf{This paper}                  & \cmark                              & \multicolumn{1}{l|}{$K\geq 1$}                  & \multicolumn{1}{l|}{MA}                  &      SA             & \multicolumn{1}{l|}{Nakagami}                  &  \cmark                     &         One-shot                             \\ \hline
\end{tabular}
\begin{tablenotes}\footnotesize
\item SA - Single antenna \qquad MA - Multiple  antenna %\qquad ML - Maximum-likelihood \qquad DL - Deep learning   
\item[$\dagger$] The reader is a full-duplex multi-antenna AP and the emitter is the AP as well as a single antenna user.  
\item[$*$] LIS-assisted \abc.
\end{tablenotes}
\end{threeparttable}
\end{center}
\vspace{-0mm}
\end{table*}}

\textit{To summarize, the current approach of using the silent protocol is highly inefficient. Another shortcoming is the focus on the single-tag case, with a lack of solutions for multiple tags  ($K>1$), likely due to the poor scaling of the silent protocol as $K$ increases. Furthermore, some of the proposed solutions are developed for specific configurations such as \mbc, \bbc, \abc, and SR, which limits their generality. There have been only a few studies, such as \cite{Abdallah2023, Darsena2018} and \cite{Mishra2019, Yerzhanova2021}, which have investigated setups like \abc and \mbc, respectively.}

\subsection{Problem Statement and Contributions}

To remedy these issues, we present a versatile channel estimation method for any  \bc configuration. 
 Unlike prior solutions, our method is a
 {one-shot estimate} of all channel gains (both direct and cascaded) for an arbitrary number of tags (Fig.~\ref{fig_SystemModel}). {Here, the term ``one-shot estimation" refers to the simultaneous (parallel) estimation of the direct channel and all cascaded channels during pilot transmission.}
Unlike the silent protocol   \cite{Zhao2018, Zhao2019, Ma2018, Mishra2019, Yerzhanova2021, Liu2021},  we develop the time spreading of tag
pilots over the entire estimation interval, i.e., all tags modulate and reflect pilots simultaneously during the estimation interval. Thus, the reader sees a   received signal that is a combination of pilots spread over the direct and the backscatter channels. That ensures that spatial and time resources are fully utilized for channel estimation, which enhances the estimation accuracy remarkably. We also ensure the orthogonality among tag transmissions through optimally designed pilot sequences. Hence, our method avoids pilot contamination and error propagation.   

To provide practical insights, we consider both large-scale fading and small-scale fading (via the versatile Nakagami-$m$ fading model)  to model the direct channel and the cascaded channels.

The contribution of this work is summarized as follows:

\begin{itemize}
    \item  
    {We first design an efficient pilot transmission protocol where all the tags reflect pilots during the estimation interval. We identify that the RF source acts as a hidden tag with an all-$\mathbf{1}$ pilot sequence. Accordingly, we derive the necessary conditions on the tag sequences to avoid pilot contamination among the tags and the RF source.}

    \item We reformulate the received signal at the reader into a linear equation to derive the minimum variance unbiased (MVU) estimator.  The MVU estimator, which is the linear least squares (LS) estimator for the linear model, is efficient and attains the Cram\'{e}r-Rao lower bound (CRLB). It thus provides a sufficient criterion to design tag sequences such that the variance of the estimation is minimized.

    \item {We next propose and investigate novel tag pilot sequence designs, i.e., (i) rows of Hadamard matrix, (ii) modified Zadoff-Chu (ZC) sequences, and (iii) rows of DFT matrix. We find that ZC sequences do not meet the design criteria. We thus propose a method for modifying any set of orthogonal sequences for {\bc} channel estimation.}

    \item {We derive classical deterministic estimators, namely   LS and scaled LS, and Bayesian estimators optimal minimum mean square error (MMSE) and linear MMSE (LMMSE). The latter two exploit the knowledge of channel statistics.}
       
    \item {Additionally, we generalize our method for cellular-based passive IoT networks with concurrent multiple tags and cellular users. We give extensive numerical and simulation results to demonstrate the gains of our approach compared to the silent protocol.   For instance, to estimate the direct and cascaded channels,  our method, compared to the silent protocol,  saves  {\num{10}} and {\num{12}} {\qty{}{\dB m}}. These are remarkable gains.}
\end{itemize}

{Our proposal is notably versatile, as it supports different rate inequalities between the RF source and the tags, encompassing scenarios where $R_b = R_s$ or $R_b < R_s$,  where $R_b$ and $R_s$ are the data rates of the tag and the RF source, respectively \cite{Long2020}. As a result, it is applicable to different communication networks, such as cellular-based passive IoT where the rate of the user is typically higher than the rate of the tags.}

\subsection{Structure and Notations}The paper is structured as follows. In Section \ref{system_modelA}, we introduce the system model and discuss the tag operations.
Section \ref{sec:channel_estimation} reviews prior research on the channel estimation problem, and  Section \ref{sec:Proposed_method}  presents our proposed method. 
{Section {\ref{Multiple_user}} extends the proposed channel estimate approach to a generalized network with multiple tags and multiple users.}
To assess the effectiveness of our approach, we provide extensive simulation examples in Section \ref{sim}. Lastly, we summarize the key findings of the paper and suggest future research directions in Section \ref{conclusion}.

\textit{Notation}:  $\mathbf{A}^\mathrm{T}$ and $\mathbf{A}^\mathrm{H}$, denote transpose and  Hermitian transpose,  respectively.  $\mathbb{E} \{ \cdot\}$ denotes the statistical expectation. $\mathcal{CN}(\bm{\mu},\mathbf C ) $ is a complex Gaussian  vector with  mean   $\bm \mu$ and co-variance matrix  $\mathbf C$. 
 {The operations $\text{vec}(\mathbf{X})$, $\otimes$, and $\odot$ are respectively the column-wise stacked version of $\mathbf{X}$, Kronecker product,  and Hadamard product, which refers to component-wise multiplication of two vectors of the same dimension.}

%Kronecker product
% The operations $\text{vec}(\mathbf{X})$ and $\otimes$ are respectively the column-wise stacked version of $\mathbf{X}$ and Kronecker product.
Besides, $\text{diag}(\mathbf{x})$ with $\mathbf{x} \in \mathbb{C}^{1 \times N}$ returns the matrix $\mathbf{X} \in \mathbb{C}^{N \times N}$ with $\mathbf{x}$ on the diagonal, $\mathbf{C}_{\mathbf{x}} = \rm{Cov}(\mathbf{x}, \mathbf{x})$ is the covariance matrix of $\mathbf{x}$, and $\mathbf{I}_N \in \mathbb{R}^{N \times N}$ is an identity matrix. Finally, $\mathcal{M} \triangleq \{1,\ldots,M\}$, $\mathcal{K} \triangleq \{1,\ldots,K\}$, $\mathcal{K}_0\triangleq \{0,1,\ldots,K\}$, $\mathcal{K}_k\triangleq \mathcal{K}/k$, and $\mathcal{N} \triangleq \{1,\ldots,N\}$.

% ====================================== %
\section{System, Channel, and Signal Models}\label{system_modelA}

% ================================== %
 \begin{figure}[!t]\centering \vspace{-0mm}
 	\def\svgwidth{220pt} 
 	\fontsize{8}{8}\selectfont 
 	\graphicspath{{Figures/}}
    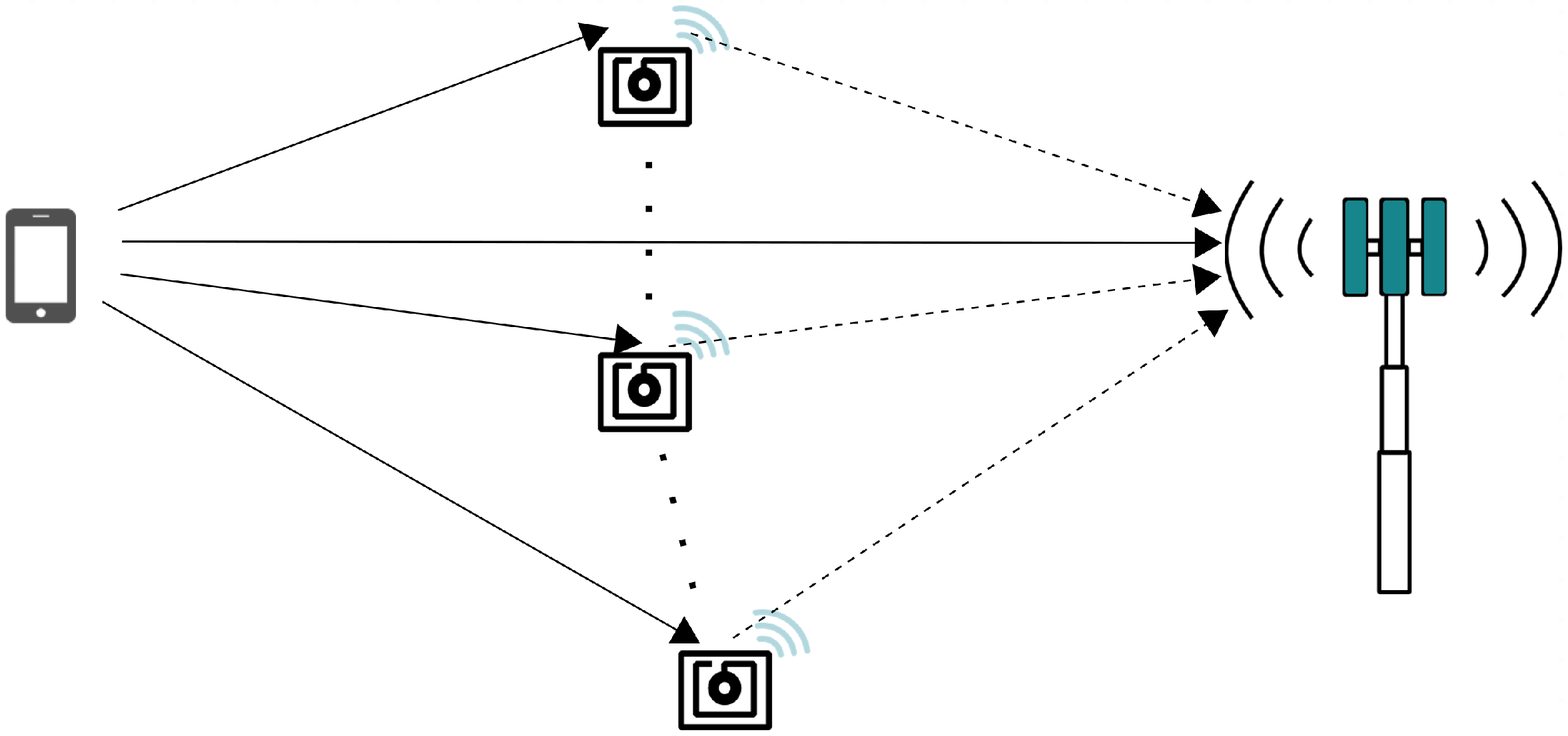 %\vspace{-0mm}
 	\caption{ {{\abc} system setup.}} \label{fig_SystemModel}
  \vspace{-0mm}
 \end{figure}
% ===================================== %

\subsection{System and Channel Models}
We consider an \abc network  comprising a single-antenna RF source {(or user)}, $K$ single-antenna passive tags ({$k$-th} tag is denoted by $T_k$), and a reader {(or BS/AP)} with $M$ antennas (Fig. \ref{fig_SystemModel}). The network operates on a block flat-fading channel model. During each fading block, $\mathbf{h}_0 = [h_{1,0}, \ldots, h_{M,0}]^{\rm{T}} \in \mathbb{C}^{M \times 1}$ is the direct channel response vector from the RF source to the reader. Moreover, $\mathbf{h}_k = f_k \mathbf{g}_k \in \mathbb{C}^{M \times 1}$ is the effective backscatter (cascaded) channel through $T_k$, which is the product of the forward-link channel from the RF source to $T_k$, i.e., $f_k \in \mathbb{C}$, and the backscatter channel from $T_k$ to the reader, i.e.,  $\mathbf{g}_k = [g_{1,k}, \ldots, g_{M,k}]^{\rm{T}}  \in \mathbb{C}^{M\times 1}$. A unified representation of all channels is given as

\begin{eqnarray} \label{channel_model}
    {a} = \alpha_{a} \exp({j \phi_{a}}),
\end{eqnarray}
where ${a} \in \mathcal{A} \triangleq \{ f_k,{g}_{m,k},{h}_{m,0} \}$ for $ m \in \mathcal{M}$ and $k \in \mathcal{K}$,  and  $\phi_{a} \in [-\pi,\pi]$ is the phase of $a$. In \eqref{channel_model}, $\alpha_{a}$ is the envelope of $a$, which is assumed to be Nakagami-$\bar{m}_{a}$ distributed with  $\bar{m}_{a}$ shape and $\Omega_{a}= \bar{m}_{a} \zeta_{a}$ scaling parameters. Here, ${\zeta}_{{a}}$ accounts for the large-scale path loss and shadowing.

\begin{rem}\label{Rayl}
Nakagami-${m}$ is a versatile model, which can represent a variety of propagation environments. For instance, when ${m} = 1$, it represents  Rayleigh fading, and when ${m} \rightarrow \infty,$ it represents the no fading scenario. Hence, our proposed channel estimation can be applied to any fading channels, e.g., Rayleigh, Rician, and others  \cite{Zhao2020Backscatter, Yang2020}. 
\end{rem}

\subsection{Tag's Data Transmission}\label{EH_data}

Passive tags are electronic devices that operate their essential circuits and processing solely through {energy harvesting (EH)}, eliminating the need for batteries. As a result, they face strict energy and power limitations. These tags modulate and reflect their data to the reader and carry out EH using two methods: time-switching and power-splitting. In the time-switching approach, the tasks are performed in separate time slots, while in the power-splitting approach, both tasks occur simultaneously, utilizing a fraction of the incident RF power \cite{Zhang2013}. Given the inability of passive tags to store energy, the power-splitting mode is generally preferred over the time-switching mode \cite{Ping2022, Galappaththige2023}.

We assume all tags have the same power reflection coefficient, $ \alpha\in (0,1)$. In the power-splitting mode, each tag reflects back   $\alpha P_{\rm{rf}}$ for data communication and absorbs  $(1-\alpha) P_{\rm{rf}}$ for EH where $P_{\rm{rf}}$ is the incident RF power.   The EH process is described in \cite{Diluka2022}. 

Tags employ a communication technique known as load modulation to modulate data. Load modulation involves encoding digital information into the amplitude, phase, or frequency of the backscattered signal. This is achieved by the tag dynamically switching between a predefined set of load impedances connected to its antenna. These impedance variations create mismatches, resulting in the generation of complex reflection coefficients. The collection of these reflection coefficients forms a multi-level ($\tilde{M}$-ary) signal constellation, enabling the tag to effectively convey data.  In this paper, we consider the phases of the reflection coefficients to create the modulation alphabet.  For details of such modulation methods, we refer the reader to  \cite{Diluka2022, Rezaei2023} and the references therein.  
Thus, the  tag symbol   $c(n)$ $(  \mathbb{E}\{\vert c(n) \vert^2\} = 1)$, is  selected from a  $\tilde{M}$-ary constant-envelope modulation \cite{Rezaei2023}.

The EH  process at the tags is not within the scope of this paper. However, it is worth noting that the integration of EH into channel estimation is an intriguing topic that has not been extensively explored in existing literature. This aspect will be addressed in our future research endeavors. For readers interested in further understanding EH in \bc networks, we recommend referring to \cite{Diluka2022} and the references cited therein.

\section{Channel Estimation}\label{sec:channel_estimation}

\begin{figure}[!tbp]\vspace{-0mm}
  \centering
   \def\svgwidth{220pt} 
 	\fontsize{8}{8}\selectfont 
 	\graphicspath{{Figures/}}
 	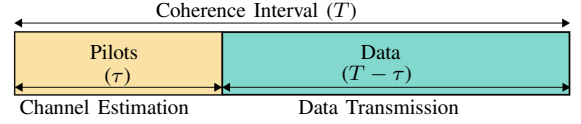 %\vspace{-0mm}
 	\caption{Coherence interval.}\vspace{-0mm} \label{fig:coherence}
\end{figure}

The goal is to estimate $\mathbf{H} = [\mathbf{h}_0, \mathbf{h}_1, \ldots, \mathbf{h}_K]$ using training based channel estimation method. To this end, each coherence interval comprises two phases, i.e., channel estimation and data transmission (Fig. \ref{fig:coherence}). During the first phase, the RF source transmits a pilot sequence with length $\tau$, i.e., $\sqrt{p}\mathbf{s} \in \mathbb{C}^{1 \times \tau}$, where $\mathbf{s}=[s_1,\ldots,s_{\tau}]$ and $s_i$  satisfies $|{s}_i|^2 = 1$ for $ i = \{1.\ldots, \tau\}$, {$p$ is the RF source transmit power}, and the tags backscatter, $c_k,$ to the reader. 
{In the following, we consider the same data rate for the tags and the source, i.e., $R_b = R_s$. We also explore the scenario where $R_b < R_s$ in Section {\ref{Multiple_user}}, demonstrating the versatility and generality of our proposed scheme.}

The two types of \bc channels are:   i) the direct-link channel from the RF source to the reader ($\mathbf{h}_0$), ii) the cascaded channels, i.e., RF source-to-tag-to-reader channels ($f_k \mathbf{g}_k$, for  $k \in \mathcal{K}$). These two types are estimated slightly differently.

Next, we first describe the existing pilot-based channel estimators and reiterate their limitations.  We then present our approach.

\subsection{Prior Art}\label{Prior_art}

\begin{figure}[!tbp]\vspace{-0mm}
  \centering
    \def\svgwidth{210pt} 
 	\fontsize{8}{8}\selectfont 
 	\graphicspath{{Figures/}}
 	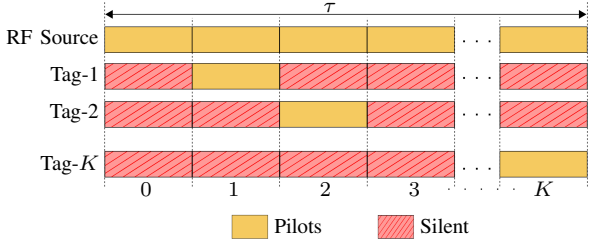 %\vspace{-0mm}
 	\caption{{Pilot transmission in the prior art.}}\vspace{-0mm} \label{fig_TagTrans}
\end{figure}

{The prior art uses what can be described as the silent protocol.  It divides the  channel estimation time, $\tau$, into $K+1$ slots, each with $t$ pilot symbols (for equal division) \cite{Zhao2018, Zhao2019, Ma2018, Mishra2019, Yerzhanova2021, Liu2021} (Fig. {\ref{fig_TagTrans}}).
In the first slot, all the tags are silent with no reflections, i.e.,  $c_k = 0$ for $k \in \mathcal{K}$. Thus, the reader computes the direct channel estimate  $\hat {\mathbf{h}}_0$ cleanly. In each of subsequent $K$ slots, one tag backscatters the pilot symbol $c_k=1$ while the others do not reflect ($c_k=0$) (Fig.{\ref{fig_TagTrans}}).  The reader then estimates  $\mathbf{h}_0+ \sqrt{\alpha}  f_k \mathbf{g}_k $, where $\alpha \in (0,1) $ is the tag reflection coefficient, which is constant for all tags, and subtracts $\hat {\mathbf{h}}_0. $ This yields  the cascaded channel estimate, $ \hat{\mathbf{h}}_k = \widehat{f_k \mathbf{g}_k}$ for $k \in \mathcal{K}$.  We now explain this process with a specific example.}

\begin{exmp}\label{example1} {For two tags ($K=2$), the silent protocol needs three time slots. If the tags use pilots 0 and 1, and the RF signal is $\sqrt{p}$ in the $i$-th time slot, the received signals at the reader  are given as}
\begin{eqnarray}\label{first_approach}
\mathbf{Y}_i =
    \begin{cases}
     & \sqrt{p}\mathbf{h}_0  + \mathbf{n}_0, \quad \text{for} \quad i=0, \\
     & \sqrt{p}(\mathbf{h}_0 + \sqrt{\alpha} f_i \mathbf{g}_i)  + \mathbf{n}_i,  \quad \text{for} \quad i\in \{1,2\},    
    \end{cases}
\end{eqnarray}
{where $\mathbf{n}_i $ is an additive white Gaussian noise (AWGN) term.  From the  received signal {\eqref{first_approach}}, the reader estimates $\mathbf{h}_0$ from  $\mathbf{Y}_0$. The  reader then  estimates $\mathbf{h}_0 + \sqrt{\alpha} f_i \mathbf{g}_i$ and subtracts  $\hat{\mathbf{h}}_0$ to obtain the cascaded channel estimates. }
\end{exmp}

\subsubsection{{Limitations of the Prior Art}} 
{Although the above approach may seem simple,  several critical drawbacks result in poor performance.   As such, it is important to identify these weaknesses in order to develop efficient and reliable alternatives. Here is a list of the limitations: }
\begin{itemize}
    \item  {The one-by-one estimation of the direct channel and the cascaded channels requires the estimation interval to increase with $K+1.$ }   
      
    \item {Estimating cascaded channels through subtracting $\hat{\mathbf{h}}_0$ leads to error propagation, making the MSE of cascaded channel estimates reliant on the MSE of $\hat{\mathbf{h}}_0$. Thus,  highly accurate $\hat{\mathbf{h}}_0$ estimation requires longer pilot sequences, consequently increasing pilot training overhead.}
    
    \item {Each channel is estimated using a few pilot symbols, i.e., $\mathbf{Y}_i$ for $ i \in \mathcal{K}_0,$ Eq. {\eqref{first_approach}}, not all the pilots.  This forces using longer pilot sequences to compensate for this design defect, increasing channel estimation overhead.}
    
    \item {Much of the coherence interval is used for channel estimation for large $K$, degrading the spectral and energy efficiencies.} 

\end{itemize}

\subsection{Proposed Method}\label{sec:Proposed_method}

For the efficient use of resources and exploiting the spatial diversity, unlike the silent protocol \cite{Zhao2018, Zhao2019, Ma2018, Mishra2019, Yerzhanova2021, Liu2021},  we assume that all the tags simultaneously reflect pilot symbols during the estimation phase. For the considered network (Fig.~\ref{fig_SystemModel}), we  design a pilot protocol to estimate $\bar{\mathbf{H}} = [\mathbf{h}_0, \sqrt{\alpha} \mathbf{h}_1, \ldots, \sqrt{\alpha} \mathbf{h}_K]$ in one-shot. {The tags are simple, passive devices with a  fixed reflection coefficient, {$\alpha.$} We assume that the reader knows the value of $\alpha.$ This assumption is consistent with the prior studies  \cite{Zhao2018, Zhao2019, Ma2018, Mishra2019, Yerzhanova2021, Liu2021}.
However, it is noteworthy to mention the pilot sequence design is independent of factors such as  $\alpha $ or the channel statistical model. Moreover, if tags have  unequal reflection coefficients, denoted as $\alpha_k$ for $k \in \mathcal{K}$, the columns of $\bar{\mathbf{H}}$ to be estimated will be $\sqrt{\alpha_k} \mathbf{h}_k$ for $k \in \mathcal{K}$. As long as the reader has the knowledge $\alpha_k$, it can estimate the cascaded channels, $\mathbf{h}_k$, by removing the effect of $\alpha_k$.}

Consequently,  the received signal at the reader over $\tau$ pilot symbols, $\mathbf{Y} \in \mathbb{C}^{M \times \tau}$, is given as 
\begin{eqnarray}\label{eqn_rx_reader}
     \mathbf{Y} = \sqrt{p} \mathbf{h}_0 \mathbf{s} + \sqrt{p \alpha} \sum\nolimits_{i\in \mathcal{K}} \mathbf{h}_i (\mathbf{s} \odot  \mathbf{c}_i) + \mathbf{N},
\end{eqnarray}
where  $\mathbf{c}_i = [c_{i1},\ldots c_{i \tau}]$ is the $i$-th tag pilot sequence, and $c_{ij}$ is the modulated symbol of the $i$-th tag  in the $j$-th slot. Moreover, $\mathbf{N} \in \mathbb{C}^{M \times \tau}$ is the noise matrix  with i.i.d $\mathcal{CN}(0, \sigma^2)$ elements. By defining $\mathbf{S}\triangleq \rm{diag}(\mathbf{s})$, we can {rewrite} the received signal at the reader \eqref{eqn_rx_reader} as follows:
\begin{eqnarray}\label{eqn_rx_reader_re}
    \mathbf{Y} = \sqrt{p} \mathbf{h}_0 \mathbf{1}_{\tau} \mathbf{S} + \sqrt{p \alpha} \sum\nolimits_{i\in \mathcal{K}} \mathbf{h}_i \mathbf{c}_i \mathbf{S} + \mathbf{N},
\end{eqnarray}
where $\mathbf{1}_{\tau} \triangleq [1,\ldots,1] \in \mathbb{R}^{1\times \tau}$ is the all-$\mathbf{1}$ sequence. From \eqref{eqn_rx_reader_re}, we can see that the RF source acts as a hidden tag with an all-$\mathbf{1}$ pilot sequence. Hence, the simple assignment of orthogonal pilot sequences to the tags is not sufficient to prevent pilot contamination. Specifically, even for a mutually set of orthogonal tag pilot sequences, if they are not orthogonal to the all-$\mathbf{1}$ sequence, the direct channel and the cascaded channels can not be separated at the reader. Hence, this is somewhat akin to pilot contamination, which appears in multi-cell networks when pilots are reused in adjacent cells. As in conventional networks, this leads to corrupted channel estimates (Section~\ref{Prior_art}). 

\begin{theorem}\label{thrm_condition}
To avoid pilot contamination,  $T_k$'s pilot sequence must satisfy the following conditions:
\begin{enumerate}
    \item To avoid interference from the RF source
    \begin{eqnarray}
    \mathbf{1}_{\tau} \mathbf{c}_k^{\rm{H}} = 0,   
    \end{eqnarray}

    \item To avoid interference from the other tags
    \begin{eqnarray}
    \mathbf{c}_{i} \mathbf{c}_k^{\rm{H}} = 0,  
    \end{eqnarray}
    where $i\in \mathcal{K}_k$.
\end{enumerate}
{The RF source serves a dual role: it transmits the RF signal allowing tags to backscatter data/pilots to the reader while also behaving as a concealed tag itself. Essentially, it functions as an additional tag emitting an all-$\mathbf{1}$ pilot sequence. It is essential to consider this aspect when designing optimal pilot schemes.} 
\end{theorem}

Next, the received signal \eqref{eqn_rx_reader_re} is transformed into a matrix form as
\begin{eqnarray}\label{eqn_estimate}
    \mathbf{Y} = \sqrt{p} \bar{\mathbf{H} } \mathbf{X} \mathbf{S} +  \mathbf{N},
\end{eqnarray}
where 
\begin{eqnarray}\label{eqn_X}
    \mathbf{X} = \begin{bmatrix}
1 & 1  & 1 & \cdots & 1_{\tau} \\
c_{11} & c_{12}  & c_{13} & \cdots & c_{1\tau} \\
c_{21} & c_{22}  & c_{23} & \cdots & c_{2\tau} \\
\vdots & \vdots  & \vdots & \ddots & \vdots \\
c_{K1} & c_{K2}  & c_{K3} & \cdots & c_{K\tau}\end{bmatrix}.
\end{eqnarray}
This signal,  $\mathbf{Y}$, can be further written as
\begin{eqnarray}\label{linearModel}
   \mathbf{y} = \mathbf{A} \bar{\mathbf{h}} + \mathbf{n},
\end{eqnarray}
where $\mathbf{y} \in \mathbb{C}^{M\tau \times 1}$, $\bar{\mathbf{h}} = \text{vec}(\bar{\mathbf{H}}) \in \mathbb{C}^{M(K+1) \times 1}$, and $\mathbf{A}  = \text{diag}{([\sqrt{p}s_1 \mathbf{1}_M, \ldots, \sqrt{p}s_{\tau}\mathbf{1}_M])} (\mathbf{X}^{\rm{T}} \otimes \mathbf{I}_M) \in \mathbb{C}^{M\tau \times M(K+1)}$. Besides, $\mathbf{n} \sim \mathcal{CN}(\mathbf{0}, \sigma^2 \mathbf{I}_{M\tau})$.  

Based upon the  linear model \eqref{linearModel}, we  derive the MVU estimator of  $\bar{\mathbf{h}}$, which provides the sufficient design criterion for $\mathbf{X}$, such that the variance of the estimator is minimized. 

\begin{theorem}\label{main_theorem}
    For the linear model  \eqref{linearModel}, with unknown deterministic variable $\bar{\mathbf{h}}$, complex Gaussian noise, i.e., $\mathbf{n} \sim \mathcal{CN}(\mathbf{0},\sigma^2 \mathbf{I}_{M\tau})$, and $\tau \ge K+1$,  the MVU estimator of the channel state $\bar{\mathbf{h}}$ is obtained as
\begin{eqnarray}\label{MVU_estimator_channel}
\hat{\bar{\mathbf{h}}} = ( \mathbf{A}^{\rm{H}}  \mathbf{A})^{-1}  \mathbf{A}^{\rm{H}} \mathbf{y},
\end{eqnarray}
which is efficient and attains the CRLB, which is a lower bound for the variance of any unbiased estimator \cite{kay1993fundamentals}. And the estimation variance is minimized when $\mathbf{X}\mathbf{X}^{\rm{H}} = \tau \mathbf{I}_{K+1}$. The covariance matrix is thus obtained as

\begin{eqnarray}\label{estimation_covariance_min}
    \mathbf{C}_{\hat{\bar{\mathbf{h}}}} = \frac{\sigma^2}{p\tau} \mathbf{I}_{M (K+1)},
\end{eqnarray}
where the estimation variance for each element of $\hat{\bar{\mathbf{h}}}$ is $\sigma^2/(p\tau)$. Therefore, increasing $\tau$ will decrease the variance of the estimates which in turn improves the estimation accuracy. 

\begin{proof}
See Appendix \ref{eqn_MVU_estimator}.
\end{proof}

\end{theorem}

\begin{rem}\label{Rem_general}
For the linear model, \eqref{linearModel}, the MVU estimator \eqref{MVU_estimator_channel} has the identical functional form as the LS estimator,  which attempts to minimize the squared distance between the given data $\mathbf{y}$ and the unknown variable $\bar{\mathbf{h}}$\cite[Section 14.3.5]{kay1993fundamentals} (Section \ref{performance_metric}).
\end{rem}

Theorem \ref{main_theorem} leads to the following theorem for a general \bc network.

\begin{theorem}\label{Theorem_backscatter}
For the  \bc network  with $K$ tags (Fig. \ref{fig_SystemModel}),  in order to  estimate $\bar{\mathbf{H}} = [\mathbf{h}_0, \sqrt{\alpha} \mathbf{h}_1, \ldots, \sqrt{\alpha} \mathbf{h}_K]$, by projecting the received signal \eqref{eqn_estimate} onto $\mathbf{S}^{\rm{H}}$, the post-processed received signal is written as
\begin{eqnarray}\label{linear_final}
    \mathbf{Y}'= \sqrt{p} \bar{\mathbf{H}}  \mathbf{X}+ \mathbf{N}', 
\end{eqnarray}
where $\mathbf{N}'= \mathbf{N} \mathbf{S}^{\rm{H}}$ and $\mathbf{X} \in \mathbb{C}^{(K+1) \times \tau}$ contains the signals of the  tags with the first row  $\mathbf{1}_{\tau}$, satisfying $\mathbf{X}\mathbf{X}^{\rm{H}} = \tau \mathbf{I}_{K+1}$, and $\mathbf{N}' \in \mathbb{C}^{M \times \tau}$ is the noise matrix  with i.i.d $\mathcal{CN}(0, \sigma^2)$ elements.

\end{theorem}

\begin{rem}\label{Remark}
  Theorem \ref{Theorem_backscatter} suggests that a  tag should not be assigned with sequence  $\mathbf{1}_{\tau}, $ which must be always assigned to the RF source.  By doing that,  the RF source is essentially treated as a hidden tag. The tags are then assigned mutual orthogonal sequences that are also orthogonal to $\mathbf{1}_{\tau}$. This strategy is different from pilot assignments to the users in conventional links, where any set of mutually orthogonal sequences can be used by the given users. Clearly, that does not work with the estimation of the direct link and the cascaded channels in \bc networks. 
\end{rem}

\begin{rem}\label{Rem_general_different_rate}
{When $R_b < R_s$, i.e., $R_b = \tau' R_s$, each tag backscatter a symbol $c_k, \forall k$ over $\tau'$ symbols of the RF source pilot signal, where $\tau'$ is an integer and $\tau' > 1$ \cite{Long2020}. Therefore, the effect of the rate inequality only appears in the matrix $\mathbf{S}$ {\eqref{eqn_estimate}}. Thus, in order to create matrix $\mathbf{X}$, $\tau \tau'$ samples of the RF source signal will be required which ultimately increases the received power, thereby improving the MSEs.}
\end{rem}

\subsection{Tag Pilot Sequence Designs}\label{sec_pilot_design}
We now present several  $\mathbf{X}$ choices that meet the design criteria in Theorem \ref{Theorem_backscatter}. These pilot sequence sets have varying levels of complexity. Depending on the processing capability of the tag and the application environment, any of the following sequences may be adopted for \bc channel estimation. 

\subsubsection{\textbf{Hadamard Matrix}}
One choice for $\mathbf{X}$ to satisfy the design constraints is the rows of the Hadamard matrix. Specifically, $\mathbf{X}$ can be the first $K+1$ rows of a Hadamard matrix of order $m$, i.e.,  $\mathbf{H}_m^{\text{h}} \in \{1,-1\}^{m \times m}$, where $m = 2^q $ and $q \ge 1$, satisfying $m \ge K+1$ \cite{horadam2012hadamard}. Thus, for the channel estimation, $\tau = m$.   Note that the assigned pilot sequences of the tags must not contain the all-$\mathbf{1}$ sequence. Hence, when assigning pilot sequences for the tag using the Hadamard matrix, the first raw should be excluded. For instance, when there are two tags, i.e.,  $K+1 =3$, we have  $m=\tau = 4$ and $\mathbf{X}$ is the first three rows of the Hadamard matrix with order 4 ($\mathbf{H}^{\rm{h}}_4$), given as \eqref{hadamard_mat}. Thereby, the second and third rows of $\mathbf{H}^{\rm{h}}_4$ are assigned to the tags. 
\begin{eqnarray}\label{hadamard_mat}
    \mathbf{X} = \begin{bmatrix}
1 & 1  & 1 & 1 \\
1 & -1  & 1 & -1 \\
1 &  1 & -1 & -1 \end{bmatrix}
\end{eqnarray}

\subsubsection{\textbf{Modified ZC Sequences}}\label{sec_ZC_sequence}
ZC sequences have/are been used for downlink (synchronization) and uplink (random access) and reference symbols (pilots)  for channel estimation in Long-Term Evolution (LTE) and is a fifth-generation (5G) new radio (NR) \cite{andrews2022primer}. 
A ZC sequence has two key parameters, (i) the root index $q= \{1,\ldots, \tau-1\}$, and (ii) the length of the sequence, $\tau$, which must be odd and often prime. The  $q$-th ZC sequence, $\mathbf{z}_q=[z_q(0),\ldots,z_q(\tau-1)]$, is defined as
\begin{eqnarray}
   z_q(n) = \exp \left(-j\pi q \frac{n(n+1)}{\tau} \right),  \quad \text{for} \quad 0\le n \le \tau-1. \quad
\end{eqnarray}

 These have the desirable constant amplitude and zero autocorrelation properties \cite{andrews2022primer}, i.e., (i)  $|z_q(n)|=1$ and (ii) each is orthogonal with cyclically-shifted versions of itself. 

However, the ZC sequences do not satisfy the design criteria for backscatter tags in Theorem \ref{thrm_condition}, i.e., $\mathbf{1}_{\tau} \mathbf{z}_q^{\rm{H}} \neq 0$ for $q= \{0,\ldots, \tau-1\}$. Hence, we next propose modified ZC sequences.  

\begin{theorem}\label{optimal_gen_theore}
Let $\mathbf{z}_k \in \mathbb{C}^{1\times \tau}$ for $k\in \mathcal{K}_0$ be any set of orthogonal sequences, i.e., $ \mathbf{z}_{k} \mathbf{z}_{k'}^{\rm{H}}  =0$ for $k\neq k'$. To use these in \bc channel estimation, all the sequences must be orthogonal to the all-$\mathbf{1}$ sequence, i.e., $\mathbf{1}_{\tau} \mathbf{z}_k^{\rm{H}} =0$ for $k \in \mathcal{K}_0$. However, since this condition does not hold generally for any set of orthogonal sequences, we modify them to satisfy the constraint as follows:
\begin{itemize}
    \item \textbf{Step 1}: Select, say,  $\mathbf{z}_0$, for convenience. 
    \item \textbf{Step 2}: Construct the matrix $\mathbf{Q}$ as follows:
    \begin{eqnarray}
        \mathbf{Q} \triangleq (\text{diag}(\mathbf{z}_0))^{-1}.
    \end{eqnarray}
    \item \textbf{Step 3}: Use the $\mathbf{Q}$ matrix to modify all other sequences as 
    \begin{eqnarray}
        \mathbf{c}_k =\mathbf{z}_k \mathbf{Q}, \quad \text{for} \quad k \in \mathcal{K}.
    \end{eqnarray}  
\end{itemize}
The modified version of $\mathbf{z}_k$, i.e., $\mathbf{c}_k$, satisfies the necessary condition $\mathbf{1}_{\tau} \mathbf{c}_k^{\rm{H}} =0$ for $k \in \mathcal{K}$. Hence, any set of orthogonal sequences can be modified to satisfy the optimal design criteria (Theorem \ref{thrm_condition}) for  \bc channel estimation. 

\end{theorem}

Following Theorem \ref{optimal_gen_theore}, we adopt modified ZC sequences for \bc channel estimation. 

\begin{exmp}\label{ZC}
 \textbf{Length \num{5} ZC sequences:}  For $K=\num{3}$, we consider $\tau = 5$ and $q = 1$. This gives a ZC sequence
\begin{eqnarray}\label{eqn_ZC_sequence}
    \mathbf{z}_0 = [1, e^{-j\frac{2\pi}{5}}, e^{-j\frac{6\pi}{5}}, e^{-j\frac{2\pi}{5}}, 1].
\end{eqnarray}
One can readily generate the sequences, i.e., $\mathbf{z}_1, \ldots, \mathbf{z}_5$, by performing cyclically-shift as follows:
\begin{subequations}
\begin{eqnarray}
    \mathbf{z}_1 &=& [1, 1, e^{-j\frac{2\pi}{5}}, e^{-j\frac{6\pi}{5}}, e^{-j\frac{2\pi}{5}} ],  \\
    \mathbf{z}_2 &=& [e^{-j\frac{2\pi}{5}}, 1, 1, e^{-j\frac{2\pi}{5}}, e^{-j\frac{6\pi}{5}} ],  \\
    \mathbf{z}_3 &=& [e^{-j\frac{6\pi}{5}}, e^{-j\frac{2\pi}{5}}, 1, 1, e^{-j\frac{2\pi}{5}} ],  \\
    \mathbf{z}_4 &=& [e^{-j\frac{2\pi}{5}}, e^{-j\frac{6\pi}{5}}, e^{-j\frac{2\pi}{5}}, 1, 1],
\end{eqnarray}
\end{subequations}
which are mutually orthogonal.
Next, we define $\mathbf{Q} = (\text{diag}(\mathbf{z}_0))^{-1}$. We then construct the pilot sequence of $T_k$ as
\begin{eqnarray}
   \mathbf{c}_k =\mathbf{z}_k \mathbf{Q}, \quad \text{for}  \quad k = \{1 , \ldots, 4\}. 
\end{eqnarray}
Therefore, the matrix $\mathbf{X}$ is given as
\begin{eqnarray}
\mathbf{X} = \begin{bmatrix}
    1 & 1  & 1 & 1 & 1\\
    e^{-j\frac{2\pi}{5}} & e^{-j\frac{4\pi}{5}}  & e^{j\frac{4\pi}{5}} & e^{j\frac{2\pi}{5}} & 1\\
    e^{-j\frac{6\pi}{5}} &  1 & e^{j\frac{6\pi}{5}} & e^{j\frac{2\pi}{5}} & e^{-j\frac{2\pi}{5}}\\
    e^{-j\frac{2\pi}{5}} &  e^{j\frac{2\pi}{5}} & e^{j\frac{6\pi}{5}} &  1 & e^{-j\frac{6\pi}{5}}\end{bmatrix},
\end{eqnarray} 
and  it can be readily checked that $\mathbf{1}_5 (\mathbf{c}_k)^{\rm{H}} = 0.$

\end{exmp}

\subsubsection{\textbf{DFT Matrix}}
The rows of DFT matrix \cite{biguesh2006training} can also be used to design the mutual orthogonal sequences of the tags, where $\tau \ge K+1$.

\begin{eqnarray}
    \mathbf{X} = \begin{bmatrix}
1 & 1  & \cdots & 1_{\tau} \\
1 & W_{\tau}  &  \cdots & W_{\tau}^{\tau-1} \\
\vdots & \vdots  &  \ddots & \vdots \\
1 & W_{\tau}^K  &  \cdots & W_{\tau}^{(\tau-1)K}\end{bmatrix},
\end{eqnarray}
where $W_{\tau} = e^{j2\pi/\tau}$.

{The Hadamard matrix is a fundamental orthogonal sequence design, comprised solely of 1s and -1s. Its versatile applications span across modern telecommunications and digital signal processing, encompassing error control coding, Walsh functions, and the generation of spread spectrum signals in CDMA {\cite{seberry2005some}}. In contrast to Hadamard sequences, ZC sequences possess a more intricate structure, characterized by unit amplitude and arbitrary phase components. These sequences are crafted from specific phase shifts of unit amplitude complex exponentials  {\cite{andrews2022primer}}. Their prominence arises from a set of highly desirable properties,  Section  II-B  of {\cite{andrews2022primer}}. ZC sequences find pivotal roles in the LTE standard and 5G NR, contributing significantly to crucial functions such as initial downlink synchronization, random access, uplink control information, and the generation of uplink reference signals (pilot signals). Similarly, the Discrete Fourier Transform (DFT) matrix also exhibits a complex-valued structure with unit amplitude, and it finds wide-ranging applications in the realms of frequency analysis, rapid convolution, and image processing {\cite{rao2018transform}}. Furthermore, the DFT matrix plays a pivotal role in designing pilot signals for channel estimation purposes  {\cite{biguesh2004downlink}} and {\cite{biguesh2006training}}. }

{Compared to the Hadamard matrix, modified ZC sequences and DFT designs can potentially reduce channel estimation duration ({$\tau$}), contingent upon the number of devices. However, this reduction comes at the cost of increased complexity for tags, owing to the intricate impedance switching necessary for load modulation \cite{Diluka2022, Rezaei2023}. Consequently, these orthogonal sequence types necessitate more intricate RF designs. While passive tags, constrained by energy and processing limitations, may not support such complex pilot sequence designs, tags equipped with greater processing capabilities and sophisticated front-end RF circuits-namely, active and semi-passive tags -- can leverage these sequences for diverse applications.}

In summary, Table \ref{tab:sequences} provides an overview of the properties and requirements associated with the proposed sequence designs.

{\linespread{1.0}
\begin{table*}[]\vspace{-0mm}
    \centering
    \caption{Pilot sequence designs.}\vspace{-0mm}
    % \scalebox{0.9}{
    \begin{tabular}{|l|l|l|}
    \hline 
     Sequence design ($\mathbf{X}$)    &  Duration ($\tau$) &  Tag's complexity \\
     \hline \hline
      Hadamard matrix   & $\tau = m (\ge K+1)$  (order of Hadamard matrix) & Simple RF design\\
      \hline
      Modified ZC sequences & $\tau \ge K+1$  (an odd and often prime number) &  Complex impedance switching\\
      \hline
      DFT matrix & $\tau \ge K+1$ & Complex impedance switching\\
      \hline
    \end{tabular}
    \label{tab:sequences}%\vspace{-0mm}
\end{table*}}

\subsection{Advantages of the Proposed Method}
\begin{itemize}
    \item The cascade channels $f_k \mathbf{g}_k$ for $ k \in \mathcal{K}$ and direct channel  $\mathbf{h}_0$ are   estimated simultaneously. 
    
    \item The cascaded channels are estimated directly, avoiding error propagation.
    
    \item  The orthogonality across the $K+1$  estimations is ensured. This is done by treating the   RF source as a hidden tag to whom an all-$\mathbf{1}$ sequence is assigned.  
    
    \item  Each channel estimate is computed over all time slots, $\tau$. Following Theorem \ref{main_theorem}, this approach will thus achieve  $\sigma^2/(\tau p)$ variance for direct and cascaded channel coefficients.  Whereas,  in the silent protocol,  the variance is  $\sigma^2 (K+1)/(\tau p)$.  Hence, our method can use shorter pilot sequences compared to the silent protocol and yet achieve the performance of long pilot sequences. 
\end{itemize}

\begin{figure*}[!t]\centering \vspace{-0mm}
 	\def\svgwidth{410pt} 
 	\fontsize{8}{8}\selectfont 
 	\graphicspath{{Figures/}}
 	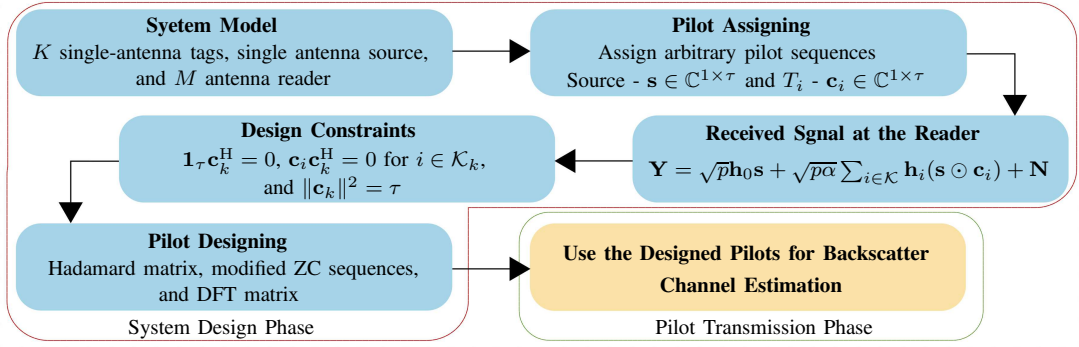 %\vspace{-0mm}
 	\caption{{The proposed channel estimation protocol.}}\label{fig_Methodology}%\vspace{-0mm} 
 \end{figure*}

{Fig.~{\ref{fig_Methodology}} provides a summary of our proposed scheme.} 

% ================================== %
% ===================================

\subsection{Deterministic \& Bayesian Channel Estimators} \label{performance_metric}

{Next, we derive multiple channel estimators, encompassing deterministic methods (LS and scaled LS) and Bayesian approaches (MMSE and LMMSE). It is important to note that the deterministic estimators assume an unknown yet deterministic channel matrix, while Bayesian estimators require the knowledge of prior probability distribution of the channels \cite{kay1993fundamentals}. Thus, MMSE and LMMSE estimators leverage prior knowledge of channel statistics to enhance estimation accuracy.}
  We define the pilot-symbol matrix as $\bar{\mathbf{X}} = \sqrt{p} \mathbf{X}$,  which is predetermined and known to the reader. These estimators of  ${\mathbf{H}}$ 
 can then be expressed  as follows\cite{biguesh2006training}: 
\begin{enumerate}
    \item \textbf{LS channel estimator:}
    This  can be obtained by using \eqref{linear_final} as follows:
        \begin{eqnarray}\label{LS_estimate}
         \hat{\bar{\mathbf{H}}}_{\rm{LS}} =  \mathbf{Y}' \bar{\mathbf{X}}^{\dagger}, 
        \end{eqnarray}
        where
        $\bar{\mathbf{X}}^{\dagger} = {\bar{\mathbf{X}}}^{\rm{H}}(\bar{\mathbf{X}}{\bar{\mathbf{X}}}^{\rm{H}})^{-1}$.
        Its versatility and effectiveness, along with its ability to function without prior channel knowledge, make this suitable for various applications.
        
    \item \textbf{Scaled LS channel estimator:} The  LS estimate \eqref{LS_estimate} may  not always  yield the minimum MSE \cite{biguesh2006training}. A solution is to optimally scale the LS channel estimate to further reduce the MSE  by allowing for a bias. This   estimator  is given as (Appendix \ref{SLS_estimator})
        \begin{eqnarray}\label{SLS}
           \hat{\bar{\mathbf{H}}}_{\rm{SLS}} = \gamma_0   \hat{\mathbf{H}}_{\rm{LS}}, 
        \end{eqnarray}
       where $\gamma_0 \!=\! \Tr\{\mathbf{R}_{\hat{\bar{\mathbf{H}}}_{\rm{LS}}}\!\}/(\sigma^2 M \Tr\{(\bar{\mathbf{X}}{\bar{\mathbf{X}}}^{\rm{H}})^{-1}\!\}+ \Tr \{\mathbf{R}_{\hat{\bar{\mathbf{H}}}_{\rm{LS}}}\!\})$,  in which $\Tr\{\!\mathbf{R}_{\hat{\bar{\mathbf{H}}}_{\rm{LS}}}\!\} \!=\! \Tr \{\!\hat{\bar{\mathbf{H}}}_{\rm{LS}}^{\rm{H}} \hat{\bar{\mathbf{H}}}_{\rm{LS}} \!\}$.
        This achieves a   lower MSE than the LS estimator  \cite{biguesh2006training}.  
    
    \item \textbf{MMSE channel estimator:}    
This  is designed to minimize the MSE. The  reader impletements it by correlating the post-processed  received  signal in \eqref{linear_final} with $\mathbf{X}$ \cite{Rezaei2020NOMA, Diluka2021}, which results in a de-spreading operation.  The output signal of this operation is thus given as
      \begin{eqnarray}\label{post_procss_Y}
    \mathbf{Y}_p = \mathbf{Y}' \mathbf{X}^{\rm{H}}/\tau = \sqrt{p} \bar{\mathbf{H}} + \mathbf{N}_p, 
\end{eqnarray}
where $\mathbf{N}_p = \mathbf{N}' \mathbf{X}^{\rm{H}}/\tau$ having i.i.d $\mathcal{CN}(0, \sigma_p^2)$ elements, where $\sigma_p^2=\sigma^2/{\tau}$. Given independent Nakagami-$\bar{m}$ fading, the elements of the channel matrix and the noise matrix are statistically independent. Next, the $(m,k)$-th element of \eqref{post_procss_Y} is given as
\begin{eqnarray}\label{post_procss_mk}
    [\mathbf{Y}_p]_{m,k} \triangleq y_{m,k} = \sqrt{p} \bar{h}_{m,k} + n_{m,k}, 
\end{eqnarray}
{where $\bar{h}_{m,k} =[\bar{\mathbf{H}}]_{m,k}$, and $n_{m,k} = [N_p]_{m,k}$.} This estimator thus becomes  {(Appendix {\ref{MMSE_estimator}})}
\begin{eqnarray}\label{mmse_h}
 \hat{\bar{h}}_{m,k} &=& \mathbb{E}\{\bar{h}_{m,k} \vert  y_{m,k} \} = \frac{\mathbb{E}\{\bar{h}_{m,k}  y_{m,k}^* \}}{ \mathbb{E}\{ \vert  y_{m,k} \vert^2 \}} y_{mk} \nonumber \\ &=&  \begin{cases}
     \frac{\sqrt{p} \beta_{m0}}{ p \beta_{m0} + \sigma_p^2} y_{m0}, \quad \text{for} \quad k=0, \\
     \frac{\sqrt{\alpha p} \beta_{mk}}{\alpha p \beta_{mk} + \sigma_p^2} y_{mk}, \quad \text{for} \quad k\in \mathcal{K}, \\
 \end{cases}
\end{eqnarray}
{where $\beta_{m,k}$ is given as}
\begin{subequations}
\begin{eqnarray}\label{eqn_beta_mk}
    \beta_{m0} &=& \frac{\Gamma(\bar{m}_{h_{0}}+1)}{\Gamma(\bar{m}_{h_{0}}) } \frac{\Omega_{h_{0}} }{\bar{m}_{h_{0}} }, \\
    \beta_{mk} &=& \frac{\Gamma(\bar{m}_{f_{k}}+1) \Gamma(\bar{m}_{g_{mk}}+1)}{\Gamma(\bar{m}_{f_{k}}) \Gamma(\bar{m}_{g_{mk}})} \frac{\Omega_{f_{k}} \Omega_{g_{mk}} }{\bar{m}_{f_{k}} \bar{m}_{g_{m k}}}.
\end{eqnarray}
\end{subequations}

The estimate of $T_k$'s effective channel, i.e., $\hat{\bar{\mathbf{h}}}_k$, is then given as $\hat{\bar{\mathbf{h}}}_k = \sqrt{\gamma_k} \mathbf{y}_k$, where $\mathbf{y}_k$ is the $k$-th column of $\mathbf{Y}_p$, and  $\gamma_k = \frac{\alpha p \beta_{k}^2}{\alpha p \beta_{k} + \sigma_p^2}$\footnote{Note that since the reader has co-located antennas, $\beta_{mk} = \beta_k$ for $m \in \mathcal{M}$.} \cite{marzetta2016fundamentals}. Thus, the estimate of the complete channel matrix, $\bar{\mathbf{H}}$,  is given as
    \begin{eqnarray}\label{MMSE}
        \hat{\bar{\mathbf{H}}}_{\rm{MMSE}} =  \mathbf{Y}_p \mathbf{D}_{\gamma}^{1/2},
    \end{eqnarray}
    where $\mathbf{D}_{\gamma} = \rm{diag}([\gamma_0, \gamma_1, \ldots, \gamma_K ])$ and $\gamma_0 = \frac{ p \beta_{0}^2}{ p \beta_{0} + \sigma_p^2}$.

       \item \textbf{LMMSE channel estimator:}
This estimator is sub-optimal but offers an ease of implementation.   Using \eqref{post_procss_Y}, the received signal from the direct channel and the cascaded channels will take the form
\begin{eqnarray}
    \mathbf{y}_{p,k} = \sqrt{p} \bar{\mathbf{h}}_k + \mathbf{n}_{p,k}, \quad \text{for} \quad k \in \mathcal{K}_0,
\end{eqnarray}
where $ \mathbf{y}_{p,k}, \bar{\mathbf{h}}_k$ and $\mathbf{n}_{p,k}$ are respectively the $k$-th column of $ \mathbf{Y}_{p}, \bar{\mathbf{H}}$ and $\mathbf{N}_{p}$.  Therefore, the LMMSE estimation of $\hat{\mathbf{h}}_k$ is expressed as (Proof \cite[Theorem 12.1]{kay1993fundamentals})
\begin{eqnarray}\label{LMMSE}    
    \hat{\bar{\mathbf{h}}}_{k} &=& \mathbb{E}\{\bar{\mathbf{h}}_k\} + \sqrt{p} \mathbf{C}_{\bar{\mathbf{h}}_k} \left( p \mathbf{C}_{\bar{\mathbf{h}}_k}  + \sigma_p^2 \mathbf{I}_{M}  \right)^{-1} \nonumber \\
    &&\times \left(\mathbf{y}_{p,k} -\sqrt{p} \mathbf{I}_M \mathbb{E}\{\bar{\mathbf{h}}_k\} \right).
\end{eqnarray}

\end{enumerate}

\begin{rem}\label{Rem_general_conf}
While this focuses on \abc, we stress that our method can also be applied to  \mbc,  \bbc, and integrated \bc that combines legacy networks (such as SR \cite{Long2020}) with newer ones. Our method can accommodate these different configurations in Figure \ref{fig:config}.  Specifically, for  \mbc, with co-located RF source and reader, the direct link disappears, removing the constraints due to all-$\mathbf{1}$ sequence.
\end{rem}

 \section{Multiple-User Multiple-Tag Scenario} \label{Multiple_user}
\begin{figure}[!t]\centering \vspace{-0mm}
 	\def\svgwidth{210pt} 
 	\fontsize{8}{8}\selectfont 
 	\graphicspath{{Figures/}}
 	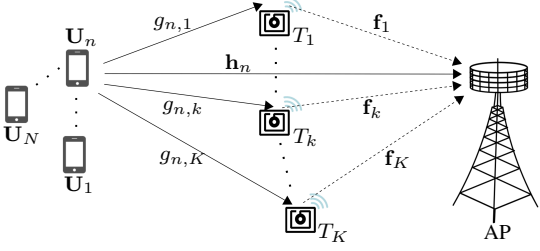 %\vspace{-0mm}
 	\caption{Cellular-based passive/ambient IoT network.} \label{fig_general_sys}
  %\vspace{-0mm}
 \end{figure}

We next present an extension of our proposed channel estimation scheme to handle multiple tags and multiple users (Fig.~\ref{fig_general_sys}). This setup comprises cellular-based passive/ambient IoT networks where multiple tags operate in the presence of multiple users. It can apply to SR networks, ambient IoT, passive IoT, and others. Such networks may find applications in smart homes, smart cities, industrial IoT, and healthcare \cite{Van2018, Toro2022, Diluka2022 }.

% ================================== %
% ===================================== %

We thus  consider a network comprising  an $M$-antenna RF source, e.g., AP, $N$ single-antenna users, i.e., $\mathbf{U}_n$ for $n \in \mathcal{N}$, and $K$ single-antenna tags, i.e., $T_k$ for $k \in \mathcal{K}$, as shown in Fig.~\ref{fig_general_sys}. Here, we will have one set of orthogonal sequences assigned to the users (e.g., pilots used for cellular users) and a subset of orthogonal sequences for a given user, which are assigned to the tags, treating the user as a hidden tag. We denote the channels between $\mathbf{U}_n$ and the AP,  $\mathbf{U}_n$ and $T_k$, and $T_k$ and the AP as $\mathbf{h}_n \in \mathbb{C}^{M \times 1}$, $g_{n,k} \in \mathbb{C}$, and $\mathbf{f_{k}} \in \mathbb{C}^{M \times 1}$, respectively, for $n \in \mathcal{N}$ and $k \in \mathcal{K}$.

{In the following, we assume that the data rate of a tag  is lower than that of the AP, i.e., the tag backscatter  one symbol over $\tau'$ symbols of the AP ($R_s = \tau' R_b$ - Remark {\ref{Rem_general_different_rate}}).} 

{Suppose that the  channel estimation phase comprises $Q$  time slots. In the $i$-th time slot, the users send mutual orthogonal pilot sequences, $\sqrt{\tau'} \mathbf{s}_n \in \mathbb{C}^{1 \times \tau'}$, where $\Vert \mathbf{s}_n \Vert^2 = 1$ and $\tau' \ge N$. Besides,  $\mathbf{s}_{n} \mathbf{s}_{n'}^{\mathrm{H}} = 0$  for $(n\neq n')$.  Meanwhile, each tag backscatters a symbol, i.e., $c_k, \forall k,$ over the users' pilot signals, $\mathbf{s}_n, \forall n$.}

The received signal at the AP over the $i$-th time slot is thus 
\begin{eqnarray}\label{received_AP}
 \mathbf{Y}_i   = \sqrt{\tau'p} \sum \nolimits_{n \in \mathcal{N} } {\bar{\mathbf{H}}_n} \mathbf{x}_i   \mathbf{s}_n + \mathbf{N}_i,
\end{eqnarray}
 where ${\bar{\mathbf{H}}_n} = [\mathbf{h}_n, \alpha \mathbf{f}_1 g_{n,1}, \ldots, \alpha \mathbf{f}_K g_{n,K}] $, $\mathbf{x}_i =[1, c_1, c_2, \ldots,c_K]^{T} \in \mathbb{C}^{(K+1)\times 1}$, and $\mathbf{N}_i \in \mathbb{C}^{M \times \tau'}$ is the AWGN matrix at the AP with i.i.d $\mathcal{CN}(0, \sigma^2)$ elements.

% where ${\bar{\mathbf{H}}_n} = [\mathbf{h}_n, \alpha \mathbf{f}_1 g_{n,1}, \ldots, \alpha \mathbf{f}_K g_{n,K}] $, $\mathbf{x}_i =$ is the $i$-th column of matrix $\mathbf{X}$, and $\mathbf{N}_i \in \mathbb{C}^{M \times \tau'}$ is the AWGN matrix at the AP with i.i.d $\mathcal{CN}(0, \sigma^2)$ elements.
 
%Meanwhile, the tags adopt the entries of the $i$-th column of matrix $\mathbf{X} \in \mathbb{C}^{(K+1) \times Q}$ \eqref{eqn_X}, excluding first element.

To estimate channels, the received pilot signal at the  AP \eqref{received_AP} is projected onto ${\bm{s}}_{n}^{\rm{H}}$ for $n \in \mathcal{N}$ which yields \cite{Rezaei2020NOMA,Diluka2021}
\begin{eqnarray}\label{projection}
  \mathbf{\tilde{y}}_{n,i}^p = \sqrt{\tau' p} {\bar{\mathbf{H}}_n} \mathbf{x}_i+ \tilde{\mathbf{n}}_{i,n},  
\end{eqnarray}
 where $\tilde{\mathbf{n}}_{i,n} = \mathbf{N}_i {\bm{s}}_{n}^{\rm{H}} \sim \mathcal{CN}(\mathbf{0}, \sigma^2\mathbf{I}_M)$.

{Using {\eqref{projection}}, it is crucial to create matrix $\mathbf{X} = [\mathbf{x}_1, \ldots, \mathbf{x}_i, \ldots, \mathbf{x}_Q] \in {\mathbb{C}}^{(K+1)Q}$ (Theorem {\ref{Theorem_backscatter}})  to enable the reader to estimate the direct channel and cascaded channels per user. Hence,  at least $K+1$ slots are required, i.e., $Q\ge K+1$.} Therefore, using \eqref{projection}, the received signal over $Q$ time slots can be collected as
 \begin{eqnarray}\label{linear_multi}
    \mathbf{\tilde{Y}}_{n}^p = {\bar{\mathbf{H}}_n} \sqrt{\tau' p} 
 \mathbf{X}+ \tilde{\mathbf{N}}_{n},
 \end{eqnarray}
 where $\tilde{\mathbf{N}}_{n} = [\tilde{\mathbf{n}}_{1,n}, \ldots, \tilde{\mathbf{n}}_{Q,n}] \in \mathbb{C}^{M \times Q}$.

The obtained equation in \eqref{linear_multi} is similar to \eqref{linear_final}. Following the same principles, Theorem \ref{Theorem_backscatter}, and Section \ref{sec_pilot_design}, designs for $\mathbf{X}$ could the first $K+1$ rows of a Hadamard matrix of order $m\ge K+1$ with $Q = m$, modified ZC sequences, or DFT matrix. 
Moreover, using \eqref{Covariance_matrix}, the estimation variance for each element of $\bar{\mathbf{H}}_n$ is $\sigma^2/(p \tau' Q)$, and the duration of the training phase is $\tau =\tau' Q$.  Next, from  \eqref{linear_multi},  a channel estimator aims to  recover $\bar{\mathbf{H}}_n$ for $n \in \mathcal{N}$, by exploiting $\mathbf{X}$ and $ \mathbf{\tilde{Y}}_{n}^p$.  Therefore, $\hat{\bar{\mathbf{H}}}_n$ can be obtained using the estimation methods in  Section \ref{performance_metric}.

\section{Simulation Results}\label{sim}

\begin{table}\vspace{-0mm}
\caption{Simulation settings.} \vspace{-0mm} % title of Table
\label{tab:sim_para}
\centering % used for centering table
\scalebox{0.9}{
\begin{tabular}{c c c c} % centered columns (4 columns)
\hline %inserts double horizontal lines
Parameter & Value & Parameter & Value \\ [0.5ex] % inserts table
%heading
\hline \hline% inserts single horizontal line
$f_c$ & \qty{3}{\GHz}  &$d_{h_0}$ & \qty{10}{\m} \\ % inserting body of the table
$B$ & \qty{10}{\MHz} & $d_{f_k}, k \in \mathcal{K}$ & $\mathcal{U}(5,7)$\qty{ }{\m}\\
$N_f$ & \qty{20}{\dB}   & $d_{g_k}, k \in \mathcal{K}$ & \qty{6}{\m}\\
$\bar{m}$ & \num{3} & $M$ & \num{10}   \\
\hline %inserts single line
\end{tabular}}
\label{table:nonlin} % is used to refer this table in the text
\vspace{-0mm}
\end{table}

 We next present and discuss extensive simulation and numerical results to validate the proposed estimators.  To evaluate their performance, we use a  channel model that accounts for both large-scale and small-scale fading, ensuring a realistic representation of the wireless communication environment. Large-scale fading is modeled by the the  3GPP urban micro (UMi), i.e., $\zeta_{\mathbf{a}}$ for $a \in \mathcal{A}$, with carrier frequency $f_c$  \cite[Table B.1.2.1]{3GPP2010}. Moreover, the AWGN variance is modeled as $\sigma^2=10\log_{10}(N_0 B N_f)$ dBm, where $N_0=\qty{-174}{\dB m/\Hz}$, $B$ is the bandwidth, and $N_f$ is the noise figure. Table \ref{tab:sim_para} gives the simulation parameters.  Unless otherwise specified, we use the Hadamard sequences for performance evaluations.  Without loss of generality, we set $\alpha =0.6$. 

 %========================================== 
\begin{figure*}[!t]\vspace{-0mm}
\centering
\fontsize{12}{12}\selectfont 
    \begin{minipage}{.5\textwidth}
      \centering
      \resizebox{.57\totalheight}{!}{\input{FiguresTxt/DC_MSE_T_8_K_7.tex}}\vspace{-0mm}
      \subcaption{Direct channel.}
      \label{DC_MSE_T_8_K_7}
    \end{minipage}%
    \begin{minipage}{.5\textwidth}
      \centering
      \resizebox{.57\totalheight}{!}{\input{FiguresTxt/BC_MSE_T_8_K_7.tex}}\vspace{-0mm}
      \subcaption{Cascaded channel.}
      \label{BC_MSE_T_8_K_7}
    \end{minipage}\vspace{-0mm}
\caption{Channel estimation for $\tau=8$ and $K=7$.  }\label{fig_T8_K_7} \vspace{-0mm}
\end{figure*}
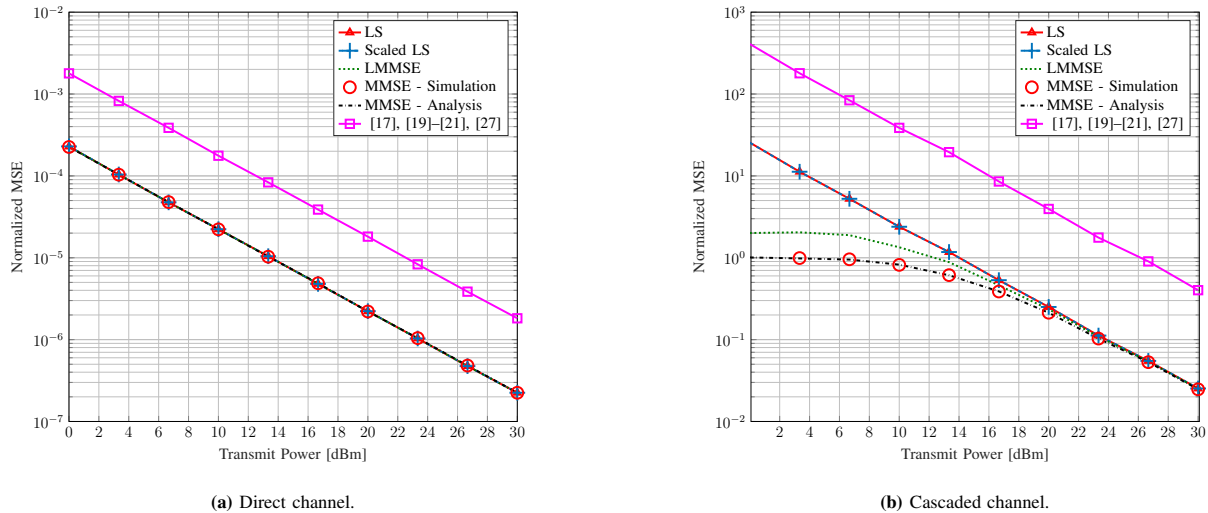
%======================================

We use the linear estimators, i.e.,  LS \eqref{LS_estimate}, scaled LS \eqref{SLS}, and LMMSE \eqref{LMMSE} to estimate the direct channel ($\mathbf{h}_0$) and the cascaded channels ($f_k \mathbf{g}_k, k \in \mathcal{K}$). 
{Note that the MVU estimator matches the LS estimator in the simulation (Remark {\ref{Rem_general}}).}  To benchmark the performance of our channel estimators,  we consider the prior art \cite{Abdallah2021,  Zhao2019, Liu2021, Zhu2018,  Yerzhanova2021}. Here, we evaluate the LMMSE estimator with phase-splitting as described in Section \ref{Prior_art}. 

The quality of channel estimators is widely assessed in terms of   normalized MSE, which  is defined as 
\begin{eqnarray}\label{eqn_mse_def}
    \text{Normalized MSE}_k = \mathbb{E} \left\{ \frac{ \left\Vert  \mathbf{h}_k - \hat{\mathbf{h}}_k \right \Vert^2_2} {\Vert \mathbf{h}_k  \Vert^2_2 }\right\}, \quad  k \in {\mathcal{K}_0},
\end{eqnarray}
where $ \hat{\mathbf{h}}_k$ is the $k$-th column of $\hat{\bar{\mathbf{H}}}_{\eta},$ where $ \eta \in \{\text{LS, scaled LS, LMMSE, MMSE} \}$. {Note that, the analytical  MSE of MMSE estimator (legend MMSE-analysis) is plotted via by substituting $\hat{\mathbf{h}}_k$ in {\eqref{mmse_h}} into {\eqref{eqn_mse_def}}, whereas the simulated MSE of MMSE estimator (legend MMSE-simulation)  is plotted by evaluating $[\hat{\bar{h}}_{m,k}]_{\rm{Simulation}} = \mathbb{E}\{\bar{h}_{m,k} \vert  y_{m,k} \}$ through Monte-Carlo simulations to validate the derived analytical expression.}

%========================================== 
\begin{figure*}[!t]%\vspace{-0mm}
\centering
\fontsize{12}{12}\selectfont 
    \begin{minipage}{.5\textwidth}
      \centering
      \resizebox{.57\totalheight}{!}{\input{FiguresTxt/DC_MSE_T_16_K_7_15.tex}}\vspace{-0mm}
      \subcaption{Direct channel.}
      \label{DC_MSE_T_16_K_7_15}
    \end{minipage}%
    \begin{minipage}{.5\textwidth}
      \centering
      \resizebox{.57\totalheight}{!}{\input{FiguresTxt/BC_MSE_T_16_K_7_15.tex}}\vspace{-0mm}
      \subcaption{Cascaded channel.}
      \label{BC_MSE_T_16_K_7_15}
    \end{minipage}\vspace{-0mm}
\caption{Channel estimation for $\tau=16$ and $K=\{7,15\}$.  }\label{fig_T16_K_7_15} %\vspace{-0mm}
\end{figure*}
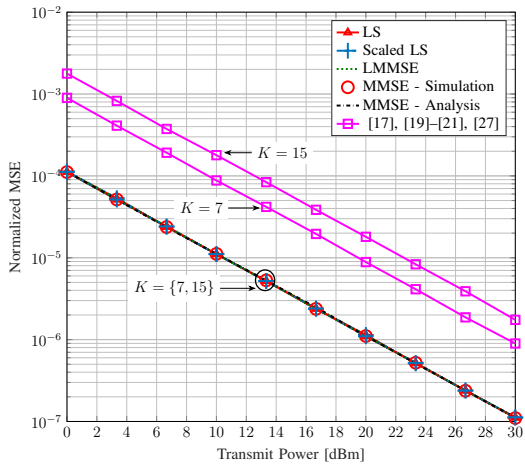
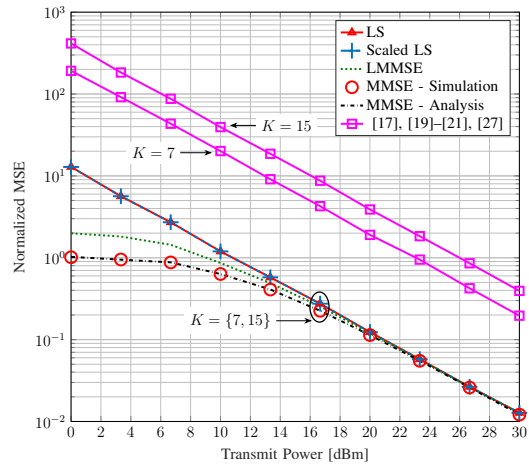
%======================================

Fig. \ref{fig_T8_K_7} shows the normalized MSE performance of different estimators versus the transmit power for the direct channel and cascaded channels, for $\tau = 8$ and $K=7$ tags. With seven tags, the silent method \cite{Abdallah2021,  Zhao2019, Liu2021, Zhu2018,  Yerzhanova2021} uses one pilot symbol ($t =1$) to estimate each channel, i.e., the direct channel $\mathbf{h}_0$ or the combined channel $\mathbf{h}_0 + \alpha f_k\mathbf{g}_k $ for $ k \in \mathcal{K}$. 

Our proposed method demonstrates remarkable accuracy in estimating both direct and cascade channels for multiple tag scenarios. It significantly outperforms the silent protocol.  
{Moreover, due to the double path loss in the cascaded channel, it excels at estimating the direct link channel more accurately than the cascaded channel (Fig. {\ref{DC_MSE_T_8_K_7}} and Fig. {\ref{BC_MSE_T_8_K_7}}).}
% Especially, it excels at accurately estimating the direct link channel (Fig. \ref{DC_MSE_T_8_K_7}). 
Additionally, our linear estimators deliver comparable performance to the optimal MMSE estimator. 

Furthermore, we achieve  a normalized MSE of $10^{-5}$ at  \qty{14}{\dB m} transmit power, whereas the silent method \cite{Abdallah2021,  Zhao2019, Liu2021, Zhu2018,  Yerzhanova2021}  requires  \qty{10}{\dB m} more to achieve the same MSE. The primary reason for this massive power savings is that our method utilizes the entire pilot length for each channel while avoiding pilot contamination, whereas the silent case allocates a portion of the pilot sequence to each channel.

It is also worth noting that the traditional LS (or scaled LS) method outperforms the LMMSE estimation method with the silent protocol. 
The estimation quality of all the cascaded channels is almost the same (Fig. \ref{BC_MSE_T_8_K_7}), independent of the number of tags. The gap between the proposed method and the silent method using the LMMSE estimator is even larger than that of the direct channel. This is because the silent method suffers from error propagation (Section \ref{Prior_art}). In particular,  compared to our method, the silent case requires $\sim \qty{13}{\dB m}$ more transmit power to achieve the normalized MSE of $10^{-1}$. Moreover, with our method, the simple LS (or scaled LS), significantly outperforms the LMMSE estimation method using the silent method. We note that LS and scale LS achieve the same performance. This is because, for the latter, the scale parameter $\gamma_0 \sim 1$ for our setup. Moreover, the proposed  MMSE estimator performs better in the low transmit power regime due to the use of second-order statistics.

Note that the cascaded channel normalized MSE has high values when compared to the direct channel normalized MSE. This is primarily due to the double path loss present in the cascaded channel and the tag's reflection coefficient  $\alpha$.

%========================================== 
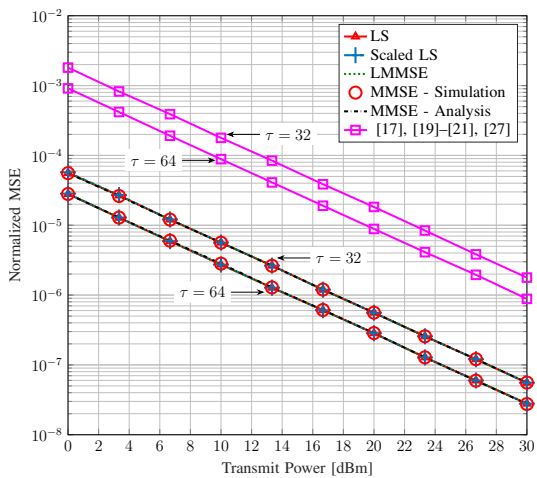
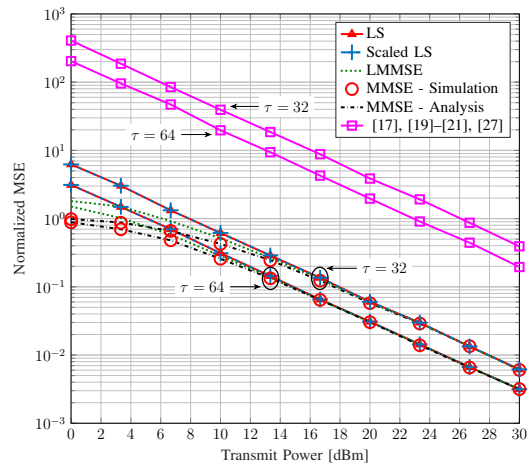
\begin{figure*}[!t]\vspace{-0mm}
\centering
\fontsize{12}{12}\selectfont 
    \begin{minipage}{.5\textwidth}
      \centering
      \resizebox{.57\totalheight}{!}{\input{FiguresTxt/DC_MSE_T_32_64_K_31.tex}}\vspace{-0mm}
      \subcaption{Direct channel.}
      \label{DC_MSE_T_32_64_K_31}
    \end{minipage}%
    \begin{minipage}{.5\textwidth}
      \centering
      \resizebox{.57\totalheight}{!}{\input{FiguresTxt/BC_MSE_T_32_64_K_31.tex}}\vspace{-0mm}
      \subcaption{Cascaded channel.}
      \label{BC_MSE_T_32_64_K_31}
    \end{minipage}\vspace{-0mm}
\caption{Channel estimation for $\tau=\{32,64\}$ and $K=31$. }\label{fig_T_32_64_K_31} %\vspace{-0mm}
\end{figure*}
%======================================

We anticipate a certain relationship between the channel estimation quality and the pilot length, as well as the number of tags.  These relationships are uncovered in Fig. \ref{fig_T16_K_7_15} and Fig. \ref{fig_T_32_64_K_31}.  In Fig. \ref{fig_T16_K_7_15}, we set $\tau = 16$ and plot the normalized MSE for $K = 7$ and $K=15.$ In particular, Fig.~\ref{DC_MSE_T_16_K_7_15} and Fig.~\ref{BC_MSE_T_16_K_7_15} investigate the normalized MSE for the direct channel and cascaded channel, respectively. As expected, increasing the number of tags adversely affects the performance of the silent case. In particular, for $K = 15$, the silent case uses one pilot symbol to estimate each of the direct channel and the cascaded ones, while for $K=7$, it allocated two pilots per link which results in better channel estimates. Therefore, with a large number of tags, the prior art has to use an excessive amount of the channel coherence interval to accurately estimate the channels. 
On the other hand, our channel estimation proposal achieves the same performance independent of the number of tags for fixed $\tau$, and significantly outperforms the silent case. In particular, for $K=15$, we observe the power gain of $\sim$\qty{13}{\dB m} and $\sim$\qty{16}{\dB m} to achieve the normalized MSE of $10^{-5}$ and $10^{-1}$ for respectively the direct link and cascaded links. Therefore, as the network scales up, our proposed scheme can efficiently handle CSI estimation.

Fig.~\ref{DC_MSE_T_32_64_K_31} and Fig.~\ref{BC_MSE_T_32_64_K_31} examine the impact of pilot sequence length on the normalized MSE of the channel estimates for a fixed number of tags, $K=32$. The results show that longer pilot sequences lead to better estimates of the direct and cascaded channels. For example, increasing the pilot sequence length from $\tau=32$ to $\tau=64$ results in a transmit power gain of approximately $\sim\qty{2.5}{\dB m}$ for the cascaded channel in the mid-to-high transmit power regime. Furthermore, the proposed scheme outperforms the silent case in terms of normalized MSE, even though both show a similar power gain.

%========================================== 
\begin{figure*}[!t]\vspace{-0mm}
\centering
\fontsize{12}{12}\selectfont 
    \begin{minipage}{.5\textwidth}
      \centering
      \resizebox{.57\totalheight}{!}{\input{FiguresTxt/DC_tau_K2.tex}}\vspace{-0mm}
      \subcaption{Direct channel.}
      \label{DC_tau_K2}
    \end{minipage}%
    \begin{minipage}{.5\textwidth}
      \centering
      \resizebox{.57\totalheight}{!}{\input{FiguresTxt/BC_tau_K2.tex}}\vspace{-0mm}
      \subcaption{Cascaded channel.}
      \label{BC_tau_K2}
    \end{minipage}\vspace{-0mm}
\caption{Channel estimation for $p=\qty{20}{\dB m}$,  $K=2$, and $\alpha=0.6$. }\label{fig_tau_Com} \vspace{-0mm}
\end{figure*}
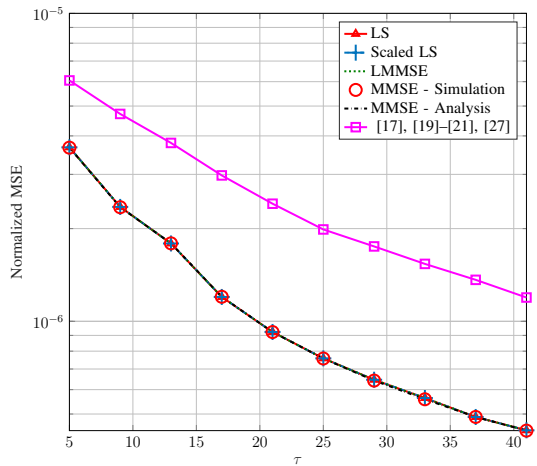
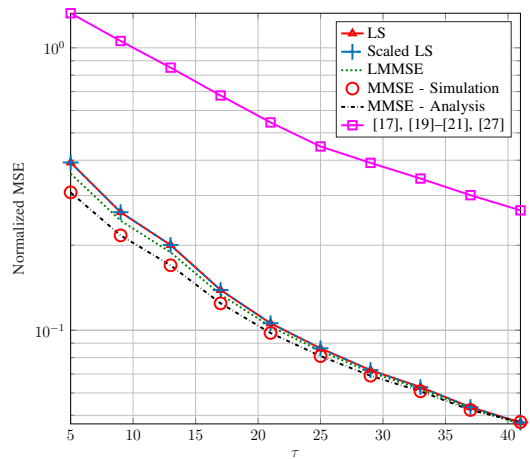
%======================================

Additionally, Fig \ref{DC_tau_K2} and Fig. \ref{BC_tau_K2} plot the normalized MSEs of direct and cascaded channels against the pilot length for $K =2$ and $p = \qty{20}{\dB m}$. As expected, increasing $\tau$ improves the channel estimation accuracy and our method significantly outperforms the prior art. It is worth noting that, unlike the prior art, our proposal achieves the same accuracy for any number of tags satisfying $\tau \ge K+1$.

 %========================================== 
\begin{figure*}[!t]\vspace{-0mm}
\centering
\fontsize{12}{12}\selectfont 
    \begin{minipage}{.5\textwidth}
      \centering
      \resizebox{.57\totalheight}{!}{\input{FiguresTxt/DC_T_63_K}}\vspace{-0mm}
      \subcaption{Direct channel.}
      \label{DC_T_63_K}
    \end{minipage}%
    \begin{minipage}{.5\textwidth}
      \centering
      \resizebox{.57\totalheight}{!}{\input{FiguresTxt/BC_T_63_K}}\vspace{-0mm}
      \subcaption{Cascaded channel.}
      \label{BC_T_63_K}
    \end{minipage}\vspace{-0mm}
\caption{Channel estimation for $p=\qty{20}{\dB m}$,  $\tau=64$, and $\alpha=0.6$.}\label{fig_K_Com} %\vspace{-0mm}
\end{figure*}
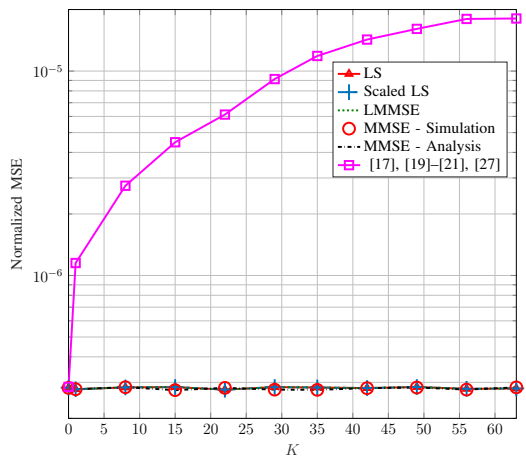
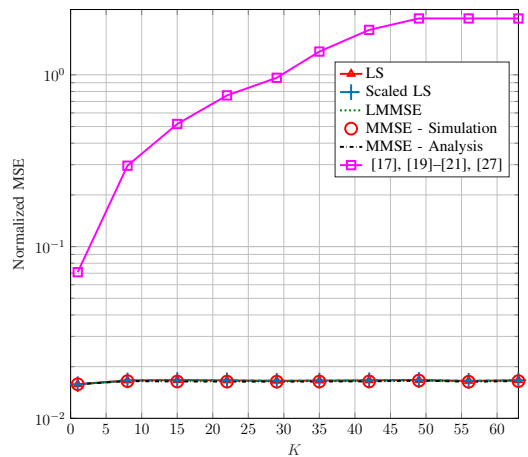
%====================================== 

In Fig.~\ref{fig_K_Com}, we plot the normalized MSE as a function of the number of tags, $K$, for $\tau=64$ and $p=  \qty{20}{\dB m}$. Since our method is independent of the number of tags, the normalized MSE remains constant, whereas the performance of the silent protocol depends on $K$. When there is no tag ($K=0$), our protocol and the silent one have the same normalized MSE because the entire pilot duration is allocated to the direct channel estimation (Fig.~\ref{DC_T_63_K}). However, as $K$  increases, the silent protocol assigns $ \tau/(K+1)$ estimation time for each channel gain and is plagued by error propagation.  For these reasons,  the performance gap, $G_p$, increases (Fig.~\ref{DC_T_63_K} and Fig.~\ref{BC_T_63_K}). From numerical experiments, we find that  $G_p$ can be modeled as
\begin{eqnarray}
    G_p = \lambda_G \tau \left(1-\frac{1}{K+1} \right), \quad \text{for} \quad 0\leq K \leq \tau-1,
\end{eqnarray}
where $\lambda_G$ accounts for the other system parameters, i.e., transmit power, noise variance,  tag's reflection coefficient, etc.

%========================================== 
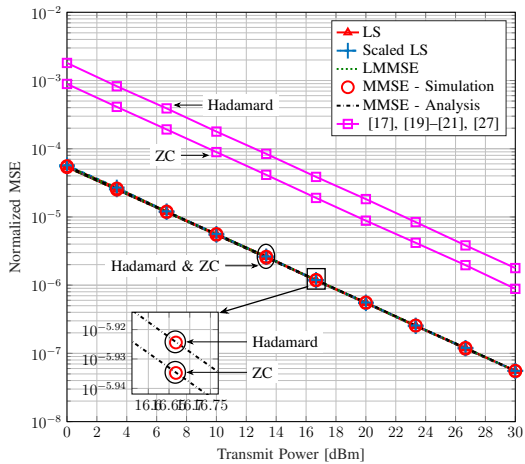
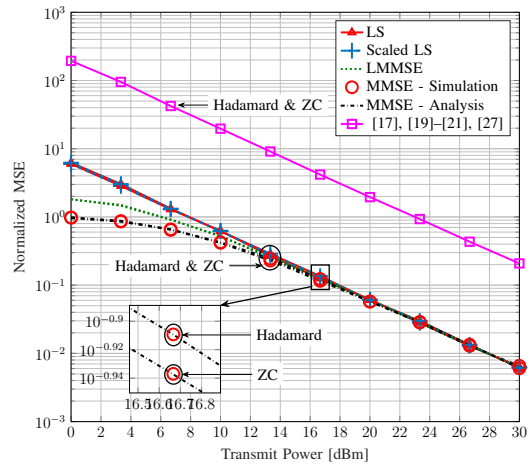
\begin{figure*}[!t]\vspace{-0mm}
\centering
\fontsize{12}{12}\selectfont 
    \begin{minipage}{.5\textwidth}
      \centering
      \resizebox{.57\totalheight}{!}{\input{FiguresTxt/DC_MSE_T_32_33_K_31_Had_ZC.tex}}\vspace{-0mm}
      \subcaption{Direct channel.}
      \label{DC_MSE_T_32_33_K_31_Had_ZC}
    \end{minipage}%
    \begin{minipage}{.5\textwidth}
      \centering
      \resizebox{.57\totalheight}{!}{\input{FiguresTxt/BC_MSE_T_32_33_K_31_Had_ZC.tex}}\vspace{-0mm}
      \subcaption{Cascaded channel.}
      \label{BC_MSE_T_32_33_K_31_Had_ZC}
    \end{minipage}\vspace{-0mm}
\caption{Performance comparison between Hadamard and ZC sequences for $K=32$. For Hadamard, $\tau=32$, and for ZC, $\tau=33$. }\label{fig_Had_ZC_Com} \vspace{-0mm}
\end{figure*}
%======================================

 It is interesting to see if different pilot sequences make a difference. To this end, Fig.~\ref{fig_Had_ZC_Com} compares   Hadamard and ZC sequences for  $K=32$. To ensure a fair comparison, we choose $\tau=32$ for  Hadamard and $\tau=33$ for  ZC. Consequently, when using the latter in the silent method, the additional pilot is used for direct channel estimation (Fig.~\ref{DC_MSE_T_32_33_K_31_Had_ZC}), while each cascaded channel is estimated using a single pilot (Fig.~\ref{BC_MSE_T_32_33_K_31_Had_ZC}). As a result, with the silent protocol, the ZC sequence outperforms the Hadamard sequence in terms of normalized MSE for the direct channel. The proposed channel estimation scheme with the ZC sequence, on the other hand, has slightly better performance in both the direct and cascaded channels, due to the use of one additional pilot symbol (enlarged portions of Fig.~\ref{DC_MSE_T_32_33_K_31_Had_ZC} and Fig.~\ref{BC_MSE_T_32_33_K_31_Had_ZC}). Accordingly, the performance of any set of pilot sequences that meet the optimal design criterion - Theorem \ref{Theorem_backscatter} and Theorem  \ref{optimal_gen_theore} -  will be the same for fixed $\tau$.

%========================================== 
\begin{figure*}[!t]\vspace{-0mm}
\centering
\fontsize{12}{12}\selectfont 
    \begin{minipage}{.5\textwidth}
      \centering
      \resizebox{.57\totalheight}{!}{\input{FiguresTxt/DC_MSE_T_16_K_7_A_6_8.tex}}\vspace{-0mm}
      \subcaption{Direct channel.}
      \label{DC_MSE_T_16_K_7_A_6_8}
    \end{minipage}%
    \begin{minipage}{.5\textwidth}
      \centering
      \resizebox{.57\totalheight}{!}{\input{FiguresTxt/BC_MSE_T_16_K_7_A_6_8.tex}}\vspace{-0mm}
      \subcaption{Cascaded channel.}
      \label{BC_MSE_T_16_K_7_A_6_8}
    \end{minipage}\vspace{-0mm}
\caption{Channel estimation for $\tau=16$, $K=7$, and $\alpha=\{0.6,0.8\}$.  }\label{fig_alpha_Com} \vspace{-0mm}
\end{figure*}
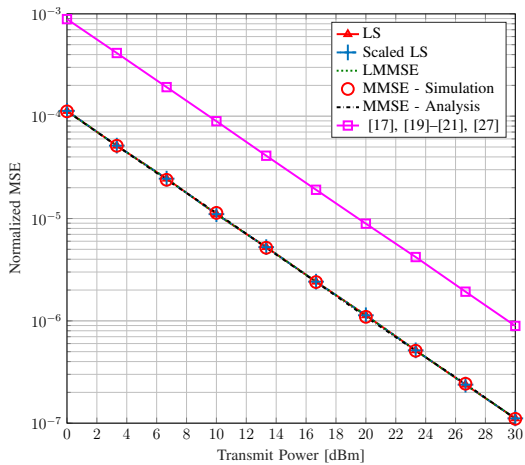
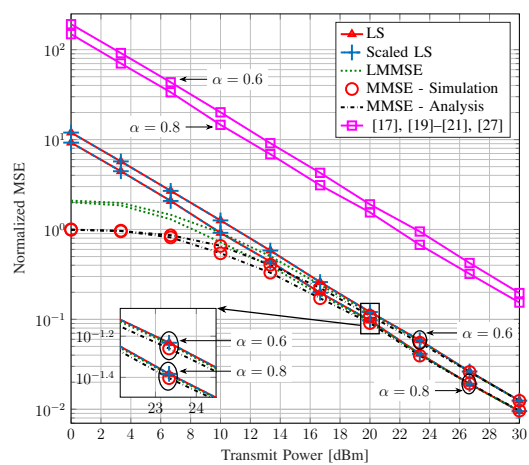
%======================================

It is interesting to see if the amount of power reflected by the tags has an impact on the channel estimation quality. We test this possibility in  Fig. \ref{fig_alpha_Com} and consider $\tau =16$, $K = 7$, with  $\alpha = \{0.6, 0.8 \}$. The estimation quality of the direct link is independent of $\alpha$ (Fig. \ref{DC_MSE_T_16_K_7_A_6_8}). However, for the cascaded links, higher $\alpha$ increases the reflected power at each tag, improving the estimation quality. Nonetheless, $\alpha$ should be set carefully to keep the passive tag functional and maintain a balance between the harvested and the reflected power (Section \ref{EH_data}).

\section{Conclusion}\label{conclusion}
The prior solutions to the \bc channel estimation problem use the silent protocol, where the tags cycle through active and silent periods. It has several, critical drawbacks (Section \ref{Prior_art}). {In this paper, we have developed a new protocol for the time spreading of tag pilots over the entire estimation interval.  It enables the simultaneous estimation of both direct and cascaded links, utilizing the whole training phase for each channel, and avoids error propagation inherent in prior sequential estimation methods.    It thus offers remarkable power reductions of at least {\qty{10}{\dB m}} (direct) and {\qty{12}{\dB m}} (cascaded) compared to the silent protocol, respectively (Fig. {\ref{fig_T8_K_7}}). This gap can be attributed to the following factors: (i) direct estimation of cascaded channels avoids error propagation, (ii) treating the RF source as a hidden tag avoids pilot contamination,  and (iii)  time-spread pilots utilize time and energy resources efficiently.   In contrast to the prior art, the proposed method can readily accommodate any number of tags. }    

Before discussing the future directions, we summarize the contributions of this study. Having set up the time-spread pilots, we derived the MVU  channel estimator and developed the pilot designs to avoid pilot contamination.  Using the designs, an analytical MMSE estimator was also derived and exploited for performance evaluations.  The proposed method was also extended for cellular-based passive IoT, where multiple tags operate along with multiple cellular users.

The future extensions of this work are multi-faceted and have the potential to significantly enhance the current state of the research. Firstly, it is essential to perform separate forward, $f_k$, and backscatter, $\mathbf{g}_k$, channel estimations to facilitate energy beamformer designs for waking up tags and other related applications. While prior work has been limited in investigating this issue, recent research such as \cite{Shuo2018} has made progress in obtaining the modulus values  $\vert f_k \vert$ and $\vert {g}_k \vert$. Hence, our proposed technique can further improve the accuracy of these estimates.

{Secondly, the quality of channel estimate has an immense effect on backscatter performance analysis (e.g., bit-error rate/symbol error rate, outage probability, and achievable rate) and resource allocation (e.g., beamforming design, reflection coefficient optimization, and others). Hence, the future extensions of this study will include performance analysis and resource allocation frameworks based on the proposed channel estimation technique. }

Thirdly, the proposed time-spread pilot-based channel estimation technique can be utilized for RIS and integrated systems, such as RIS-assisted \bc, integrated sensing and \bc, and others. This approach is particularly useful for passive devices that do not generate RF signals, and it has the potential to enhance the efficiency and effectiveness of these systems.

{Finally,  hardware impairments can also impact the quality of channel estimation. The RF  source and reader have  RF front-ends that are vulnerable to amplifier non-linearities, in/quadrature -phase imbalance, phase noise, quantization error, etc. \cite{Schenk2008,  Xingwang2020, Xingwang2021}.  Antenna impedance mismatching at the tags affects the reflection coefficient and EH. Synchronization/timing errors and hardware impairments have been scarcely investigated. Thus, these challenges inspire further research.} {Deep learning-based techniques may help in this context \cite{Liu2021Deep}. }

Overall, our channel estimation approach shows great promise in enhancing the accuracy and reliability of \bc networks, RIS, and other integrated networks. By proposing extensions to this work, we aim to push the boundaries of the current state-of-the-art in this field and facilitate the development of passive tag-based communication solutions. 

% \vspace{-0mm}
\appendices
\section{MVU Estimator for Linear Model}\label{eqn_MVU_estimator}
\vspace{-0mm}
 For an  estimator of   $\theta,$ the MSE is  defined as \cite[Section 2.4]{kay1993fundamentals}
 
\begin{eqnarray}\label{mseError}
   \text{MSE}(\hat{\theta}) = \mathbb{E} \{ (\hat{\theta}-{\theta})^2\}= \text{Var}(\hat{\theta}) + b^2(\theta),
\end{eqnarray}
where $b(\theta) = \mathbb{E} \{\hat{\theta} \} - \theta$. Thus, the  MSE  can be decomposed into bias and variance. An estimator is unbiased if, on average, it yields the true value,  $\mathbb{E} {[ \hat{\theta}] } = \theta$. Among unbiased estimators, the one with the minimum variance (MVU estimator) is desirable since its variance, $\text{Var}(\hat{\theta})$, is the smallest among all unbiased estimators. 
\begin{theorem}\label{theorem_MVU}
   \cite[Theorem 4.1]{kay1993fundamentals} If the data observed can be modeled as 
\begin{eqnarray}\label{LinearModel_Kay}
   \mathbf{y}  = \mathbf{D}\bm{\theta} + \mathbf{w},
\end{eqnarray}
    where $\mathbf{y}$ is an $N \times 1$ vector of observations, $\mathbf{D}$ is a known $N \times p$  matrix (with $N>p$) with rank $p$, $\bm{\theta}$ is a $p \times 1$ vector of parameters to be estimated, and $\mathbf{w}$ is an $N \times 1$ noise vector with PDF $\mathcal{CN}(\mathbf{0}, \sigma^2 \mathbf{I})$, then the MVU estimator is 
\begin{eqnarray}
\hat{\bm{\theta}} =  ( \mathbf{D}^{\rm{H}}  \mathbf{D})^{-1}  \mathbf{D}^{\rm{H}} \mathbf{y},
\end{eqnarray}
And the covariance matrix of $\hat{\bm{\theta}}$ is
\begin{eqnarray}
\mathbf{C}_{\hat{\bm{\theta}}} = \sigma^2  ( \mathbf{D}^{\rm{H}}  \mathbf{D})^{-1}.
\end{eqnarray}

For the linear model \eqref{LinearModel_Kay}, the MVU estimator attains the CRLB.  Therefore, we have $\mathbf{C}_{\hat{\bm{\theta}}} = \mathcal{I}^{-1}(\bm{\theta})$, where $\mathcal{I}(\bm{\theta})$ is the Fisher information matrix, satisfying \cite[Theorem 3.2]{kay1993fundamentals}
\begin{eqnarray}\label{LowerBound}
    [\mathbf{C}_{\hat{\bm{\theta}}}]_{i,i} = [ \mathcal{I}^{-1}(\bm{\theta})]_{i,i}\ge \frac{1}{[\mathcal{I}(\bm{\theta})]_{i,i}}.
\end{eqnarray}
And equality holds  when $\mathcal{I}^{-1}(\bm{\theta})$ is diagonal \cite[Ex. 3.12]{kay1993fundamentals}.
\end{theorem}
Using Theorem \ref{theorem_MVU}, the MVU estimator of  \bc channels is obtained as \eqref{MVU_estimator_channel}, and the covariance matrix is given as 
\begin{eqnarray}\label{Covariance_matrix}
\mathbf{C}_{\hat{\bar{\mathbf{h}}}} &=&  \sigma^2 ( \mathbf{A}^{\rm{H}}  \mathbf{A})^{-1} \nonumber \\
&=& \sigma^2 \left((\mathbf{X}^{\rm{T}} \otimes \mathbf{I}_M)^{\rm{H}}  \mathbf{B}^{\rm{H}} \mathbf{B}(\mathbf{X}^{\rm{T}} \otimes \mathbf{I}_M)\right)^{-1}\nonumber \\
&=& \frac{\sigma^2}{p}  \left({\mathbf{X}^{\rm{T}}}^{\rm{H}}  {\mathbf{X}^{\rm{T}}} \otimes \mathbf{I}_M \right)^{-1} \nonumber \\
&=& \frac{\sigma^2}{p} \left({\mathbf{X}^{\rm{T}}}^{\rm{H}}  {\mathbf{X}^{\rm{T}}} \right)^{-1} \otimes \mathbf{I}_M,
\end{eqnarray}
where $\mathbf{B} = \text{diag}{([\sqrt{p}s_1 \mathbf{1}_M, \ldots, \sqrt{p}s_{\tau}\mathbf{1}_M])}$.

Using \eqref{Covariance_matrix}, the lower bound in \eqref{LowerBound} can be attained when $ {\mathbf{X}}{\mathbf{X}}^{\rm{H}} $ is diagonal. This implies ${\mathbf{X}}$ has equally scaled orthogonal rows, i.e.,  $ {\mathbf{X}} {\mathbf{X}}^{\rm{H}} = \beta \mathbf{I}_{K+1}$, such that the same variance is achieved for all unknown $\bar{\mathbf{h}}$. Therefore, to minimize the variance of the estimate, $\beta$ should be maximized under the constraint that the first row of $\mathbf{X}$ is $\mathbf{1}_{\tau}$ and the remaining elements follow the signal model that can be modulated and reflected by the tags. For $|[\mathbf{X}]i,j|^2 = 1$, $\beta$ is found as
\begin{eqnarray}
   \beta =  \frac{1}{K+1} \Tr(\mathbf{X}\mathbf{X}^{\rm{H}} ) = \frac{1}{K+1} (K+1) \tau = \tau,
\end{eqnarray}
and the estimation covariance matrix is obtained as  \eqref{estimation_covariance_min}.

\section{ Derivation of Scaled LS Estimator in \eqref{SLS}}\label{SLS_estimator}
Following  \cite[Section IV]{biguesh2006training}, we first express  channel estimation error as 
\begin{eqnarray}\label{mse_SLS}
    &\mathbb{E}& \left\{ \Vert \bar{\mathbf{H}} - \gamma \hat{\bar{\mathbf{H}}}_{\rm{LS}} \Vert_F^2 \right\}  \nonumber \\
    &&= \Tr\left( \!\mathbb{E} \left \{ (\bar{\mathbf{H}} - \gamma \hat{\bar{\mathbf{H}}}_{\rm{LS}})^{\rm{H}} (\bar{\mathbf{H}} - \gamma \hat{\bar{\mathbf{H}}}_{\rm{LS}}) \right \} \right) \nonumber \\
    &&= (1- \gamma)^2 \mathbf{R}_{\bar{\mathbf{H}}} + \gamma^2 M \sigma^2 \Tr \left( (\bar{\mathbf{X}} {\bar{\mathbf{X}}}^{\rm{H}})^{-1} \right) \nonumber \\
    &&= \left(J_{\rm{LS}} + \Tr \left( \mathbf{R}_{\bar{\mathbf{H}}} \right)\right) \left(\gamma - \frac{\Tr \left( \mathbf{R}_{\bar{\mathbf{H}}} \right)}{J_{\rm{LS}} + \Tr \left( \mathbf{R}_{\bar{\mathbf{H}}} \right)}  \right)^2 \nonumber \\
    &&\quad+ \frac{J_{\rm{LS}}  \Tr \left( \mathbf{R}_{\bar{\mathbf{H}}} \right)}{J_{\rm{LS}} + \Tr \left( \mathbf{R}_{\bar{\mathbf{H}}} \right)},
\end{eqnarray}
where 
$ J_{\rm{LS}} = \mathbb{E} \{ \Vert \bar{\mathbf{H}} - \hat{\bar{\mathbf{H}}}_{\rm{LS}} \Vert_F^2 \} 
    % &=& \Tr\left( \mathbb{E} \left \{ (\bar{\mathbf{H}} - \hat{\bar{\mathbf{H}}}_{\rm{LS}})^{\rm{H}} (\bar{\mathbf{H}} - \hat{\bar{\mathbf{H}}}_{\rm{LS}}) \right \} \right) \nonumber \\
    = M \sigma^2 \Tr ( (\bar{\mathbf{X}} {\bar{\mathbf{X}}}^{\rm{H}})^{-1})= M(K+1)\sigma^2/(\tau p),$ 
in which $\hat{\bar{\mathbf{H}}}_{\rm{LS}} = \bar{\mathbf{H}} + \mathbf{N}'\bar{\mathbf{X}}^{\dagger}$, and $ \mathbb{E} \{ {\mathbf{N'}}^{\rm{H}} \mathbf{N'} \} = M \sigma^2 \mathbf{I}$. Hence, \eqref{mse_SLS} is minimized when 
\begin{eqnarray}
    \gamma_0 = \frac{\Tr \left( \mathbf{R}_{\bar{\mathbf{H}}} \right)}{J_{\rm{LS}} + \Tr \left( \mathbf{R}_{\bar{\mathbf{H}}} \right)}, 
\end{eqnarray}
and by using the LS-based consistent sample estimate, i.e., $\Tr ({\mathbf{R}}_{\hat{\bar{\mathbf{H}}}})$ instead of $\Tr ({\mathbf{R}}_{{\bar{\mathbf{H}}}})$, scaled LS is obtained as \eqref{SLS}.
% \appendices
\section{ Derivation of MMSE Estimator in \eqref{mmse_h}}\label{MMSE_estimator}
We first evaluate  the expectation term in the numerator as
\begin{eqnarray}
    \mathbb{E} \{\bar{h}_{m,k} y_{m,k}^* \} &=&  \mathbb{E} \{\bar{h}_{m,k} ( \sqrt{p}\bar{h}_{m,k}^* + n_{m,k}^*) \} \nonumber \\
    &=& \sqrt{p} \mathbb{E} \{\bar{h}_{m,k}  \bar{h}_{m,k}^*\} + \mathbb{E} \{\bar{h}_{m,k} n_{m,k}^* \} \nonumber \\
    &=& \begin{cases}
        \sqrt{p} \beta_{m0}, \quad \text{for} \quad k=0, \\
         \sqrt{\alpha p} \beta_{mk}, \quad \text{for} \quad k\in \mathcal{K},     
    \end{cases}
\end{eqnarray}
where $\beta_{mk}$ is given in \eqref{eqn_beta_mk}.
Next, the expectation term in the denominator is obtained  as
\begin{eqnarray}
    \mathbb{E} \{ \vert y_{m,k} \vert^2 \} 
    % &=& \mathbb{E} \{ \vert\sqrt{p}\bar{h}_{m,k} + n_{m,k} \vert^2 \} \nonumber \\
    &=& p \mathbb{E} \{ \vert \bar{h}_{m,k} \vert^2 \} + \mathbb{E} \{ \vert n_{m,k} \vert^2 \} \nonumber \\
    &=& \begin{cases}
        p \beta_{m0} + \sigma_p^2, \quad \text{for} \quad k=0, \\
         \alpha p \beta_{mk} + \sigma_p^2, \quad \text{for} \quad k\in \mathcal{K}.     
    \end{cases}
\end{eqnarray}
Hence, the MMSE of $\bar{h}_{m,k}$ is given as \eqref{mmse_h}.

% \linespread{1.4}
\bibliographystyle{IEEEtran}
\bibliography{IEEEabrv,Ref}

\end{document}

%% file: Figures/Configurations.eps_tex
%% Creator: Inkscape 1.2.1 (9c6d41e410, 2022-07-14), www.inkscape.org
%% PDF/EPS/PS + LaTeX output extension by Johan Engelen, 2010
%% Accompanies image file 'Configurations.eps' (pdf, eps, ps)
%%
%% To include the image in your LaTeX document, write
%%   \input{<filename>.pdf_tex}
%%  instead of
%%   \includegraphics{<filename>.pdf}
%% To scale the image, write
%%   \def\svgwidth{<desired width>}
%%   \input{<filename>.pdf_tex}
%%  instead of
%%   \includegraphics[width=<desired width>]{<filename>.pdf}
%%
%% Images with a different path to the parent latex file can
%% be accessed with the `import' package (which may need to be
%% installed) using
%%   \usepackage{import}
%% in the preamble, and then including the image with
%%   \import{<path to file>}{<filename>.pdf_tex}
%% Alternatively, one can specify
%%   \graphicspath{{<path to file>/}}
%% 
%% For more information, please see info/svg-inkscape on CTAN:
%%   http://tug.ctan.org/tex-archive/info/svg-inkscape
%%
\begingroup%
  \makeatletter%
  \providecommand\color[2][]{%
    \errmessage{(Inkscape) Color is used for the text in Inkscape, but the package 'color.sty' is not loaded}%
    \renewcommand\color[2][]{}%
  }%
  \providecommand\transparent[1]{%
    \errmessage{(Inkscape) Transparency is used (non-zero) for the text in Inkscape, but the package 'transparent.sty' is not loaded}%
    \renewcommand\transparent[1]{}%
  }%
  \providecommand\rotatebox[2]{#2}%
  \newcommand*\fsize{\dimexpr\f@size pt\relax}%
  \newcommand*\lineheight[1]{\fontsize{\fsize}{#1\fsize}\selectfont}%
  \ifx\svgwidth\undefined%
    \setlength{\unitlength}{1274.20870972bp}%
    \ifx\svgscale\undefined%
      \relax%
    \else%
      \setlength{\unitlength}{\unitlength * \real{\svgscale}}%
    \fi%
  \else%
    \setlength{\unitlength}{\svgwidth}%
  \fi%
  \global\let\svgwidth\undefined%
  \global\let\svgscale\undefined%
  \makeatother%
  \begin{picture}(1,0.20055744)%
    \lineheight{1}%
    \setlength\tabcolsep{0pt}%
    \put(0,0){\includegraphics[width=\unitlength]{Configurations.eps}}%
    \put(0.17530886,0.10768558){\makebox(0,0)[t]{\lineheight{1.25}\smash{\begin{tabular}[t]{c}Reader\end{tabular}}}}%
    \put(0.2371645,0.05819351){\makebox(0,0)[t]{\lineheight{1.25}\smash{\begin{tabular}[t]{c}Emitter\end{tabular}}}}%
    \put(0.08783472,0.01038659){\makebox(0,0)[t]{\lineheight{1.25}\smash{\begin{tabular}[t]{c}(a) MoBC\end{tabular}}}}%
    \put(0.43016321,0.08947139){\makebox(0,0)[t]{\lineheight{1.25}\smash{\begin{tabular}[t]{c}Reader\end{tabular}}}}%
    \put(0.71200414,0.06350546){\makebox(0,0)[t]{\lineheight{1.25}\smash{\begin{tabular}[t]{c}Reader\end{tabular}}}}%
    \put(0.50630017,0.04756005){\makebox(0,0)[t]{\lineheight{1.25}\smash{\begin{tabular}[t]{c}RF Source\end{tabular}}}}%
    \put(0.78529466,0.04836952){\makebox(0,0)[t]{\lineheight{1.25}\smash{\begin{tabular}[t]{c}RF Source\end{tabular}}}}%
    \put(0.70355387,0.15106596){\makebox(0,0)[t]{\lineheight{1.25}\smash{\begin{tabular}[t]{c}User\end{tabular}}}}%
    \put(0.97581824,0.11220127){\makebox(0,0)[t]{\lineheight{1.25}\smash{\begin{tabular}[t]{c}User\end{tabular}}}}%
    \put(0.32091794,0.01038659){\makebox(0,0)[t]{\lineheight{1.25}\smash{\begin{tabular}[t]{c}(b) BiBC\end{tabular}}}}%
    \put(0.59285411,0.01038659){\makebox(0,0)[t]{\lineheight{1.25}\smash{\begin{tabular}[t]{c}(c) AmBC\end{tabular}}}}%
    \put(0.87421009,0.01038659){\makebox(0,0)[t]{\lineheight{1.25}\smash{\begin{tabular}[t]{c}(d) SR\end{tabular}}}}%
  \end{picture}%
\endgroup%

%% file: Figures/SystemFigure.eps_tex
%% Creator: Inkscape 1.2.2 (732a01da63, 2022-12-09), www.inkscape.org
%% PDF/EPS/PS + LaTeX output extension by Johan Engelen, 2010
%% Accompanies image file 'SystemFigure.eps' (pdf, eps, ps)
%%
%% To include the image in your LaTeX document, write
%%   \input{<filename>.pdf_tex}
%%  instead of
%%   \includegraphics{<filename>.pdf}
%% To scale the image, write
%%   \def\svgwidth{<desired width>}
%%   \input{<filename>.pdf_tex}
%%  instead of
%%   \includegraphics[width=<desired width>]{<filename>.pdf}
%%
%% Images with a different path to the parent latex file can
%% be accessed with the `import' package (which may need to be
%% installed) using
%%   \usepackage{import}
%% in the preamble, and then including the image with
%%   \import{<path to file>}{<filename>.pdf_tex}
%% Alternatively, one can specify
%%   \graphicspath{{<path to file>/}}
%% 
%% For more information, please see info/svg-inkscape on CTAN:
%%   http://tug.ctan.org/tex-archive/info/svg-inkscape
%%
\begingroup%
  \makeatletter%
  \providecommand\color[2][]{%
    \errmessage{(Inkscape) Color is used for the text in Inkscape, but the package 'color.sty' is not loaded}%
    \renewcommand\color[2][]{}%
  }%
  \providecommand\transparent[1]{%
    \errmessage{(Inkscape) Transparency is used (non-zero) for the text in Inkscape, but the package 'transparent.sty' is not loaded}%
    \renewcommand\transparent[1]{}%
  }%
  \providecommand\rotatebox[2]{#2}%
  \newcommand*\fsize{\dimexpr\f@size pt\relax}%
  \newcommand*\lineheight[1]{\fontsize{\fsize}{#1\fsize}\selectfont}%
  \ifx\svgwidth\undefined%
    \setlength{\unitlength}{700.81114197bp}%
    \ifx\svgscale\undefined%
      \relax%
    \else%
      \setlength{\unitlength}{\unitlength * \real{\svgscale}}%
    \fi%
  \else%
    \setlength{\unitlength}{\svgwidth}%
  \fi%
  \global\let\svgwidth\undefined%
  \global\let\svgscale\undefined%
  \makeatother%
  \begin{picture}(1,0.4699568)%
    \lineheight{1}%
    \setlength\tabcolsep{0pt}%
    \put(0,0){\includegraphics[width=\unitlength]{SystemFigure.eps}}%
    \put(0.89502985,0.0517184){\makebox(0,0)[t]{\lineheight{1.25}\smash{\begin{tabular}[t]{c}Reader\end{tabular}}}}%
    \put(0.54561331,0.01533199){\makebox(0,0)[t]{\lineheight{1.25}\smash{\begin{tabular}[t]{c}$T_K$\end{tabular}}}}%
    \put(0.49176145,0.19726403){\makebox(0,0)[t]{\lineheight{1.25}\smash{\begin{tabular}[t]{c}$T_k$\end{tabular}}}}%
    \put(0.48962108,0.38775758){\makebox(0,0)[t]{\lineheight{1.25}\smash{\begin{tabular}[t]{c}$T_1$\end{tabular}}}}%
    \put(0.04820117,0.22639492){\makebox(0,0)[t]{\lineheight{1.25}\smash{\begin{tabular}[t]{c}RF Source\end{tabular}}}}%
    \put(0.35833889,0.32766984){\makebox(0,0)[t]{\lineheight{1.25}\smash{\begin{tabular}[t]{c}$\mathbf{h}_0$\end{tabular}}}}%
    \put(0.63810114,0.40985235){\makebox(0,0)[t]{\lineheight{1.25}\smash{\begin{tabular}[t]{c}$\mathbf{g}_1$\end{tabular}}}}%
    \put(0.61929811,0.25159148){\makebox(0,0)[t]{\lineheight{1.25}\smash{\begin{tabular}[t]{c}$\mathbf{g}_k$\end{tabular}}}}%
    \put(0.66707513,0.14668875){\makebox(0,0)[t]{\lineheight{1.25}\smash{\begin{tabular}[t]{c}$\mathbf{g}_K$\end{tabular}}}}%
    \put(0.24055503,0.40234484){\makebox(0,0)[t]{\lineheight{1.25}\smash{\begin{tabular}[t]{c}$f_1$\end{tabular}}}}%
    \put(0.25934038,0.15531933){\makebox(0,0)[t]{\lineheight{1.25}\smash{\begin{tabular}[t]{c}$f_K$\end{tabular}}}}%
    \put(0.25855309,0.24948505){\makebox(0,0)[t]{\lineheight{1.25}\smash{\begin{tabular}[t]{c}$f_k$\end{tabular}}}}%
  \end{picture}%
\endgroup%

%% file: Figures/CoheranceTime.eps_tex
%% Creator: Inkscape 1.2.1 (9c6d41e410, 2022-07-14), www.inkscape.org
%% PDF/EPS/PS + LaTeX output extension by Johan Engelen, 2010
%% Accompanies image file 'CoheranceTime.eps' (pdf, eps, ps)
%%
%% To include the image in your LaTeX document, write
%%   \input{<filename>.pdf_tex}
%%  instead of
%%   \includegraphics{<filename>.pdf}
%% To scale the image, write
%%   \def\svgwidth{<desired width>}
%%   \input{<filename>.pdf_tex}
%%  instead of
%%   \includegraphics[width=<desired width>]{<filename>.pdf}
%%
%% Images with a different path to the parent latex file can
%% be accessed with the `import' package (which may need to be
%% installed) using
%%   \usepackage{import}
%% in the preamble, and then including the image with
%%   \import{<path to file>}{<filename>.pdf_tex}
%% Alternatively, one can specify
%%   \graphicspath{{<path to file>/}}
%% 
%% For more information, please see info/svg-inkscape on CTAN:
%%   http://tug.ctan.org/tex-archive/info/svg-inkscape
%%
\begingroup%
  \makeatletter%
  \providecommand\color[2][]{%
    \errmessage{(Inkscape) Color is used for the text in Inkscape, but the package 'color.sty' is not loaded}%
    \renewcommand\color[2][]{}%
  }%
  \providecommand\transparent[1]{%
    \errmessage{(Inkscape) Transparency is used (non-zero) for the text in Inkscape, but the package 'transparent.sty' is not loaded}%
    \renewcommand\transparent[1]{}%
  }%
  \providecommand\rotatebox[2]{#2}%
  \newcommand*\fsize{\dimexpr\f@size pt\relax}%
  \newcommand*\lineheight[1]{\fontsize{\fsize}{#1\fsize}\selectfont}%
  \ifx\svgwidth\undefined%
    \setlength{\unitlength}{462.54052734bp}%
    \ifx\svgscale\undefined%
      \relax%
    \else%
      \setlength{\unitlength}{\unitlength * \real{\svgscale}}%
    \fi%
  \else%
    \setlength{\unitlength}{\svgwidth}%
  \fi%
  \global\let\svgwidth\undefined%
  \global\let\svgscale\undefined%
  \makeatother%
  \begin{picture}(1,0.21481724)%
    \lineheight{1}%
    \setlength\tabcolsep{0pt}%
    \put(0,0){\includegraphics[width=\unitlength]{CoheranceTime.eps}}%
    \put(0.60798618,0.18433885){\color[rgb]{0,0,0}\makebox(0,0)[rt]{\lineheight{1.25}\smash{\begin{tabular}[t]{r}Coherence Interval ($T$)\end{tabular}}}}%
    \put(0.2197437,0.07341254){\color[rgb]{0,0,0}\makebox(0,0)[rt]{\lineheight{1.25}\smash{\begin{tabular}[t]{r}($\tau$)\end{tabular}}}}%
    \put(0.70294457,0.07665549){\color[rgb]{0,0,0}\makebox(0,0)[rt]{\lineheight{1.25}\smash{\begin{tabular}[t]{r}($T-\tau$)\end{tabular}}}}%
    \put(0.32027543,0.01828223){\color[rgb]{0,0,0}\makebox(0,0)[rt]{\lineheight{1.25}\smash{\begin{tabular}[t]{r}Channel Estimation\end{tabular}}}}%
    \put(0.78401855,0.01828223){\color[rgb]{0,0,0}\makebox(0,0)[rt]{\lineheight{1.25}\smash{\begin{tabular}[t]{r}Data Transmission\end{tabular}}}}%
    \put(0.23271553,0.11232804){\color[rgb]{0,0,0}\makebox(0,0)[rt]{\lineheight{1.25}\smash{\begin{tabular}[t]{r}Pilots\end{tabular}}}}%
    \put(0.68348682,0.11232804){\color[rgb]{0,0,0}\makebox(0,0)[rt]{\lineheight{1.25}\smash{\begin{tabular}[t]{r}Data\end{tabular}}}}%
  \end{picture}%
\endgroup%

%% file: Figures/TagTransNew.eps_tex
%% Creator: Inkscape 1.2.2 (732a01da63, 2022-12-09), www.inkscape.org
%% PDF/EPS/PS + LaTeX output extension by Johan Engelen, 2010
%% Accompanies image file 'TagTransNew.eps' (pdf, eps, ps)
%%
%% To include the image in your LaTeX document, write
%%   \input{<filename>.pdf_tex}
%%  instead of
%%   \includegraphics{<filename>.pdf}
%% To scale the image, write
%%   \def\svgwidth{<desired width>}
%%   \input{<filename>.pdf_tex}
%%  instead of
%%   \includegraphics[width=<desired width>]{<filename>.pdf}
%%
%% Images with a different path to the parent latex file can
%% be accessed with the `import' package (which may need to be
%% installed) using
%%   \usepackage{import}
%% in the preamble, and then including the image with
%%   \import{<path to file>}{<filename>.pdf_tex}
%% Alternatively, one can specify
%%   \graphicspath{{<path to file>/}}
%% 
%% For more information, please see info/svg-inkscape on CTAN:
%%   http://tug.ctan.org/tex-archive/info/svg-inkscape
%%
\begingroup%
  \makeatletter%
  \providecommand\color[2][]{%
    \errmessage{(Inkscape) Color is used for the text in Inkscape, but the package 'color.sty' is not loaded}%
    \renewcommand\color[2][]{}%
  }%
  \providecommand\transparent[1]{%
    \errmessage{(Inkscape) Transparency is used (non-zero) for the text in Inkscape, but the package 'transparent.sty' is not loaded}%
    \renewcommand\transparent[1]{}%
  }%
  \providecommand\rotatebox[2]{#2}%
  \newcommand*\fsize{\dimexpr\f@size pt\relax}%
  \newcommand*\lineheight[1]{\fontsize{\fsize}{#1\fsize}\selectfont}%
  \ifx\svgwidth\undefined%
    \setlength{\unitlength}{571.61206055bp}%
    \ifx\svgscale\undefined%
      \relax%
    \else%
      \setlength{\unitlength}{\unitlength * \real{\svgscale}}%
    \fi%
  \else%
    \setlength{\unitlength}{\svgwidth}%
  \fi%
  \global\let\svgwidth\undefined%
  \global\let\svgscale\undefined%
  \makeatother%
  \begin{picture}(1,0.43873724)%
    \lineheight{1}%
    \setlength\tabcolsep{0pt}%
    \put(0,0){\includegraphics[width=\unitlength]{TagTransNew.eps}}%
    \put(0.09223423,0.35432111){\color[rgb]{0,0,0}\makebox(0,0)[rt]{\lineheight{1.25}\smash{\begin{tabular}[t]{r}RF Source\end{tabular}}}}%
    \put(0.09354631,0.28871718){\color[rgb]{0,0,0}\makebox(0,0)[rt]{\lineheight{1.25}\smash{\begin{tabular}[t]{r}Tag-1\end{tabular}}}}%
    \put(0.09223423,0.22180117){\color[rgb]{0,0,0}\makebox(0,0)[rt]{\lineheight{1.25}\smash{\begin{tabular}[t]{r}Tag-2\end{tabular}}}}%
    \put(0.09354631,0.12864358){\color[rgb]{0,0,0}\makebox(0,0)[rt]{\lineheight{1.25}\smash{\begin{tabular}[t]{r}Tag-$K$\end{tabular}}}}%
    \put(0.18807991,0.08169504){\color[rgb]{0,0,0}\makebox(0,0)[rt]{\lineheight{1.25}\smash{\begin{tabular}[t]{r}$0$\end{tabular}}}}%
    \put(0.34552936,0.08169504){\color[rgb]{0,0,0}\makebox(0,0)[rt]{\lineheight{1.25}\smash{\begin{tabular}[t]{r}$1$\end{tabular}}}}%
    \put(0.51085127,0.08169504){\color[rgb]{0,0,0}\makebox(0,0)[rt]{\lineheight{1.25}\smash{\begin{tabular}[t]{r}$2$\end{tabular}}}}%
    \put(0.67092487,0.08169504){\color[rgb]{0,0,0}\makebox(0,0)[rt]{\lineheight{1.25}\smash{\begin{tabular}[t]{r}$3$\end{tabular}}}}%
    \put(0.91234731,0.08169504){\color[rgb]{0,0,0}\makebox(0,0)[rt]{\lineheight{1.25}\smash{\begin{tabular}[t]{r}$K$\end{tabular}}}}%
    \put(0.51973972,0.41376854){\color[rgb]{0,0,0}\makebox(0,0)[rt]{\lineheight{1.25}\smash{\begin{tabular}[t]{r}$\tau$\end{tabular}}}}%
    \put(0.49532635,0.01370135){\color[rgb]{0,0,0}\makebox(0,0)[rt]{\lineheight{1.25}\smash{\begin{tabular}[t]{r}Pilots\end{tabular}}}}%
    \put(0.76036624,0.01370135){\color[rgb]{0,0,0}\makebox(0,0)[rt]{\lineheight{1.25}\smash{\begin{tabular}[t]{r}Silent\end{tabular}}}}%
  \end{picture}%
\endgroup%

%% file: Figures/Methodology.eps_tex
%% Creator: Inkscape 1.2.2 (732a01da63, 2022-12-09), www.inkscape.org
%% PDF/EPS/PS + LaTeX output extension by Johan Engelen, 2010
%% Accompanies image file 'Methodology.eps' (pdf, eps, ps)
%%
%% To include the image in your LaTeX document, write
%%   \input{<filename>.pdf_tex}
%%  instead of
%%   \includegraphics{<filename>.pdf}
%% To scale the image, write
%%   \def\svgwidth{<desired width>}
%%   \input{<filename>.pdf_tex}
%%  instead of
%%   \includegraphics[width=<desired width>]{<filename>.pdf}
%%
%% Images with a different path to the parent latex file can
%% be accessed with the `import' package (which may need to be
%% installed) using
%%   \usepackage{import}
%% in the preamble, and then including the image with
%%   \import{<path to file>}{<filename>.pdf_tex}
%% Alternatively, one can specify
%%   \graphicspath{{<path to file>/}}
%% 
%% For more information, please see info/svg-inkscape on CTAN:
%%   http://tug.ctan.org/tex-archive/info/svg-inkscape
%%
\begingroup%
  \makeatletter%
  \providecommand\color[2][]{%
    \errmessage{(Inkscape) Color is used for the text in Inkscape, but the package 'color.sty' is not loaded}%
    \renewcommand\color[2][]{}%
  }%
  \providecommand\transparent[1]{%
    \errmessage{(Inkscape) Transparency is used (non-zero) for the text in Inkscape, but the package 'transparent.sty' is not loaded}%
    \renewcommand\transparent[1]{}%
  }%
  \providecommand\rotatebox[2]{#2}%
  \newcommand*\fsize{\dimexpr\f@size pt\relax}%
  \newcommand*\lineheight[1]{\fontsize{\fsize}{#1\fsize}\selectfont}%
  \ifx\svgwidth\undefined%
    \setlength{\unitlength}{875.28735352bp}%
    \ifx\svgscale\undefined%
      \relax%
    \else%
      \setlength{\unitlength}{\unitlength * \real{\svgscale}}%
    \fi%
  \else%
    \setlength{\unitlength}{\svgwidth}%
  \fi%
  \global\let\svgwidth\undefined%
  \global\let\svgscale\undefined%
  \makeatother%
  \begin{picture}(1,0.32786336)%
    \lineheight{1}%
    \setlength\tabcolsep{0pt}%
    \put(0,0){\includegraphics[width=\unitlength]{Methodology.eps}}%
    \put(0.2243679,0.19575202){\makebox(0,0)[lt]{\lineheight{1.25}\smash{\begin{tabular}[t]{l}\textbf{Design Constraints}\end{tabular}}}}%
    \put(0.137007,0.29264068){\makebox(0,0)[lt]{\lineheight{1.25}\smash{\begin{tabular}[t]{l}\textbf{Syetem Model}\end{tabular}}}}%
    \put(0.52132871,0.08223799){\makebox(0,0)[lt]{\lineheight{1.25}\smash{\begin{tabular}[t]{l}\textbf{Use the Designed Pilots for Backscatter }\end{tabular}}}}%
    \put(0.03109876,0.26778467){\makebox(0,0)[lt]{\lineheight{1.25}\smash{\begin{tabular}[t]{l}$K$ single-antenna tags, single antenna source,\end{tabular}}}}%
    \put(0.13840069,0.09345527){\makebox(0,0)[lt]{\lineheight{1.25}\smash{\begin{tabular}[t]{l}\textbf{Pilot Designing}\end{tabular}}}}%
    \put(0.04620223,0.06859925){\makebox(0,0)[lt]{\lineheight{1.25}\smash{\begin{tabular}[t]{l}Hadamard matrix, modified ZC sequences, \end{tabular}}}}%
    \put(0.12706724,0.24379255){\makebox(0,0)[lt]{\lineheight{1.25}\smash{\begin{tabular}[t]{l}and $M$ antenna reader\end{tabular}}}}%
    \put(0.11999103,0.01420233){\makebox(0,0)[lt]{\lineheight{1.25}\smash{\begin{tabular}[t]{l}System Design Phase\end{tabular}}}}%
    \put(0.14830258,0.0442276){\makebox(0,0)[lt]{\lineheight{1.25}\smash{\begin{tabular}[t]{l}and DFT matrix\end{tabular}}}}%
    \put(0.62199058,0.29264068){\makebox(0,0)[lt]{\lineheight{1.25}\smash{\begin{tabular}[t]{l}\textbf{Pilot Assigning}\end{tabular}}}}%
    \put(0.55378422,0.26778467){\makebox(0,0)[lt]{\lineheight{1.25}\smash{\begin{tabular}[t]{l}Assign arbitrary pilot sequences\end{tabular}}}}%
    \put(0.52465093,0.24207883){\makebox(0,0)[lt]{\lineheight{1.25}\smash{\begin{tabular}[t]{l}Source - $\mathbf{s}\in \mathbb{C}^{1 \times \tau}$ and $T_i$ - $\mathbf{c}_i \in \mathbb{C}^{1 \times \tau}$\end{tabular}}}}%
    \put(0.65283759,0.19324475){\makebox(0,0)[lt]{\lineheight{1.25}\smash{\begin{tabular}[t]{l}\textbf{Received Sgnal at the Reader}\end{tabular}}}}%
    \put(0.60005474,0.15982013){\makebox(0,0)[lt]{\lineheight{1.25}\smash{\begin{tabular}[t]{l}$\mathbf{Y} = \sqrt{p} \mathbf{h}_0 \mathbf{s} + \sqrt{p \alpha} \sum\nolimits_{i\in \mathcal{K}} \mathbf{h}_i (\mathbf{s} \odot  \mathbf{c}_i) + \mathbf{N}$\end{tabular}}}}%
    \put(0.6104423,0.0531047){\makebox(0,0)[lt]{\lineheight{1.25}\smash{\begin{tabular}[t]{l}\textbf{Channel Estimation}\end{tabular}}}}%
    \put(0.60668835,0.01420233){\makebox(0,0)[lt]{\lineheight{1.25}\smash{\begin{tabular}[t]{l}Pilot Transmission Phase\end{tabular}}}}%
    \put(0.16987133,0.1691823){\makebox(0,0)[lt]{\lineheight{1.25}\smash{\begin{tabular}[t]{l}$\mathbf{1}_{\tau} \mathbf{c}_k^{\rm{H}} = 0$, $\mathbf{c}_{i} \mathbf{c}_k^{\rm{H}} = 0$ for $i\in \mathcal{K}_k$,\end{tabular}}}}%
    \put(0.24539094,0.1431674){\makebox(0,0)[lt]{\lineheight{1.25}\smash{\begin{tabular}[t]{l} and $\Vert\mathbf{c}_k\Vert^2=\tau$ \end{tabular}}}}%
  \end{picture}%
\endgroup%

%% file: Figures/SystemFigureGen.eps_tex
%% Creator: Inkscape 1.2.2 (732a01da63, 2022-12-09), www.inkscape.org
%% PDF/EPS/PS + LaTeX output extension by Johan Engelen, 2010
%% Accompanies image file 'SystemFigureGen.eps' (pdf, eps, ps)
%%
%% To include the image in your LaTeX document, write
%%   \input{<filename>.pdf_tex}
%%  instead of
%%   \includegraphics{<filename>.pdf}
%% To scale the image, write
%%   \def\svgwidth{<desired width>}
%%   \input{<filename>.pdf_tex}
%%  instead of
%%   \includegraphics[width=<desired width>]{<filename>.pdf}
%%
%% Images with a different path to the parent latex file can
%% be accessed with the `import' package (which may need to be
%% installed) using
%%   \usepackage{import}
%% in the preamble, and then including the image with
%%   \import{<path to file>}{<filename>.pdf_tex}
%% Alternatively, one can specify
%%   \graphicspath{{<path to file>/}}
%% 
%% For more information, please see info/svg-inkscape on CTAN:
%%   http://tug.ctan.org/tex-archive/info/svg-inkscape
%%
\begingroup%
  \makeatletter%
  \providecommand\color[2][]{%
    \errmessage{(Inkscape) Color is used for the text in Inkscape, but the package 'color.sty' is not loaded}%
    \renewcommand\color[2][]{}%
  }%
  \providecommand\transparent[1]{%
    \errmessage{(Inkscape) Transparency is used (non-zero) for the text in Inkscape, but the package 'transparent.sty' is not loaded}%
    \renewcommand\transparent[1]{}%
  }%
  \providecommand\rotatebox[2]{#2}%
  \newcommand*\fsize{\dimexpr\f@size pt\relax}%
  \newcommand*\lineheight[1]{\fontsize{\fsize}{#1\fsize}\selectfont}%
  \ifx\svgwidth\undefined%
    \setlength{\unitlength}{739.896698bp}%
    \ifx\svgscale\undefined%
      \relax%
    \else%
      \setlength{\unitlength}{\unitlength * \real{\svgscale}}%
    \fi%
  \else%
    \setlength{\unitlength}{\svgwidth}%
  \fi%
  \global\let\svgwidth\undefined%
  \global\let\svgscale\undefined%
  \makeatother%
  \begin{picture}(1,0.46561424)%
    \lineheight{1}%
    \setlength\tabcolsep{0pt}%
    \put(0,0){\includegraphics[width=\unitlength]{SystemFigureGen.eps}}%
    \put(0.90397294,0.02486893){\makebox(0,0)[t]{\lineheight{1.25}\smash{\begin{tabular}[t]{c}AP\end{tabular}}}}%
    \put(0.60950616,0.02689624){\makebox(0,0)[t]{\lineheight{1.25}\smash{\begin{tabular}[t]{c}$T_K$\end{tabular}}}}%
    \put(0.55849906,0.19921759){\makebox(0,0)[t]{\lineheight{1.25}\smash{\begin{tabular}[t]{c}$T_k$\end{tabular}}}}%
    \put(0.55647175,0.37964818){\makebox(0,0)[t]{\lineheight{1.25}\smash{\begin{tabular}[t]{c}$T_1$\end{tabular}}}}%
    \put(0.43820657,0.33084385){\makebox(0,0)[t]{\lineheight{1.25}\smash{\begin{tabular}[t]{c}$\mathbf{h}_n$\end{tabular}}}}%
    \put(0.69710825,0.4046304){\makebox(0,0)[t]{\lineheight{1.25}\smash{\begin{tabular}[t]{c}$\mathbf{f}_1$\end{tabular}}}}%
    \put(0.67929851,0.25067515){\makebox(0,0)[t]{\lineheight{1.25}\smash{\begin{tabular}[t]{c}$\mathbf{f}_k$\end{tabular}}}}%
    \put(0.72455167,0.15942321){\makebox(0,0)[t]{\lineheight{1.25}\smash{\begin{tabular}[t]{c}$\mathbf{f}_K$\end{tabular}}}}%
    \put(0.15084444,0.1123186){\makebox(0,0)[t]{\lineheight{1.25}\smash{\begin{tabular}[t]{c}$\mathbf{U}_1$\end{tabular}}}}%
    \put(0.15513315,0.37801865){\makebox(0,0)[t]{\lineheight{1.25}\smash{\begin{tabular}[t]{c}$\mathbf{U}_n$\end{tabular}}}}%
    \put(0.32056279,0.40968334){\makebox(0,0)[t]{\lineheight{1.25}\smash{\begin{tabular}[t]{c}$g_{n,1}$\end{tabular}}}}%
    \put(0.33835579,0.16759788){\makebox(0,0)[t]{\lineheight{1.25}\smash{\begin{tabular}[t]{c}$g_{n,K}$\end{tabular}}}}%
    \put(0.33761009,0.25678924){\makebox(0,0)[t]{\lineheight{1.25}\smash{\begin{tabular}[t]{c}$g_{n,k}$\end{tabular}}}}%
    \put(0.0477931,0.20179944){\makebox(0,0)[t]{\lineheight{1.25}\smash{\begin{tabular}[t]{c}$\mathbf{U}_N$\end{tabular}}}}%
  \end{picture}%
\endgroup%

%% file: FiguresTxt/DC_MSE_T_8_K_7.tex
% This file was created by matlab2tikz.
%
%The latest updates can be retrieved from
%  http://www.mathworks.com/matlabcentral/fileexchange/22022-matlab2tikz-matlab2tikz
%where you can also make suggestions and rate matlab2tikz.
%
\definecolor{mycolor1}{rgb}{0.00000,0.45000,0.74000}%
\definecolor{mycolor2}{rgb}{1.00000,0.00000,1.00000}%
\begin{tikzpicture}

\begin{axis}[%
width=4.739in,
height=4.33in,
at={(0.795in,0.584in)},
scale only axis,
xmin=0,
xmax=30,
xlabel style={font=\color{white!15!black}},
xlabel={Transmit Power [dBm]},
ymode=log,
ymin=1e-07,
ymax=0.01,
yminorticks=true,
ylabel style={font=\color{white!15!black}},
ylabel={Normalized MSE},
axis background/.style={fill=white},
xmajorgrids,
ymajorgrids,
yminorgrids,
legend style={legend cell align=left, align=left, draw=white!15!black}
]
\addplot [color=red, line width=1.5pt, mark size=3.0pt, mark=triangle, mark options={solid, red}]
  table[row sep=crcr]{%
0	0.000228219603206201\\
3.33333333333333	0.000103747680729938\\
6.66666666666667	4.74872626063954e-05\\
10	2.23260851024894e-05\\
13.3333333333333	1.04329485066717e-05\\
16.6666666666667	4.7959915789334e-06\\
20	2.22593543145878e-06\\
23.3333333333333	1.03284825990583e-06\\
26.6666666666667	4.74862393296697e-07\\
30	2.24352443977694e-07\\
};
\addlegendentry{LS}

\addplot [color=mycolor1, dashed, line width=1.5pt, mark size=6.0pt, mark=+, mark options={solid, mycolor1}]
  table[row sep=crcr]{%
0	0.000229234292007876\\
3.33333333333333	0.000103977690167214\\
6.66666666666667	4.75553821837783e-05\\
10	2.23291508017332e-05\\
13.3333333333333	1.04353537633577e-05\\
16.6666666666667	4.79658608554812e-06\\
20	2.22601816996794e-06\\
23.3333333333333	1.03286587256909e-06\\
26.6666666666667	4.74850817315217e-07\\
30	2.24368726845061e-07\\
};
\addlegendentry{Scaled LS}

\addplot [color=black!50!green, dotted, line width=1.5pt]
  table[row sep=crcr]{%
0	0.000228231771537853\\
3.33333333333333	0.000103748930170719\\
6.66666666666667	4.74867348732637e-05\\
10	2.23258516349323e-05\\
13.3333333333333	1.04327245489693e-05\\
16.6666666666667	4.79598584161153e-06\\
20	2.22595168147922e-06\\
23.3333333333333	1.03285323970137e-06\\
26.6666666666667	4.74861282908647e-07\\
30	2.24352690840691e-07\\
};
\addlegendentry{LMMSE}

\addplot [color=red, line width=1.5pt, only marks, mark size=4.5pt, mark=o, mark options={solid, red}]
  table[row sep=crcr]{%
0	0.000225490548122364\\
3.33333333333333	0.000103479923950376\\
6.66666666666667	4.78632918792694e-05\\
10	2.2224193114683e-05\\
13.3333333333333	1.03049921413288e-05\\
16.6666666666667	4.88273717925281e-06\\
20	2.20452677409336e-06\\
23.3333333333333	1.04356154800624e-06\\
26.6666666666667	4.81837410014985e-07\\
30	2.24195639327465e-07\\
};
\addlegendentry{MMSE - Simulation}

\addplot [color=black, dashdotted, line width=1.5pt]
  table[row sep=crcr]{%
0	0.000225564742427533\\
3.33333333333333	0.000103528930527433\\
6.66666666666667	4.78889113984706e-05\\
10	2.22272895013101e-05\\
13.3333333333333	1.03073576312348e-05\\
16.6666666666667	4.88352026700048e-06\\
20	2.20518266876907e-06\\
23.3333333333333	1.04407622517825e-06\\
26.6666666666667	4.81992489032777e-07\\
30	2.24295822299896e-07\\
};
\addlegendentry{MMSE - Analysis}

\addplot [color=mycolor2, line width=1.5pt, mark size=3.2pt, mark=square, mark options={solid, mycolor2}]
  table[row sep=crcr]{%
0	0.0017882191200692\\
3.33333333333333	0.000824000666307637\\
6.66666666666667	0.00038656381990644\\
10	0.000176378583816632\\
13.3333333333333	8.33666772684238e-05\\
16.6666666666667	3.87821537270964e-05\\
20	1.81241161978438e-05\\
23.3333333333333	8.31914737750099e-06\\
26.6666666666667	3.85404249838829e-06\\
30	1.82206742669045e-06\\
};
\addlegendentry{\cite{Abdallah2021,  Zhao2019, Liu2021, Zhu2018,  Yerzhanova2021}}

\end{axis}

\begin{axis}[%
width=6.115in,
height=5.312in,
at={(0in,0in)},
scale only axis,
xmin=0,
xmax=1,
ymin=0,
ymax=1,
axis line style={draw=none},
ticks=none,
axis x line*=bottom,
axis y line*=left
]
\end{axis}
\end{tikzpicture}%

%% file: FiguresTxt/BC_MSE_T_8_K_7.tex
% This file was created by matlab2tikz.
%
%The latest updates can be retrieved from
%  http://www.mathworks.com/matlabcentral/fileexchange/22022-matlab2tikz-matlab2tikz
%where you can also make suggestions and rate matlab2tikz.
%
\definecolor{mycolor1}{rgb}{0.00000,0.45000,0.74000}%
\definecolor{mycolor2}{rgb}{1.00000,0.00000,1.00000}%
\begin{tikzpicture}

\begin{axis}[%
width=4.739in,
height=4.33in,
at={(0.795in,0.584in)},
scale only axis,
xmin=0.0796495420151355,
xmax=30.0796495420151,
xlabel style={font=\color{white!15!black}},
xlabel={Transmit Power [dBm]},
ymode=log,
ymin=0.01,
ymax=999.999999999999,
yminorticks=true,
ylabel style={font=\color{white!15!black}},
ylabel={Normalized MSE},
axis background/.style={fill=white},
xmajorgrids,
ymajorgrids,
yminorgrids,
legend style={legend cell align=left, align=left, draw=white!15!black}
]
\addplot [color=red, line width=1.5pt, mark size=3.0pt, mark=triangle, mark options={solid, red}]
  table[row sep=crcr]{%
0	25.5283493951504\\
3.33333333333333	11.2500646420983\\
6.66666666666667	5.25221128325411\\
10	2.39206296535903\\
13.3333333333333	1.17447357854156\\
16.6666666666667	0.532229763320395\\
20	0.250505303458384\\
23.3333333333333	0.112042998892933\\
26.6666666666667	0.0549441995802716\\
30	0.0252903955151987\\
};
\addlegendentry{LS}

\addplot [color=mycolor1, dashed, line width=1.5pt, mark size=6.0pt, mark=+, mark options={solid, mycolor1}]
  table[row sep=crcr]{%
0	25.4649598751894\\
3.33333333333333	11.2370928696059\\
6.66666666666667	5.24942004407427\\
10	2.39146990401332\\
13.3333333333333	1.17433910802856\\
16.6666666666667	0.532201383521115\\
20	0.250499073840229\\
23.3333333333333	0.112041738421127\\
26.6666666666667	0.0549438998553502\\
30	0.0252903349092799\\
};
\addlegendentry{Scaled LS}

\addplot [color=black!50!green, dotted, line width=1.5pt]
  table[row sep=crcr]{%
0	2.00529770103497\\
3.33333333333333	2.04436048327627\\
6.66666666666667	1.88744216933534\\
10	1.34297998195753\\
13.3333333333333	0.882945442386253\\
16.6666666666667	0.46075177121435\\
20	0.234692541931191\\
23.3333333333333	0.108617105000923\\
26.6666666666667	0.0541729195735438\\
30	0.0251158150459198\\
};
\addlegendentry{LMMSE}

\addplot [color=red, line width=1.5pt, only marks, mark size=4.5pt, mark=o, mark options={solid, red}]
  table[row sep=crcr]{%
0	1.00525548104885\\
3.33333333333333	0.991179914722824\\
6.66666666666667	0.957889347505229\\
10	0.818159531374291\\
13.3333333333333	0.613306664646567\\
16.6666666666667	0.387117811208909\\
20	0.213023804512869\\
23.3333333333333	0.10293685009079\\
26.6666666666667	0.0530083132786235\\
30	0.0247001412730216\\
};
\addlegendentry{MMSE - Simulation}

\addplot [color=black, dashdotted, line width=1.5pt]
  table[row sep=crcr]{%
0	1.00910825834124\\
3.33333333333333	0.981099928452984\\
6.66666666666667	0.95040659680094\\
10	0.823996375035416\\
13.3333333333333	0.611152624806943\\
16.6666666666667	0.38793196540234\\
20	0.212944076906934\\
23.3333333333333	0.102869639595569\\
26.6666666666667	0.0530654771612169\\
30	0.0247192236342851\\
};
\addlegendentry{MMSE - Analysis}

\addplot [color=mycolor2, line width=1.5pt, mark size=3.2pt, mark=square, mark options={solid, mycolor2}]
  table[row sep=crcr]{%
0	408.283073811985\\
3.33333333333333	179.379054242006\\
6.66666666666667	83.9377084759443\\
10	38.6076264776538\\
13.3333333333333	19.4916664403803\\
16.6666666666667	8.53912527573522\\
20	3.94685566894131\\
23.3333333333333	1.76286124175417\\
26.6666666666667	0.902770192566808\\
30	0.401758584130444\\
};
\addlegendentry{\cite{Abdallah2021,  Zhao2019, Liu2021, Zhu2018,  Yerzhanova2021}}

\end{axis}

\begin{axis}[%
width=6.115in,
height=5.312in,
at={(0in,0in)},
scale only axis,
xmin=0,
xmax=1,
ymin=0,
ymax=1,
axis line style={draw=none},
ticks=none,
axis x line*=bottom,
axis y line*=left
]
\end{axis}
\end{tikzpicture}%

%% file: FiguresTxt/DC_MSE_T_16_K_7_15.tex
% This file was created by matlab2tikz.
%
%The latest updates can be retrieved from
%  http://www.mathworks.com/matlabcentral/fileexchange/22022-matlab2tikz-matlab2tikz
%where you can also make suggestions and rate matlab2tikz.
%
\definecolor{mycolor1}{rgb}{0.00000,0.45000,0.74000}%
\definecolor{mycolor2}{rgb}{1.00000,0.00000,1.00000}%
\begin{tikzpicture}

\begin{axis}[%
width=4.739in,
height=4.33in,
at={(0.795in,0.584in)},
scale only axis,
xmin=0,
xmax=30,
xlabel style={font=\color{white!15!black}},
xlabel={Transmit Power [dBm]},
ymode=log,
ymin=1e-07,
ymax=0.01,
yminorticks=true,
ylabel style={font=\color{white!15!black}},
ylabel={Normalized MSE},
axis background/.style={fill=white},
xmajorgrids,
ymajorgrids,
yminorgrids,
legend style={legend cell align=left, align=left, draw=white!15!black}
]
\addplot [color=red, line width=1.5pt, mark size=3.0pt, mark=triangle, mark options={solid, red}]
  table[row sep=crcr]{%
0	0.000111991580056265\\
3.33333333333333	5.31776108916926e-05\\
6.66666666666667	2.3762127085084e-05\\
10	1.10772949035022e-05\\
13.3333333333333	5.19794589181381e-06\\
16.6666666666667	2.37285901605676e-06\\
20	1.14013648258537e-06\\
23.3333333333333	5.12334650207316e-07\\
26.6666666666667	2.39333990031922e-07\\
30	1.10826188252024e-07\\
};
\addlegendentry{LS}

\addplot [color=mycolor1, dashed, line width=1.5pt, mark size=6.0pt, mark=+, mark options={solid, mycolor1}]
  table[row sep=crcr]{%
0	0.000113419396989558\\
3.33333333333333	5.35001156556158e-05\\
6.66666666666667	2.38371124888789e-05\\
10	1.10866264840239e-05\\
13.3333333333333	5.20117152719492e-06\\
16.6666666666667	2.37333206227178e-06\\
20	1.14036611170982e-06\\
23.3333333333333	5.12316682921305e-07\\
26.6666666666667	2.39344618473133e-07\\
30	1.10830953461913e-07\\
};
\addlegendentry{Scaled LS}

\addplot [color=black!50!green, dotted, line width=1.5pt]
  table[row sep=crcr]{%
0	0.00011198667681813\\
3.33333333333333	5.31774412431473e-05\\
6.66666666666667	2.37617430314564e-05\\
10	1.10771769349934e-05\\
13.3333333333333	5.19788928402305e-06\\
16.6666666666667	2.37288980795169e-06\\
20	1.14013169870526e-06\\
23.3333333333333	5.12336966049395e-07\\
26.6666666666667	2.39334410575094e-07\\
30	1.10826084109788e-07\\
};
\addlegendentry{LMMSE}

\addplot [color=red, line width=1.5pt, only marks, mark size=4.5pt, mark=o, mark options={solid, red}]
  table[row sep=crcr]{%
0	0.000111215710150184\\
3.33333333333333	5.15365526494848e-05\\
6.66666666666667	2.37517401062638e-05\\
10	1.11540058283242e-05\\
13.3333333333333	5.2948098247366e-06\\
16.6666666666667	2.3752505541247e-06\\
20	1.10705406040965e-06\\
23.3333333333333	5.21546166056394e-07\\
26.6666666666667	2.3747673262739e-07\\
30	1.10083963726731e-07\\
};
\addlegendentry{MMSE - Simulation}

\addplot [color=black, dashdotted, line width=1.5pt]
  table[row sep=crcr]{%
0	0.000111301367212935\\
3.33333333333333	5.15522532699111e-05\\
6.66666666666667	2.37592323619241e-05\\
10	1.11575714363398e-05\\
13.3333333333333	5.29659046348157e-06\\
16.6666666666667	2.37599741140098e-06\\
20	1.10747644124543e-06\\
23.3333333333333	5.21787542750704e-07\\
26.6666666666667	2.37505282055604e-07\\
30	1.10131034886882e-07\\
};
\addlegendentry{MMSE - Analysis}

\addplot [color=mycolor2, line width=1.5pt, mark size=3.2pt, mark=square, mark options={solid, mycolor2}]
  table[row sep=crcr]{%
0	0.00177717689581859\\
3.33333333333333	0.000824523878448636\\
6.66666666666667	0.000374440230139226\\
10	0.000179076097585952\\
13.3333333333333	8.38655274137051e-05\\
16.6666666666667	3.85119138788897e-05\\
20	1.80251775132832e-05\\
23.3333333333333	8.32143013304576e-06\\
26.6666666666667	3.89564627986438e-06\\
30	1.74401332294989e-06\\
};
\addlegendentry{\cite{Abdallah2021,  Zhao2019, Liu2021, Zhu2018,  Yerzhanova2021}}

\addplot [color=mycolor2, line width=1.5pt, mark size=3.2pt, mark=square, mark options={solid, mycolor2}, forget plot]
  table[row sep=crcr]{%
0	0.000897322486760192\\
3.33333333333333	0.000410956564600314\\
6.66666666666667	0.000191950611296035\\
10	8.79479550798722e-05\\
13.3333333333333	4.18117421559669e-05\\
16.6666666666667	1.95445445359272e-05\\
20	8.85869854220279e-06\\
23.3333333333333	4.12689885866772e-06\\
26.6666666666667	1.87098990651117e-06\\
30	8.96964315932428e-07\\
};
\addplot [color=red, line width=1.5pt, mark size=3.0pt, mark=triangle, mark options={solid, red}, forget plot]
  table[row sep=crcr]{%
0	0.000111413081957948\\
3.33333333333333	5.12090562802674e-05\\
6.66666666666667	2.41824615800974e-05\\
10	1.10753850406318e-05\\
13.3333333333333	5.19871243256552e-06\\
16.6666666666667	2.40666415028786e-06\\
20	1.09795412494449e-06\\
23.3333333333333	5.19283882801183e-07\\
26.6666666666667	2.37877621789477e-07\\
30	1.12807981042943e-07\\
};
\addplot [color=mycolor1, dashed, line width=1.5pt, mark size=6.0pt, mark=+, mark options={solid, mycolor1}, forget plot]
  table[row sep=crcr]{%
0	0.000111675023852909\\
3.33333333333333	5.12482167558093e-05\\
6.66666666666667	2.42016349701718e-05\\
10	1.10768765609877e-05\\
13.3333333333333	5.2001282864322e-06\\
16.6666666666667	2.4068042102437e-06\\
20	1.09801660636878e-06\\
23.3333333333333	5.19285630040072e-07\\
26.6666666666667	2.37880072507802e-07\\
30	1.12808525094203e-07\\
};
\addplot [color=black!50!green, dotted, line width=1.5pt, forget plot]
  table[row sep=crcr]{%
0	0.000111408701421836\\
3.33333333333333	5.12109227304295e-05\\
6.66666666666667	2.41825076866922e-05\\
10	1.10752916703356e-05\\
13.3333333333333	5.19864963612312e-06\\
16.6666666666667	2.40665920882614e-06\\
20	1.09795840283797e-06\\
23.3333333333333	5.19283988341541e-07\\
26.6666666666667	2.37877550160913e-07\\
30	1.12807985721959e-07\\
};
\addplot [color=black, dashdotted, line width=1.5pt, forget plot]
  table[row sep=crcr]{%
0	0.000112086844987913\\
3.33333333333333	5.15444429148439e-05\\
6.66666666666667	2.35441707781749e-05\\
10	1.10938894540713e-05\\
13.3333333333333	5.22819061459234e-06\\
16.6666666666667	2.49015856880445e-06\\
20	1.10318990795468e-06\\
23.3333333333333	5.13321475420132e-07\\
26.6666666666667	2.39167881601132e-07\\
30	1.12068173529667e-07\\
};
\end{axis}

\begin{axis}[%
width=6.115in,
height=5.312in,
at={(0in,0in)},
scale only axis,
xmin=0,
xmax=1,
ymin=0,
ymax=1,
axis line style={draw=none},
ticks=none,
axis x line*=bottom,
axis y line*=left
]
\draw [black, line width=1.0pt] (axis cs:0.472361,0.391176) ellipse [x radius=0.0165661, y radius=0.0205882];

\draw (0.31552361,0.3751176) node[fill=white] {$K=\{7,15\}$} edge[-Stealth] (0.4552361,0.3751176);

\draw (0.362361,0.5351176) node[fill=white] {$K=7$} edge[-Stealth] (0.462361,0.5351176);

\draw (0.502361,0.6451176) node[fill=white] {$K=15$} edge[-Stealth] (0.402361,0.6451176);

\end{axis}

\end{tikzpicture}%

%% file: FiguresTxt/BC_MSE_T_16_K_7_15.tex
% This file was created by matlab2tikz.
%
%The latest updates can be retrieved from
%  http://www.mathworks.com/matlabcentral/fileexchange/22022-matlab2tikz-matlab2tikz
%where you can also make suggestions and rate matlab2tikz.
%
\definecolor{mycolor1}{rgb}{0.00000,0.45000,0.74000}%
\definecolor{mycolor2}{rgb}{1.00000,0.00000,1.00000}%
\begin{tikzpicture}

\begin{axis}[%
width=4.739in,
height=4.33in,
at={(0.795in,0.584in)},
scale only axis,
xmin=0,
xmax=30,
xlabel style={font=\color{white!15!black}},
xlabel={Transmit Power [dBm]},
ymode=log,
ymin=0.01,
ymax=1000,
yminorticks=true,
ylabel style={font=\color{white!15!black}},
ylabel={Normalized MSE},
axis background/.style={fill=white},
xmajorgrids,
ymajorgrids,
yminorgrids,
legend style={legend cell align=left, align=left, draw=white!15!black}
]
\addplot [color=red, line width=1.5pt, mark size=3.0pt, mark=triangle, mark options={solid, red}]
  table[row sep=crcr]{%
0	12.8953009750275\\
3.33333333333333	5.64777145796632\\
6.66666666666667	2.71677770401422\\
10	1.19585663690648\\
13.3333333333333	0.576318352018041\\
16.6666666666667	0.274955169050615\\
20	0.123650899191898\\
23.3333333333333	0.0575780780982064\\
26.6666666666667	0.0265442500248414\\
30	0.0127425332104926\\
};
\addlegendentry{LS}

\addplot [color=mycolor1, dashed, line width=1.5pt, mark size=6.0pt, mark=+, mark options={solid, mycolor1}]
  table[row sep=crcr]{%
0	12.8636259528839\\
3.33333333333333	5.64130190323658\\
6.66666666666667	2.71536300641281\\
10	1.19556318746892\\
13.3333333333333	0.576251964650035\\
16.6666666666667	0.274940637092499\\
20	0.123647879688582\\
23.3333333333333	0.0575773981044815\\
26.6666666666667	0.0265441168168754\\
30	0.0127425030696627\\
};
\addlegendentry{Scaled LS}

\addplot [color=black!50!green, dotted, line width=1.5pt]
  table[row sep=crcr]{%
0	1.98956899687851\\
3.33333333333333	1.8178331522497\\
6.66666666666667	1.43959992800084\\
10	0.864113444406961\\
13.3333333333333	0.490349121632064\\
16.6666666666667	0.254545777065064\\
20	0.119398696286945\\
23.3333333333333	0.0566598058864022\\
26.6666666666667	0.0263207600760819\\
30	0.0126925457855159\\
};
\addlegendentry{LMMSE}

\addplot [color=red, line width=1.5pt, only marks, mark size=4.5pt, mark=o, mark options={solid, red}]
  table[row sep=crcr]{%
0	1.0146968980452\\
3.33333333333333	0.951972690193764\\
6.66666666666667	0.877641943725905\\
10	0.6332065431742\\
13.3333333333333	0.40894925706608\\
16.6666666666667	0.224821076131298\\
20	0.113607315643135\\
23.3333333333333	0.0551070506458687\\
26.6666666666667	0.02600337435528\\
30	0.0121357484280178\\
};
\addlegendentry{MMSE - Simulation}

\addplot [color=black, dashdotted, line width=1.5pt]
  table[row sep=crcr]{%
0	1.02642964615778\\
3.33333333333333	0.947330314869305\\
6.66666666666667	0.879256286144802\\
10	0.632397253283271\\
13.3333333333333	0.410082173481651\\
16.6666666666667	0.225259426296528\\
20	0.113581395345147\\
23.3333333333333	0.0551404372834079\\
26.6666666666667	0.0260104022024637\\
30	0.0121390490656383\\
};
\addlegendentry{MMSE - Analysis}

\addplot [color=mycolor2, line width=1.5pt, mark size=3.2pt, mark=square, mark options={solid, mycolor2}]
  table[row sep=crcr]{%
0	415.104674161436\\
3.33333333333333	183.208900455013\\
6.66666666666667	87.4548541322389\\
10	39.444833442787\\
13.3333333333333	18.6391769424135\\
16.6666666666667	8.72949248882125\\
20	3.88884290070274\\
23.3333333333333	1.8342712671831\\
26.6666666666667	0.856161456167047\\
30	0.393730124914617\\
};
\addlegendentry{\cite{Abdallah2021,  Zhao2019, Liu2021, Zhu2018,  Yerzhanova2021}}

\addplot [color=mycolor2, line width=1.5pt, mark size=3.2pt, mark=square, mark options={solid, mycolor2}, forget plot]
  table[row sep=crcr]{%
0	192.060935321087\\
3.33333333333333	91.6485068736334\\
6.66666666666667	43.3682526771007\\
10	20.1124133769014\\
13.3333333333333	9.11210346450977\\
16.6666666666667	4.26612603383399\\
20	1.90296031603817\\
23.3333333333333	0.948846398159574\\
26.6666666666667	0.423770511217203\\
30	0.196117078671952\\
};
\end{axis}

\begin{axis}[%
width=6.115in,
height=5.312in,
at={(0in,0in)},
scale only axis,
xmin=0,
xmax=1,
ymin=0,
ymax=1,
axis line style={draw=none},
ticks=none,
axis x line*=bottom,
axis y line*=left
]
\draw [black, line width=1.0pt] (axis cs:0.559349,0.337752) ellipse [x radius=0.01646, y radius=0.029909];

\draw (0.405452361,0.31051176) node[fill=white] {$K=\{7,15\}$} edge[-Stealth] (0.5452361,0.31051176);

\draw (0.272361,0.6451176) node[fill=white] {$K=7$} edge[-Stealth] (0.372361,0.6451176);

\draw (0.502361,0.69951176) node[fill=white] {$K=15$} edge[-Stealth] (0.40552361,0.69951176);

\end{axis}
\end{tikzpicture}%

%% file: FiguresTxt/DC_MSE_T_32_64_K_31.tex
% This file was created by matlab2tikz.
%
%The latest updates can be retrieved from
%  http://www.mathworks.com/matlabcentral/fileexchange/22022-matlab2tikz-matlab2tikz
%where you can also make suggestions and rate matlab2tikz.
%
\definecolor{mycolor1}{rgb}{0.00000,0.45000,0.74000}%
\definecolor{mycolor2}{rgb}{1.00000,0.00000,1.00000}%
\begin{tikzpicture}

\begin{axis}[%
width=4.739in,
height=4.33in,
at={(0.795in,0.584in)},
scale only axis,
xmin=0,
xmax=30,
xlabel style={font=\color{white!15!black}},
xlabel={Transmit Power [dBm]},
ymode=log,
ymin=1e-08,
ymax=0.01,
yminorticks=true,
ylabel style={font=\color{white!15!black}},
ylabel={Normalized MSE},
axis background/.style={fill=white},
xmajorgrids,
ymajorgrids,
yminorgrids,
legend style={legend cell align=left, align=left, draw=white!15!black}
]
\addplot [color=red, line width=1.5pt, mark size=3.0pt, mark=triangle, mark options={solid, red}]
  table[row sep=crcr]{%
0	2.80424219883453e-05\\
3.33333333333333	1.28103805838562e-05\\
6.66666666666667	5.89999671286508e-06\\
10	2.7351857893696e-06\\
13.3333333333333	1.27547110544943e-06\\
16.6666666666667	6.06926248271838e-07\\
20	2.84189072692703e-07\\
23.3333333333333	1.27999915796357e-07\\
26.6666666666667	5.98106793843006e-08\\
30	2.78580719106946e-08\\
};
\addlegendentry{LS}

\addplot [color=mycolor1, dashed, line width=1.5pt, mark size=6.0pt, mark=+, mark options={solid, mycolor1}]
  table[row sep=crcr]{%
0	2.83647962325866e-05\\
3.33333333333333	1.29032028688515e-05\\
6.66666666666667	5.92058256250522e-06\\
10	2.73912782662864e-06\\
13.3333333333333	1.27627623550382e-06\\
16.6666666666667	6.06884589053181e-07\\
20	2.84246905455733e-07\\
23.3333333333333	1.28015644496349e-07\\
26.6666666666667	5.98046553884069e-08\\
30	2.78587042391587e-08\\
};
\addlegendentry{Scaled LS}

\addplot [color=black!50!green, dotted, line width=1.5pt]
  table[row sep=crcr]{%
0	2.80422676871122e-05\\
3.33333333333333	1.28100957959134e-05\\
6.66666666666667	5.89990915710593e-06\\
10	2.73517395926867e-06\\
13.3333333333333	1.2754717812128e-06\\
16.6666666666667	6.06927811395653e-07\\
20	2.84189852852914e-07\\
23.3333333333333	1.27999800449102e-07\\
26.6666666666667	5.98107232283341e-08\\
30	2.78580574939524e-08\\
};
\addlegendentry{LMMSE}

\addplot [color=red, line width=1.5pt, only marks, mark size=4.5pt, mark=o, mark options={solid, red}]
  table[row sep=crcr]{%
0	2.80175657403348e-05\\
3.33333333333333	1.28819750780957e-05\\
6.66666666666667	6.01617635021156e-06\\
10	2.82017865045688e-06\\
13.3333333333333	1.28615861475438e-06\\
16.6666666666667	6.11492835214379e-07\\
20	2.85434036691479e-07\\
23.3333333333333	1.28176603984623e-07\\
26.6666666666667	5.9514383627468e-08\\
30	2.75048395371442e-08\\
};
\addlegendentry{MMSE - Simulation}

\addplot [color=black, dashdotted, line width=1.5pt]
  table[row sep=crcr]{%
0	2.80272689238385e-05\\
3.33333333333333	1.28865467024951e-05\\
6.66666666666667	6.01812434526501e-06\\
10	2.82079719196186e-06\\
13.3333333333333	1.28664364363777e-06\\
16.6666666666667	6.11612188673941e-07\\
20	2.85507680710155e-07\\
23.3333333333333	1.28204794985129e-07\\
26.6666666666667	5.9530433780455e-08\\
30	2.75193228876602e-08\\
};
\addlegendentry{MMSE - Analysis}

\addplot [color=mycolor2, line width=1.5pt, mark size=3.2pt, mark=square, mark options={solid, mycolor2}]
  table[row sep=crcr]{%
0	0.000912536940816029\\
3.33333333333333	0.000418627536046655\\
6.66666666666667	0.000191333711725371\\
10	8.83809851537384e-05\\
13.3333333333333	4.10465077998548e-05\\
16.6666666666667	1.90432435672718e-05\\
20	8.85031929698057e-06\\
23.3333333333333	4.11846857770758e-06\\
26.6666666666667	1.94999894675846e-06\\
30	8.82775846344744e-07\\
};
\addlegendentry{\cite{Abdallah2021,  Zhao2019, Liu2021, Zhu2018,  Yerzhanova2021}}

\addplot [color=mycolor2, line width=1.5pt, mark size=3.2pt, mark=square, mark options={solid, mycolor2}, forget plot]
  table[row sep=crcr]{%
0	0.00180889138698362\\
3.33333333333333	0.000823732914818367\\
6.66666666666667	0.000390156221033539\\
10	0.000178420367746946\\
13.3333333333333	8.41781515840953e-05\\
16.6666666666667	3.86553755367512e-05\\
20	1.82738555715378e-05\\
23.3333333333333	8.40770134202041e-06\\
26.6666666666667	3.82974598766426e-06\\
30	1.77996885069339e-06\\
};
\addplot [color=red, line width=1.5pt, mark size=3.0pt, mark=triangle, mark options={solid, red}, forget plot]
  table[row sep=crcr]{%
0	5.59818161503824e-05\\
3.33333333333333	2.67185106005759e-05\\
6.66666666666667	1.20532908819536e-05\\
10	5.63721210235588e-06\\
13.3333333333333	2.63894078745201e-06\\
16.6666666666667	1.19330172774836e-06\\
20	5.58679146228745e-07\\
23.3333333333333	2.56759559560305e-07\\
26.6666666666667	1.20678231644166e-07\\
30	5.55059175461986e-08\\
};
\addplot [color=mycolor1, dashed, line width=1.5pt, mark size=6.0pt, mark=+, mark options={solid, mycolor1}, forget plot]
  table[row sep=crcr]{%
0	5.73817560967429e-05\\
3.33333333333333	2.70015498734097e-05\\
6.66666666666667	1.21126241700949e-05\\
10	5.64928450391178e-06\\
13.3333333333333	2.64315938908479e-06\\
16.6666666666667	1.19427226695374e-06\\
20	5.5874855199459e-07\\
23.3333333333333	2.56730994352761e-07\\
26.6666666666667	1.2066763932996e-07\\
30	5.55115620186401e-08\\
};
\addplot [color=black!50!green, dotted, line width=1.5pt, forget plot]
  table[row sep=crcr]{%
0	5.59821466608033e-05\\
3.33333333333333	2.67183799364627e-05\\
6.66666666666667	1.2053128299174e-05\\
10	5.63710721575707e-06\\
13.3333333333333	2.63893236855064e-06\\
16.6666666666667	1.19329718182731e-06\\
20	5.58678908310791e-07\\
23.3333333333333	2.5675984537603e-07\\
26.6666666666667	1.20678366382662e-07\\
30	5.55059315544341e-08\\
};
\addplot [color=red, line width=1.5pt, only marks, mark size=4.5pt, mark=o, mark options={solid, red}, forget plot]
  table[row sep=crcr]{%
0	5.57352618731543e-05\\
3.33333333333333	2.61986916804316e-05\\
6.66666666666667	1.20481815036008e-05\\
10	5.63573002024684e-06\\
13.3333333333333	2.62316251550738e-06\\
16.6666666666667	1.19907662831613e-06\\
20	5.59541411550576e-07\\
23.3333333333333	2.56477540133739e-07\\
26.6666666666667	1.20800325675279e-07\\
30	5.58336956036235e-08\\
};
\addplot [color=black, dashdotted, line width=1.5pt, forget plot]
  table[row sep=crcr]{%
0	5.57553318869689e-05\\
3.33333333333333	2.6203709424389e-05\\
6.66666666666667	1.20513804237817e-05\\
10	5.63727344435202e-06\\
13.3333333333333	2.62368850205387e-06\\
16.6666666666667	1.199542322673e-06\\
20	5.59683168707178e-07\\
23.3333333333333	2.56557299513518e-07\\
26.6666666666667	1.20817896636675e-07\\
30	5.58476126279084e-08\\
};

\draw (14.4,2.0e-4) node[fill=white] {$\tau=32$} edge[-Stealth] (10.4,2.0e-4);

\draw (5.6,0.85e-4) node[fill=white] {$\tau=64$} edge[-Stealth] (9.6,0.85e-4);

\draw (17.7,3.4e-6) node[fill=white] {$\tau=32$} edge[-Stealth] (13.7,3.4e-6);

\draw (8.8,1.1e-6) node[fill=white] {$\tau=64$} edge[-Stealth] (12.8,1.1e-6);

\end{axis}
\end{tikzpicture}%

%% file: FiguresTxt/BC_MSE_T_32_64_K_31.tex
% This file was created by matlab2tikz.
%
%The latest updates can be retrieved from
%  http://www.mathworks.com/matlabcentral/fileexchange/22022-matlab2tikz-matlab2tikz
%where you can also make suggestions and rate matlab2tikz.
%
\definecolor{mycolor1}{rgb}{0.00000,0.45000,0.74000}%
\definecolor{mycolor2}{rgb}{1.00000,0.00000,1.00000}%
\begin{tikzpicture}

\begin{axis}[%
width=4.739in,
height=4.33in,
at={(0.795in,0.584in)},
scale only axis,
xmin=0,
xmax=30,
xlabel style={font=\color{white!15!black}},
xlabel={Transmit Power [dBm]},
ymode=log,
ymin=0.001,
ymax=1000,
yminorticks=true,
ylabel style={font=\color{white!15!black}},
ylabel={Normalized MSE},
axis background/.style={fill=white},
xmajorgrids,
ymajorgrids,
yminorgrids,
legend style={legend cell align=left, align=left, draw=white!15!black}
]
\addplot [color=red, line width=1.5pt, mark size=3.0pt, mark=triangle, mark options={solid, red}]
  table[row sep=crcr]{%
0	3.12171366233889\\
3.33333333333333	1.4864001837923\\
6.66666666666667	0.699241381357251\\
10	0.311010499613943\\
13.3333333333333	0.144085451268621\\
16.6666666666667	0.0653475427033149\\
20	0.030930756062937\\
23.3333333333333	0.0143430531569834\\
26.6666666666667	0.00667462109016032\\
30	0.00314850649591805\\
};
\addlegendentry{LS}

\addplot [color=mycolor1, dashed, line width=1.5pt, mark size=6.0pt, mark=+, mark options={solid, mycolor1}]
  table[row sep=crcr]{%
0	3.11783219182855\\
3.33333333333333	1.48554849731654\\
6.66666666666667	0.699056172437289\\
10	0.310972285144758\\
13.3333333333333	0.144077251991083\\
16.6666666666667	0.0653457554579973\\
20	0.0309303734455697\\
23.3333333333333	0.0143429681312916\\
26.6666666666667	0.00667460425440707\\
30	0.00314850300228554\\
};
\addlegendentry{Scaled LS}

\addplot [color=black!50!green, dotted, line width=1.5pt]
  table[row sep=crcr]{%
0	1.50565853050031\\
3.33333333333333	1.01758469585466\\
6.66666666666667	0.581401506919609\\
10	0.284854842919964\\
13.3333333333333	0.138037264530414\\
16.6666666666667	0.0640890991315909\\
20	0.0306169709795617\\
23.3333333333333	0.0142890756638657\\
26.6666666666667	0.00665792513387703\\
30	0.00314528244518056\\
};
\addlegendentry{LMMSE}

\addplot [color=red, line width=1.5pt, only marks, mark size=4.5pt, mark=o, mark options={solid, red}]
  table[row sep=crcr]{%
0	0.880909015378182\\
3.33333333333333	0.698105609501764\\
6.66666666666667	0.481486059799931\\
10	0.258126754429374\\
13.3333333333333	0.13324677533662\\
16.6666666666667	0.0643781960664835\\
20	0.0303328088416587\\
23.3333333333333	0.0139065326582473\\
26.6666666666667	0.00656439334211866\\
30	0.00316576633446087\\
};
\addlegendentry{MMSE - Simulation}

\addplot [color=black, dashdotted, line width=1.5pt]
  table[row sep=crcr]{%
0	0.883272877582272\\
3.33333333333333	0.697159337235263\\
6.66666666666667	0.480261015418137\\
10	0.257667716913845\\
13.3333333333333	0.13338646093315\\
16.6666666666667	0.0643801745793701\\
20	0.030323756012705\\
23.3333333333333	0.0139188927089202\\
26.6666666666667	0.00656753297154737\\
30	0.00316680427053457\\
};
\addlegendentry{MMSE - Analysis}

\addplot [color=mycolor2, line width=1.5pt, mark size=3.2pt, mark=square, mark options={solid, mycolor2}]
  table[row sep=crcr]{%
0	203.54970480779\\
3.33333333333333	95.7106935673226\\
6.66666666666667	47.0674012363418\\
10	19.696375210252\\
13.3333333333333	9.40410124866106\\
16.6666666666667	4.26881730373171\\
20	1.96317162956347\\
23.3333333333333	0.906658569301156\\
26.6666666666667	0.443735439390584\\
30	0.195262139085713\\
};
\addlegendentry{\cite{Abdallah2021,  Zhao2019, Liu2021, Zhu2018,  Yerzhanova2021}}

\addplot [color=mycolor2, line width=1.5pt, mark size=3.2pt, mark=square, mark options={solid, mycolor2}, forget plot]
  table[row sep=crcr]{%
0	407.395591876831\\
3.33333333333333	186.547693616661\\
6.66666666666667	84.7704060767562\\
10	39.3914633900674\\
13.3333333333333	18.5839956946989\\
16.6666666666667	8.79332466064601\\
20	3.86727170216694\\
23.3333333333333	1.91534305071144\\
26.6666666666667	0.86863349322023\\
30	0.392661660015541\\
};
\addplot [color=red, line width=1.5pt, only marks, mark size=4.5pt, mark=o, mark options={solid, red}, forget plot]
  table[row sep=crcr]{%
0	0.982049194223682\\
3.33333333333333	0.869637579543217\\
6.66666666666667	0.659653357289111\\
10	0.42556785482894\\
13.3333333333333	0.244923389292195\\
16.6666666666667	0.123181147885855\\
20	0.0576745347141401\\
23.3333333333333	0.029002630272534\\
26.6666666666667	0.0134696578673809\\
30	0.00605810475563635\\
};
\addplot [color=black, dashdotted, line width=1.5pt, forget plot]
  table[row sep=crcr]{%
0	0.986092860445488\\
3.33333333333333	0.877195490457738\\
6.66666666666667	0.659568593672646\\
10	0.426019096805044\\
13.3333333333333	0.246018320128275\\
16.6666666666667	0.123149329964929\\
20	0.0578087559997422\\
23.3333333333333	0.0290285816950707\\
26.6666666666667	0.0134774749219632\\
30	0.00605842258751411\\
};
\addplot [color=red, line width=1.5pt, mark size=3.0pt, mark=triangle, mark options={solid, red}, forget plot]
  table[row sep=crcr]{%
0	6.24619270445372\\
3.33333333333333	3.03712578715145\\
6.66666666666667	1.32397238925765\\
10	0.616271735716631\\
13.3333333333333	0.288257842440196\\
16.6666666666667	0.135741794018139\\
20	0.0603741493775073\\
23.3333333333333	0.0298777501188417\\
26.6666666666667	0.0133807144595978\\
30	0.00616100806690784\\
};
\addplot [color=black!50!green, dotted, line width=1.5pt, forget plot]
  table[row sep=crcr]{%
0	1.80842888582819\\
3.33333333333333	1.50343717860871\\
6.66666666666667	0.920419088168691\\
10	0.516029903477742\\
13.3333333333333	0.264680105002741\\
16.6666666666667	0.130598947148081\\
20	0.0593115180756356\\
23.3333333333333	0.0296237238769076\\
26.6666666666667	0.0133311651578731\\
30	0.00615313880109673\\
};
\addplot [color=mycolor1, dashed, line width=1.5pt, mark size=6.0pt, mark=+, mark options={solid, mycolor1}, forget plot]
  table[row sep=crcr]{%
0	6.23096238222911\\
3.33333333333333	3.03361742379573\\
6.66666666666667	1.32327290513871\\
10	0.616122926713299\\
13.3333333333333	0.288224618449239\\
16.6666666666667	0.135734458904987\\
20	0.0603726394814945\\
23.3333333333333	0.0298774092563748\\
26.6666666666667	0.0133806439921681\\
30	0.00616099151293285\\
};
\end{axis}

\begin{axis}[%
width=6.115in,
height=5.312in,
at={(0in,0in)},
scale only axis,
xmin=0,
xmax=1,
ymin=0,
ymax=1,
axis line style={draw=none},
ticks=none,
axis x line*=bottom,
axis y line*=left
]
\draw [black, line width=1.0pt] (axis cs:0.474741,0.398572) ellipse [x radius=0.0141855, y radius=0.0221012];
\draw [black, line width=1.0pt] (axis cs:0.55992,0.398572) ellipse [x radius=0.0141855, y radius=0.0221012];

\draw (0.502361,0.7409) node[fill=white] {$\tau=32$} edge[-Stealth] (0.40552361,0.74095);

\draw (0.272361,0.685) node[fill=white] {$\tau=64$} edge[-Stealth] (0.372361,0.685);

\draw (0.671992,0.418572) node[fill=white] {$\tau=32$} edge[-Stealth] (0.571992,0.418572);

\draw (0.3570992,0.3818572) node[fill=white] {$\tau=64$} edge[-Stealth] (0.4570992,0.3818572);

\end{axis}
\end{tikzpicture}%

%% file: FiguresTxt/DC_tau_K2.tex
% This file was created by matlab2tikz.
%
%The latest updates can be retrieved from
%  http://www.mathworks.com/matlabcentral/fileexchange/22022-matlab2tikz-matlab2tikz
%where you can also make suggestions and rate matlab2tikz.
%
\definecolor{mycolor1}{rgb}{0.00000,0.45000,0.74000}%
\definecolor{mycolor2}{rgb}{1.00000,0.00000,1.00000}%
\begin{tikzpicture}

\begin{axis}[%
width=4.739in,
height=4.33in,
at={(0.795in,0.584in)},
scale only axis,
xmin=5,
xmax=41,
xlabel style={font=\color{white!15!black}},
xlabel={$\tau$},
ymode=log,
ymin=4.41642915346036e-07,
ymax=1e-05,
yminorticks=true,
ylabel style={font=\color{white!15!black}},
ylabel={Normalized MSE},
axis background/.style={fill=white},
xmajorgrids,
ymajorgrids,
yminorgrids,
legend style={legend cell align=left, align=left, draw=white!15!black}
]
\addplot [color=red, line width=1.5pt, mark size=3.0pt, mark=triangle, mark options={solid, red}]
  table[row sep=crcr]{%
5	3.67267214484883e-06\\
9	2.35220178465723e-06\\
13	1.78981186915818e-06\\
17	1.19972207751319e-06\\
21	9.23510089516211e-07\\
25	7.57997584782586e-07\\
29	6.46120695970264e-07\\
33	5.63664049737031e-07\\
37	4.89066001542234e-07\\
41	4.42634330948679e-07\\
};
\addlegendentry{LS}

\addplot [color=mycolor1, dashed, line width=1.5pt, mark size=6.0pt, mark=+, mark options={solid, mycolor1}]
  table[row sep=crcr]{%
5	3.67270982615385e-06\\
9	2.3522165678866e-06\\
13	1.7898221579851e-06\\
17	1.19972348674819e-06\\
21	9.23503885452946e-07\\
25	7.57995165346257e-07\\
29	6.46118925470218e-07\\
33	5.63662971746008e-07\\
37	4.89065812212344e-07\\
41	4.42645484615601e-07\\
};
\addlegendentry{Scaled LS}

\addplot [color=black!50!green, dotted, line width=1.5pt]
  table[row sep=crcr]{%
5	3.67265971705065e-06\\
9	2.35219655625191e-06\\
13	1.78980875018656e-06\\
17	1.19972190891012e-06\\
21	9.23510327970525e-07\\
25	7.57997985202063e-07\\
29	6.46121149571971e-07\\
33	5.6366439305648e-07\\
37	4.8906601378479e-07\\
41	4.42634184897377e-07\\
};
\addlegendentry{LMMSE}

\addplot [color=red, line width=1.5pt, only marks, mark size=4.5pt, mark=o, mark options={solid, red}]
  table[row sep=crcr]{%
5	3.66400122140655e-06\\
9	2.34663758154655e-06\\
13	1.79145439750808e-06\\
17	1.20128294137592e-06\\
21	9.22773369480574e-07\\
25	7.57335201976581e-07\\
29	6.42048792791631e-07\\
33	5.58291249197963e-07\\
37	4.89215693370901e-07\\
41	4.41642915346036e-07\\
};
\addlegendentry{MMSE - Simulation}

\addplot [color=black, dashdotted, line width=1.5pt]
  table[row sep=crcr]{%
5	3.66515598623762e-06\\
9	2.3471780994549e-06\\
13	1.79183876603578e-06\\
17	1.20146877597139e-06\\
21	9.22910319091576e-07\\
25	7.57473998774404e-07\\
29	6.42181447275721e-07\\
33	5.5841537386255e-07\\
37	4.89333788764454e-07\\
41	4.4174746625691e-07\\
};
\addlegendentry{MMSE - Analysis}

\addplot [color=mycolor2, line width=1.5pt, mark size=3.2pt, mark=square, mark options={solid, mycolor2}]
  table[row sep=crcr]{%
5	6.05641419079623e-06\\
9	4.7139917016902e-06\\
13	3.79583201613683e-06\\
17	2.97679244682795e-06\\
21	2.40925721032642e-06\\
25	1.98759275875585e-06\\
29	1.75056220976653e-06\\
33	1.53628272078099e-06\\
37	1.36363646715961e-06\\
41	1.19471015772932e-06\\
};
\addlegendentry{\cite{Abdallah2021,  Zhao2019, Liu2021, Zhu2018,  Yerzhanova2021}}

\end{axis}
\end{tikzpicture}%

%% file: FiguresTxt/BC_tau_K2.tex
% This file was created by matlab2tikz.
%
%The latest updates can be retrieved from
%  http://www.mathworks.com/matlabcentral/fileexchange/22022-matlab2tikz-matlab2tikz
%where you can also make suggestions and rate matlab2tikz.
%
\definecolor{mycolor1}{rgb}{0.00000,0.45000,0.74000}%
\definecolor{mycolor2}{rgb}{1.00000,0.00000,1.00000}%
\begin{tikzpicture}

\begin{axis}[%
width=4.739in,
height=4.33in,
at={(0.795in,0.584in)},
scale only axis,
xmin=5,
xmax=41,
xlabel style={font=\color{white!15!black}},
xlabel={$\tau$},
ymode=log,
ymin=0.046620321487773,
ymax=1.32587812807944,
yminorticks=true,
ylabel style={font=\color{white!15!black}},
ylabel={Normalized MSE},
axis background/.style={fill=white},
xmajorgrids,
ymajorgrids,
yminorgrids,
legend style={legend cell align=left, align=left, draw=white!15!black}
]
\addplot [color=red, line width=1.5pt, mark size=3.0pt, mark=triangle, mark options={solid, red}]
  table[row sep=crcr]{%
5	0.392742803119166\\
9	0.261833252148617\\
13	0.200333929811666\\
17	0.138855982619966\\
21	0.105875102600948\\
25	0.0862960701510064\\
29	0.0722844258249287\\
33	0.062927511995481\\
37	0.0534546414731587\\
41	0.0471469518900829\\
};
\addlegendentry{LS}

\addplot [color=mycolor1, dashed, line width=1.5pt, mark size=6.0pt, mark=+, mark options={solid, mycolor1}]
  table[row sep=crcr]{%
5	0.392736946264759\\
9	0.261830361342743\\
13	0.20033202065277\\
17	0.138855193362699\\
21	0.105874659850514\\
25	0.0862957825424172\\
29	0.0722842250104226\\
33	0.0629273608444985\\
37	0.0534545322657034\\
41	0.0471468670194844\\
};
\addlegendentry{Scaled LS}

\addplot [color=black!50!green, dotted, line width=1.5pt]
  table[row sep=crcr]{%
5	0.35826971879689\\
9	0.244444969580918\\
13	0.188783878619593\\
17	0.133892252892336\\
21	0.103070097568545\\
25	0.0844579069993556\\
29	0.0710177756167147\\
33	0.0619765376875101\\
37	0.0527784329585654\\
41	0.046620321487773\\
};
\addlegendentry{LMMSE}

\addplot [color=red, line width=1.5pt, only marks, mark size=4.5pt, mark=o, mark options={solid, red}]
  table[row sep=crcr]{%
5	0.307841159675231\\
9	0.216775763444242\\
13	0.169894752068834\\
17	0.124535059065942\\
21	0.0978459996145841\\
25	0.081034033972576\\
29	0.0690617555669278\\
33	0.0608141991117861\\
37	0.0522155538520259\\
41	0.0473459048656412\\
};
\addlegendentry{MMSE - Simulation}

\addplot [color=black, dashdotted, line width=1.5pt]
  table[row sep=crcr]{%
5	0.307382634226004\\
9	0.216737552116595\\
13	0.169871214590495\\
17	0.124597134505935\\
21	0.0978441950352015\\
25	0.0810211117234305\\
29	0.069056350087928\\
33	0.0608000493317871\\
37	0.052205530420449\\
41	0.0473248747967114\\
};
\addlegendentry{MMSE - Analysis}

\addplot [color=mycolor2, line width=1.5pt, mark size=3.2pt, mark=square, mark options={solid, mycolor2}]
  table[row sep=crcr]{%
5	1.32587812807944\\
9	1.05706561193351\\
13	0.850820566357313\\
17	0.678056358156242\\
21	0.544221361289155\\
25	0.447511566386218\\
29	0.391535612290978\\
33	0.344478896162235\\
37	0.300783341179299\\
41	0.265825519144605\\
};
\addlegendentry{\cite{Abdallah2021,  Zhao2019, Liu2021, Zhu2018,  Yerzhanova2021}}

\end{axis}
\end{tikzpicture}%

%% file: FiguresTxt/DC_T_63_K.tex
% This file was created by matlab2tikz.
%
%The latest updates can be retrieved from
%  http://www.mathworks.com/matlabcentral/fileexchange/22022-matlab2tikz-matlab2tikz
%where you can also make suggestions and rate matlab2tikz.
%
\definecolor{mycolor1}{rgb}{0.00000,0.45000,0.74000}%
\definecolor{mycolor2}{rgb}{1.00000,0.00000,1.00000}%
\begin{tikzpicture}

\begin{axis}[%
width=4.739in,
height=4.33in,
at={(0.798in,0.586in)},
scale only axis,
xmin=0,
xmax=63,
xlabel style={font=\color{white!15!black}},
xlabel={$K$},
ymode=log,
ymin=2e-07,
ymax=2e-05,
yminorticks=true,
ylabel style={font=\color{white!15!black}},
ylabel={Normalized MSE},
axis background/.style={fill=white},
xmajorgrids,
ymajorgrids,
yminorgrids,
legend style={at={(0.590,0.59)}, anchor=south west, legend cell align=left, align=left, draw=white!15!black}
]
\addplot [color=red, line width=1.5pt, mark size=3.0pt, mark=triangle, mark options={solid, red}]
  table[row sep=crcr]{%
0	2.8266629241481e-07\\
1	2.77733916883078e-07\\
8	2.83416295098189e-07\\
15	2.84578268170511e-07\\
22	2.76398850332888e-07\\
29	2.84528558554295e-07\\
35	2.83117826774992e-07\\
42	2.81313503402465e-07\\
49	2.84694902200991e-07\\
56	2.7886700544636e-07\\
63	2.80065772857037e-07\\
};
\addlegendentry{LS}

\addplot [color=mycolor1, dashed, line width=1.5pt, mark size=6.0pt, mark=+, mark options={solid, mycolor1}]
  table[row sep=crcr]{%
0	2.82666949957431e-07\\
1	2.77735607119679e-07\\
8	2.83408281053729e-07\\
15	2.84600680180903e-07\\
22	2.76374453268308e-07\\
29	2.84535930907201e-07\\
35	2.83206567457092e-07\\
42	2.81419424356143e-07\\
49	2.84776710829358e-07\\
56	2.79050357799055e-07\\
63	2.80201988298491e-07\\
};
\addlegendentry{Scaled LS}

\addplot [color=black!50!green, dotted, line width=1.5pt]
  table[row sep=crcr]{%
0	2.82666349027661e-07\\
1	2.77733363103329e-07\\
8	2.83417335440409e-07\\
15	2.84578089151702e-07\\
22	2.76399359688064e-07\\
29	2.8452849243774e-07\\
35	2.83117459481387e-07\\
42	2.81313495001751e-07\\
49	2.84695099906276e-07\\
56	2.78866509406292e-07\\
63	2.80066027404404e-07\\
};
\addlegendentry{LMMSE}

\addplot [color=red, line width=1.5pt, only marks, mark size=4.5pt, mark=o, mark options={solid, red}]
  table[row sep=crcr]{%
0	2.81938021496919e-07\\
1	2.77615758308588e-07\\
8	2.84063914820479e-07\\
15	2.74855820079788e-07\\
22	2.82000438974153e-07\\
29	2.7643915117608e-07\\
35	2.75792635932944e-07\\
42	2.80976309019322e-07\\
49	2.83048494662846e-07\\
56	2.76516784228307e-07\\
63	2.83479704264161e-07\\
};
\addlegendentry{MMSE - Simulation}

\addplot [color=black, dashdotted, line width=1.5pt]
  table[row sep=crcr]{%
0	2.82031677358161e-07\\
1	2.77666104765387e-07\\
8	2.84108246647175e-07\\
15	2.74923135190066e-07\\
22	2.820474317863e-07\\
29	2.76462547728079e-07\\
35	2.75869982041473e-07\\
42	2.81052685795764e-07\\
49	2.83134715770827e-07\\
56	2.76597766647501e-07\\
63	2.83609769369454e-07\\
};
\addlegendentry{MMSE - Analysis}

\addplot [color=mycolor2, line width=1.5pt, mark size=3.2pt, mark=square, mark options={solid, mycolor2}]
  table[row sep=crcr]{%
0	2.85134862455841e-07\\
1	1.15089476350733e-06\\
8	2.74752890518271e-06\\
15	4.48451647198943e-06\\
22	6.1266535810332e-06\\
29	9.14074799770822e-06\\
35	1.18540396433871e-05\\
42	1.42607709486326e-05\\
49	1.60831332702492e-05\\
56	1.79889864315286e-05\\
63	1.80818843045721e-05\\
};
\addlegendentry{\cite{Abdallah2021,  Zhao2019, Liu2021, Zhu2018,  Yerzhanova2021}}

\end{axis}

\begin{axis}[%
width=6.135in,
height=5.323in,
at={(0in,0in)},
scale only axis,
xmin=0,
xmax=1,
ymin=0,
ymax=1,
axis line style={draw=none},
ticks=none,
axis x line*=bottom,
axis y line*=left
]
\end{axis}
\end{tikzpicture}%

%% file: FiguresTxt/BC_T_63_K.tex
% This file was created by matlab2tikz.
%
%The latest updates can be retrieved from
%  http://www.mathworks.com/matlabcentral/fileexchange/22022-matlab2tikz-matlab2tikz
%where you can also make suggestions and rate matlab2tikz.
%
\definecolor{mycolor1}{rgb}{0.00000,0.45000,0.74000}%
\definecolor{mycolor2}{rgb}{1.00000,0.00000,1.00000}%
\begin{tikzpicture}

\begin{axis}[%
width=4.739in,
height=4.33in,
at={(0.798in,0.586in)},
scale only axis,
unbounded coords=jump,
xmin=0,
xmax=63,
xlabel style={font=\color{white!15!black}},
xlabel={$K$},
ymode=log,
ymin=0.01,
ymax=2.4,
yminorticks=true,
ylabel style={font=\color{white!15!black}},
ylabel={Normalized MSE},
axis background/.style={fill=white},
xmajorgrids,
ymajorgrids,
yminorgrids,
legend style={at={(0.59,0.59)}, anchor=south west, legend cell align=left, align=left, draw=white!15!black}
]
\addplot [color=red, line width=1.5pt, mark size=3.0pt, mark=triangle, mark options={solid, red}]
  table[row sep=crcr]{%
0	nan\\
1	0.0158038763211165\\
8	0.0166282933064149\\
15	0.0167003805428395\\
22	0.016633271671874\\
29	0.0165464922366121\\
35	0.016591010559543\\
42	0.0166430699086518\\
49	0.0167175853327966\\
56	0.0165251687468197\\
63	0.0166663943967686\\
};
\addlegendentry{LS}

\addplot [color=mycolor1, dashed, line width=1.5pt, mark size=6.0pt, mark=+, mark options={solid, mycolor1}]
  table[row sep=crcr]{%
0	nan\\
1	0.0158038635183263\\
8	0.0166282348751339\\
15	0.0167002782515869\\
22	0.0166331257750127\\
29	0.0165463007831873\\
35	0.0165907824270227\\
42	0.0166427952151643\\
49	0.0167172618488278\\
56	0.0165248080597315\\
63	0.0166659847274434\\
};
\addlegendentry{Scaled LS}

\addplot [color=black!50!green, dotted, line width=1.5pt]
  table[row sep=crcr]{%
0	nan\\
1	0.015744883443894\\
8	0.0165529585238826\\
15	0.0166200714709172\\
22	0.0165504733396875\\
29	0.016465174225782\\
35	0.0165076208784202\\
42	0.0165591857086401\\
49	0.0166338549673686\\
56	0.01644251517583\\
63	0.0165837346459978\\
};
\addlegendentry{LMMSE}

\addplot [color=red, line width=1.5pt, only marks, mark size=4.5pt, mark=o, mark options={solid, red}]
  table[row sep=crcr]{%
0	nan\\
1	0.0157969403675691\\
8	0.0164715617414193\\
15	0.0163898197911452\\
22	0.0163696289049961\\
29	0.0163535654717066\\
35	0.0163854472198688\\
42	0.0164076983481664\\
49	0.0165910301185001\\
56	0.0163334293556547\\
63	0.0164818570679333\\
};
\addlegendentry{MMSE - Simulation}

\addplot [color=black, dashdotted, line width=1.5pt]
  table[row sep=crcr]{%
0	nan\\
1	0.0157971437436545\\
8	0.0164762996104138\\
15	0.016390681359013\\
22	0.0163735818996608\\
29	0.0163560302853817\\
35	0.0163897450640578\\
42	0.0164120517690319\\
49	0.0165941947610414\\
56	0.0163381266650651\\
63	0.0164859653112562\\
};
\addlegendentry{MMSE - Analysis}

\addplot [color=mycolor2, line width=1.5pt, mark size=3.2pt, mark=square, mark options={solid, mycolor2}]
  table[row sep=crcr]{%
0	-0.157541228175172\\
1	0.0709804279023704\\
8	0.295820636165051\\
15	0.517195669299644\\
22	0.759184686166744\\
29	0.961181979778966\\
35	1.36599240190573\\
42	1.82629063375193\\
49	2.12957786655849\\
56	2.12922329020325\\
63	2.1285617110514\\
};
\addlegendentry{\cite{Abdallah2021,  Zhao2019, Liu2021, Zhu2018,  Yerzhanova2021}}

\end{axis}

\begin{axis}[%
width=6.135in,
height=5.323in,
at={(0in,0in)},
scale only axis,
xmin=0,
xmax=1,
ymin=0,
ymax=1,
axis line style={draw=none},
ticks=none,
axis x line*=bottom,
axis y line*=left
]
\end{axis}
\end{tikzpicture}%

%% file: FiguresTxt/DC_MSE_T_32_33_K_31_Had_ZC.tex
% This file was created by matlab2tikz.
%
%The latest updates can be retrieved from
%  http://www.mathworks.com/matlabcentral/fileexchange/22022-matlab2tikz-matlab2tikz
%where you can also make suggestions and rate matlab2tikz.
%
\definecolor{mycolor1}{rgb}{0.00000,0.45000,0.74000}%
\definecolor{mycolor2}{rgb}{1.00000,0.00000,1.00000}%
\begin{tikzpicture}

\begin{axis}[%
width=4.739in,
height=4.33in,
at={(0.795in,0.584in)},
scale only axis,
xmin=0,
xmax=30,
xlabel style={font=\color{white!15!black}},
xlabel={Transmit Power [dBm]},
ymode=log,
ymin=1e-08,
ymax=0.01,
yminorticks=true,
ylabel style={font=\color{white!15!black}},
ylabel={Normalized MSE},
axis background/.style={fill=white},
xmajorgrids,
ymajorgrids,
yminorgrids,
legend style={legend cell align=left, align=left, draw=white!15!black}
]
\addplot [color=red, line width=1.5pt, mark size=3.0pt, mark=triangle, mark options={solid, red}]
  table[row sep=crcr]{%
0	5.34831969042502e-05\\
3.33333333333333	2.50342420365356e-05\\
6.66666666666667	1.172949553748e-05\\
10	5.44341420208556e-06\\
13.3333333333333	2.5709692672392e-06\\
16.6666666666667	1.16438993688927e-06\\
20	5.43987917708529e-07\\
23.3333333333333	2.50337852155217e-07\\
26.6666666666667	1.17889251521961e-07\\
30	5.55335375271633e-08\\
};
\addlegendentry{LS}

\addplot [color=mycolor1, dashed, line width=1.5pt, mark size=6.0pt, mark=+, mark options={solid, mycolor1}]
  table[row sep=crcr]{%
0	5.49004812905737e-05\\
3.33333333333333	2.53403299165516e-05\\
6.66666666666667	1.17852466381891e-05\\
10	5.45664248178483e-06\\
13.3333333333333	2.57449303696988e-06\\
16.6666666666667	1.16480117058557e-06\\
20	5.43963852408759e-07\\
23.3333333333333	2.50328864522914e-07\\
26.6666666666667	1.17908429314884e-07\\
30	5.5538689321339e-08\\
};
\addlegendentry{Scaled LS}

\addplot [color=black!50!green, dotted, line width=1.5pt]
  table[row sep=crcr]{%
0	5.34839036822125e-05\\
3.33333333333333	2.50344309322461e-05\\
6.66666666666667	1.17296807330803e-05\\
10	5.44340072548855e-06\\
13.3333333333333	2.57096860997529e-06\\
16.6666666666667	1.16439078762846e-06\\
20	5.4398853847791e-07\\
23.3333333333333	2.50337960462922e-07\\
26.6666666666667	1.17889254068799e-07\\
30	5.55334706686973e-08\\
};
\addlegendentry{LMMSE}

\addplot [color=red, line width=1.5pt, only marks, mark size=4.5pt, mark=o, mark options={solid, red}]
  table[row sep=crcr]{%
0	5.34737076735742e-05\\
3.33333333333333	2.50821575286961e-05\\
6.66666666666667	1.17149030427623e-05\\
10	5.44593940139342e-06\\
13.3333333333333	2.51320017696775e-06\\
16.6666666666667	1.17091401854532e-06\\
20	5.46308703323777e-07\\
23.3333333333333	2.54940319317615e-07\\
26.6666666666667	1.16384308657553e-07\\
30	5.45388165437603e-08\\
};
\addlegendentry{MMSE - Simulation}

\addplot [color=black, dashdotted, line width=1.5pt]
  table[row sep=crcr]{%
0	5.34917108422132e-05\\
3.33333333333333	2.50967321190438e-05\\
6.66666666666667	1.17196218909524e-05\\
10	5.44684055412285e-06\\
13.3333333333333	2.51390541911702e-06\\
16.6666666666667	1.17131043652912e-06\\
20	5.4657773185844e-07\\
23.3333333333333	2.5503212765426e-07\\
26.6666666666667	1.16409654301618e-07\\
30	5.45522459483005e-08\\
};
\addlegendentry{MMSE - Analysis}

\addplot [color=mycolor2, line width=1.5pt, mark size=3.2pt, mark=square, mark options={solid, mycolor2}]
  table[row sep=crcr]{%
0	0.000890726638781041\\
3.33333333333333	0.000410901680103096\\
6.66666666666667	0.000191852340704107\\
10	8.92768671613451e-05\\
13.3333333333333	4.14647443722798e-05\\
16.6666666666667	1.90119836713683e-05\\
20	8.81935223885348e-06\\
23.3333333333333	4.18394854013903e-06\\
26.6666666666667	1.96565298391368e-06\\
30	8.84152362074344e-07\\
};
\addlegendentry{\cite{Abdallah2021,  Zhao2019, Liu2021, Zhu2018,  Yerzhanova2021}}

\addplot [color=mycolor2, line width=1.5pt, mark size=3.2pt, mark=square, mark options={solid, mycolor2}, forget plot]
  table[row sep=crcr]{%
0	0.00180889138698362\\
3.33333333333333	0.000823732914818367\\
6.66666666666667	0.000390156221033539\\
10	0.000178420367746946\\
13.3333333333333	8.41781515840953e-05\\
16.6666666666667	3.86553755367512e-05\\
20	1.82738555715378e-05\\
23.3333333333333	8.40770134202041e-06\\
26.6666666666667	3.82974598766426e-06\\
30	1.77996885069339e-06\\
};
\addplot [color=red, line width=1.5pt, mark size=3.0pt, mark=triangle, mark options={solid, red}, forget plot]
  table[row sep=crcr]{%
0	5.59818161503824e-05\\
3.33333333333333	2.67185106005759e-05\\
6.66666666666667	1.20532908819536e-05\\
10	5.63721210235588e-06\\
13.3333333333333	2.63894078745201e-06\\
16.6666666666667	1.19330172774836e-06\\
20	5.58679146228745e-07\\
23.3333333333333	2.56759559560305e-07\\
26.6666666666667	1.20678231644166e-07\\
30	5.55059175461986e-08\\
};
\addplot [color=mycolor1, dashed, line width=1.5pt, mark size=6.0pt, mark=+, mark options={solid, mycolor1}, forget plot]
  table[row sep=crcr]{%
0	5.73817560967429e-05\\
3.33333333333333	2.70015498734097e-05\\
6.66666666666667	1.21126241700949e-05\\
10	5.64928450391178e-06\\
13.3333333333333	2.64315938908479e-06\\
16.6666666666667	1.19427226695374e-06\\
20	5.5874855199459e-07\\
23.3333333333333	2.56730994352761e-07\\
26.6666666666667	1.2066763932996e-07\\
30	5.55115620186401e-08\\
};
\addplot [color=black!50!green, dotted, line width=1.5pt, forget plot]
  table[row sep=crcr]{%
0	5.59821466608033e-05\\
3.33333333333333	2.67183799364627e-05\\
6.66666666666667	1.2053128299174e-05\\
10	5.63710721575707e-06\\
13.3333333333333	2.63893236855064e-06\\
16.6666666666667	1.19329718182731e-06\\
20	5.58678908310791e-07\\
23.3333333333333	2.5675984537603e-07\\
26.6666666666667	1.20678366382662e-07\\
30	5.55059315544341e-08\\
};
\addplot [color=red, line width=1.5pt, only marks, mark size=4.5pt, mark=o, mark options={solid, red}, forget plot]
  table[row sep=crcr]{%
0	5.57352618731543e-05\\
3.33333333333333	2.61986916804316e-05\\
6.66666666666667	1.20481815036008e-05\\
10	5.63573002024684e-06\\
13.3333333333333	2.62316251550738e-06\\
16.6666666666667	1.19907662831613e-06\\
20	5.59541411550576e-07\\
23.3333333333333	2.56477540133739e-07\\
26.6666666666667	1.20800325675279e-07\\
30	5.58336956036235e-08\\
};
\addplot [color=black, dashdotted, line width=1.5pt, forget plot]
  table[row sep=crcr]{%
0	5.57553318869689e-05\\
3.33333333333333	2.6203709424389e-05\\
6.66666666666667	1.20513804237817e-05\\
10	5.63727344435202e-06\\
13.3333333333333	2.62368850205387e-06\\
16.6666666666667	1.199542322673e-06\\
20	5.59683168707178e-07\\
23.3333333333333	2.56557299513518e-07\\
26.6666666666667	1.20817896636675e-07\\
30	5.58476126279084e-08\\
};
\end{axis}

\begin{axis}[%
width=0.912in,
height=0.866in,
at={(1.481in,0.877in)},
scale only axis,
xmin=16.5604437138091,
xmax=16.7688493871954,
ymode=log,
ymin=1.15175942166285e-06,
ymax=1.22786097570216e-06,
yminorticks=true,
axis background/.style={fill=white},
xmajorgrids,
ymajorgrids,
yminorgrids
]
\addplot [color=black, dashdotted, line width=1.5pt, forget plot]
  table[row sep=crcr]{%
0	5.34917108422132e-05\\
3.33333333333333	2.50967321190438e-05\\
6.66666666666667	1.17196218909524e-05\\
10	5.44684055412285e-06\\
13.3333333333333	2.51390541911702e-06\\
16.6666666666667	1.17131043652912e-06\\
20	5.4657773185844e-07\\
23.3333333333333	2.5503212765426e-07\\
26.6666666666667	1.16409654301618e-07\\
30	5.45522459483005e-08\\
};
\addplot [color=red, line width=1.5pt, only marks, mark size=4.5pt, mark=o, mark options={solid, red}, forget plot]
  table[row sep=crcr]{%
0	5.34737076735742e-05\\
3.33333333333333	2.50821575286961e-05\\
6.66666666666667	1.17149030427623e-05\\
10	5.44593940139342e-06\\
13.3333333333333	2.51320017696775e-06\\
16.6666666666667	1.17091401854532e-06\\
20	5.46308703323777e-07\\
23.3333333333333	2.54940319317615e-07\\
26.6666666666667	1.16384308657553e-07\\
30	5.45388165437603e-08\\
};
\addplot [color=black, dashdotted, line width=1.5pt, forget plot]
  table[row sep=crcr]{%
0	5.57553318869689e-05\\
3.33333333333333	2.6203709424389e-05\\
6.66666666666667	1.20513804237817e-05\\
10	5.63727344435202e-06\\
13.3333333333333	2.62368850205387e-06\\
16.6666666666667	1.199542322673e-06\\
20	5.59683168707178e-07\\
23.3333333333333	2.56557299513518e-07\\
26.6666666666667	1.20817896636675e-07\\
30	5.58476126279084e-08\\
};
\addplot [color=red, line width=1.5pt, only marks, mark size=4.5pt, mark=o, mark options={solid, red}, forget plot]
  table[row sep=crcr]{%
0	5.57352618731543e-05\\
3.33333333333333	2.61986916804316e-05\\
6.66666666666667	1.20481815036008e-05\\
10	5.63573002024684e-06\\
13.3333333333333	2.62316251550738e-06\\
16.6666666666667	1.19907662831613e-06\\
20	5.59541411550576e-07\\
23.3333333333333	2.56477540133739e-07\\
26.6666666666667	1.20800325675279e-07\\
30	5.58336956036235e-08\\
};
\end{axis}

\begin{axis}[%
width=6.115in,
height=5.312in,
at={(0in,0in)},
scale only axis,
xmin=0,
xmax=1,
ymin=0,
ymax=1,
axis line style={draw=none},
ticks=none,
axis x line*=bottom,
axis y line*=left
]
\draw[-{Stealth}, color=black, line width=1.0pt] (axis cs:0.547,0.379) -- (axis cs:0.393,0.326);
\draw[line width=1.0pt, draw=black] (axis cs:0.545267489711934,0.373482277526395) rectangle (axis cs:0.576131687242798,0.41434766214178);
\draw [black, line width=1.0pt] (axis cs:0.473889,0.438768) ellipse [x radius=0.0150373, y radius=0.0239772];
\draw [black, line width=1.0pt] (axis cs:0.317129,0.268493) ellipse [x radius=0.0175506, y radius=0.0215851];
\draw [black, line width=1.0pt] (axis cs:0.317129,0.207787) ellipse [x radius=0.0175506, y radius=0.0215851];

\draw (0.43,0.74) node[fill=white] {Hadamard} edge[-Stealth] (0.315,0.74);

\draw (0.3,0.64) node[fill=white] {ZC} edge[-Stealth] (0.375,0.64);

\draw (0.3,0.42) node[fill=white] {Hadamard \& ZC} edge[-Stealth] (0.46,0.42);

\draw (0.507129,0.268493) node[fill=white] {Hadamard} edge[-Stealth] (0.337129,0.268493);

\draw (0.4680,0.207787) node[fill=white] {ZC} edge[-Stealth] (0.337129,0.207787);

\end{axis}
\end{tikzpicture}%

%% file: FiguresTxt/BC_MSE_T_32_33_K_31_Had_ZC.tex
% This file was created by matlab2tikz.
%
%The latest updates can be retrieved from
%  http://www.mathworks.com/matlabcentral/fileexchange/22022-matlab2tikz-matlab2tikz
%where you can also make suggestions and rate matlab2tikz.
%
\definecolor{mycolor1}{rgb}{0.00000,0.45000,0.74000}%
\definecolor{mycolor2}{rgb}{1.00000,0.00000,1.00000}%
\begin{tikzpicture}

\begin{axis}[%
width=4.739in,
height=4.33in,
at={(0.795in,0.584in)},
scale only axis,
xmin=0,
xmax=30,
xlabel style={font=\color{white!15!black}},
xlabel={Transmit Power [dBm]},
ymode=log,
ymin=0.001,
ymax=1000,
yminorticks=true,
ylabel style={font=\color{white!15!black}},
ylabel={Normalized MSE},
axis background/.style={fill=white},
xmajorgrids,
ymajorgrids,
yminorgrids,
legend style={legend cell align=left, align=left, draw=white!15!black}
]
\addplot [color=red, line width=1.5pt, mark size=3.0pt, mark=triangle, mark options={solid, red}]
  table[row sep=crcr]{%
0	5.98910151487224\\
3.33333333333333	2.84725780555772\\
6.66666666666667	1.2916741846848\\
10	0.627634215188177\\
13.3333333333333	0.267845895887867\\
16.6666666666667	0.125405592217005\\
20	0.0616074057714169\\
23.3333333333333	0.0281026925533481\\
26.6666666666667	0.0129628774179349\\
30	0.00628544421762222\\
};
\addlegendentry{LS}

\addplot [color=mycolor1, dashed, line width=1.5pt, mark size=6.0pt, mark=+, mark options={solid, mycolor1}]
  table[row sep=crcr]{%
0	5.97487791215968\\
3.33333333333333	2.84408562424856\\
6.66666666666667	1.29100887432287\\
10	0.627484727915401\\
13.3333333333333	0.267816138087423\\
16.6666666666667	0.125399066442599\\
20	0.0616059480348695\\
23.3333333333333	0.0281023720457834\\
26.6666666666667	0.0129628087212471\\
30	0.00628542840535921\\
};
\addlegendentry{Scaled LS}

\addplot [color=black!50!green, dotted, line width=1.5pt]
  table[row sep=crcr]{%
0	1.82102080931237\\
3.33333333333333	1.45699292567861\\
6.66666666666667	0.905826651296872\\
10	0.529144601639112\\
13.3333333333333	0.247540208124105\\
16.6666666666667	0.120893997825306\\
20	0.0605613571674423\\
23.3333333333333	0.027864758168334\\
26.6666666666667	0.0129214152707147\\
30	0.0062756732582548\\
};
\addlegendentry{LMMSE}

\addplot [color=red, line width=1.5pt, only marks, mark size=4.5pt, mark=o, mark options={solid, red}]
  table[row sep=crcr]{%
0	0.974407286845475\\
3.33333333333333	0.855424043418911\\
6.66666666666667	0.645398147228907\\
10	0.417514059439454\\
13.3333333333333	0.229116593756149\\
16.6666666666667	0.115558967513481\\
20	0.0573716734547821\\
23.3333333333333	0.0282742170117559\\
26.6666666666667	0.012851328485309\\
30	0.00655887886948375\\
};
\addlegendentry{MMSE - Simulation}

\addplot [color=black, dashdotted, line width=1.5pt]
  table[row sep=crcr]{%
0	0.96428595679887\\
3.33333333333333	0.852503226553716\\
6.66666666666667	0.649907269084702\\
10	0.415623565795558\\
13.3333333333333	0.229322423097365\\
16.6666666666667	0.115462777005396\\
20	0.0574390824122347\\
23.3333333333333	0.0282864307588103\\
26.6666666666667	0.012852265283413\\
30	0.00656015476008798\\
};
\addlegendentry{MMSE - Analysis}

\addplot [color=mycolor2, line width=1.5pt, mark size=3.2pt, mark=square, mark options={solid, mycolor2}]
  table[row sep=crcr]{%
0	194.005038786759\\
3.33333333333333	95.32361911481\\
6.66666666666667	42.2928004457782\\
10	19.6754172906286\\
13.3333333333333	9.11153171381089\\
16.6666666666667	4.18204865162887\\
20	1.95286471998543\\
23.3333333333333	0.930972361876614\\
26.6666666666667	0.433585484006395\\
30	0.208316692009276\\
};
\addlegendentry{\cite{Abdallah2021,  Zhao2019, Liu2021, Zhu2018,  Yerzhanova2021}}

\addplot [color=red, line width=1.5pt, mark size=3.0pt, mark=triangle, mark options={solid, red}, forget plot]
  table[row sep=crcr]{%
0	6.24619270445372\\
3.33333333333333	3.03712578715145\\
6.66666666666667	1.32397238925765\\
10	0.616271735716631\\
13.3333333333333	0.288257842440196\\
16.6666666666667	0.135741794018139\\
20	0.0603741493775073\\
23.3333333333333	0.0298777501188417\\
26.6666666666667	0.0133807144595978\\
30	0.00616100806690784\\
};
\addplot [color=mycolor1, dashed, line width=1.5pt, mark size=6.0pt, mark=+, mark options={solid, mycolor1}, forget plot]
  table[row sep=crcr]{%
0	6.23096238222911\\
3.33333333333333	3.03361742379573\\
6.66666666666667	1.32327290513871\\
10	0.616122926713299\\
13.3333333333333	0.288224618449239\\
16.6666666666667	0.135734458904987\\
20	0.0603726394814945\\
23.3333333333333	0.0298774092563748\\
26.6666666666667	0.0133806439921681\\
30	0.00616099151293285\\
};
\addplot [color=black!50!green, dotted, line width=1.5pt, forget plot]
  table[row sep=crcr]{%
0	1.80842888582819\\
3.33333333333333	1.50343717860871\\
6.66666666666667	0.920419088168691\\
10	0.516029903477742\\
13.3333333333333	0.264680105002741\\
16.6666666666667	0.130598947148081\\
20	0.0593115180756356\\
23.3333333333333	0.0296237238769076\\
26.6666666666667	0.0133311651578731\\
30	0.00615313880109673\\
};
\addplot [color=black, dashdotted, line width=1.5pt, forget plot]
  table[row sep=crcr]{%
0	0.986092860445488\\
3.33333333333333	0.877195490457738\\
6.66666666666667	0.659568593672646\\
10	0.426019096805044\\
13.3333333333333	0.246018320128275\\
16.6666666666667	0.123149329964929\\
20	0.0578087559997422\\
23.3333333333333	0.0290285816950707\\
26.6666666666667	0.0134774749219632\\
30	0.00605842258751411\\
};
\addplot [color=red, line width=1.5pt, only marks, mark size=4.5pt, mark=o, mark options={solid, red}, forget plot]
  table[row sep=crcr]{%
0	0.982049194223682\\
3.33333333333333	0.869637579543217\\
6.66666666666667	0.659653357289111\\
10	0.42556785482894\\
13.3333333333333	0.244923389292195\\
16.6666666666667	0.123181147885855\\
20	0.0576745347141401\\
23.3333333333333	0.029002630272534\\
26.6666666666667	0.0134696578673809\\
30	0.00605810475563635\\
};
\end{axis}

\begin{axis}[%
width=0.962in,
height=0.926in,
at={(1.412in,0.886in)},
scale only axis,
xmin=16.4599840922617,
xmax=16.8893780857642,
ymode=log,
ymin=0.112048252138089,
ymax=0.12910046007859,
yminorticks=true,
axis background/.style={fill=white},
xmajorgrids,
ymajorgrids,
yminorgrids
]
\addplot [color=red, line width=1.5pt, only marks, mark size=4.5pt, mark=o, mark options={solid, red}, forget plot]
  table[row sep=crcr]{%
0	0.974407286845475\\
3.33333333333333	0.855424043418911\\
6.66666666666667	0.645398147228907\\
10	0.417514059439454\\
13.3333333333333	0.229116593756149\\
16.6666666666667	0.115558967513481\\
20	0.0573716734547821\\
23.3333333333333	0.0282742170117559\\
26.6666666666667	0.012851328485309\\
30	0.00655887886948375\\
};
\addplot [color=black, dashdotted, line width=1.5pt, forget plot]
  table[row sep=crcr]{%
0	0.96428595679887\\
3.33333333333333	0.852503226553716\\
6.66666666666667	0.649907269084702\\
10	0.415623565795558\\
13.3333333333333	0.229322423097365\\
16.6666666666667	0.115462777005396\\
20	0.0574390824122347\\
23.3333333333333	0.0282864307588103\\
26.6666666666667	0.012852265283413\\
30	0.00656015476008798\\
};
\addplot [color=red, line width=1.5pt, only marks, mark size=4.5pt, mark=o, mark options={solid, red}, forget plot]
  table[row sep=crcr]{%
0	0.982049194223682\\
3.33333333333333	0.869637579543217\\
6.66666666666667	0.659653357289111\\
10	0.42556785482894\\
13.3333333333333	0.244923389292195\\
16.6666666666667	0.123181147885855\\
20	0.0576745347141401\\
23.3333333333333	0.029002630272534\\
26.6666666666667	0.0134696578673809\\
30	0.00605810475563635\\
};
\addplot [color=black, dashdotted, line width=1.5pt, forget plot]
  table[row sep=crcr]{%
0	0.986092860445488\\
3.33333333333333	0.877195490457738\\
6.66666666666667	0.659568593672646\\
10	0.426019096805044\\
13.3333333333333	0.246018320128275\\
16.6666666666667	0.123149329964929\\
20	0.0578087559997422\\
23.3333333333333	0.0290285816950707\\
26.6666666666667	0.0134774749219632\\
30	0.00605842258751411\\
};
\end{axis}

\begin{axis}[%
width=6.115in,
height=5.312in,
at={(0in,0in)},
scale only axis,
xmin=0,
xmax=1,
ymin=0,
ymax=1,
axis line style={draw=none},
ticks=none,
axis x line*=bottom,
axis y line*=left
]
\draw[-{Stealth}, color=black, line width=1.0pt] (axis cs:0.547,0.379) -- (axis cs:0.387,0.341);
\draw[line width=1.0pt, draw=black] (axis cs:0.545267489711934,0.372549019607843) rectangle (axis cs:0.576131687242798,0.42156862745098);
\draw [black, line width=1.0pt] (axis cs:0.473595,0.436051) ellipse [x radius=0.0187394, y radius=0.0242859];
\draw [black, line width=1.0pt] (axis cs:0.306931,0.203587) ellipse [x radius=0.0150451, y radius=0.0192732];
\draw [black, line width=1.0pt] (axis cs:0.306931,0.282245) ellipse [x radius=0.0150451, y radius=0.0214602];

\draw (0.47,0.74) node[fill=white] {Hadamard \& ZC} edge[-Stealth] (0.315,0.74);

\draw (0.3,0.42) node[fill=white] {Hadamard \& ZC} edge[-Stealth] (0.46,0.42);

\draw (0.506931,0.282245) node[fill=white] {Hadamard} edge[-Stealth] (0.321931,0.282245);

\draw (0.47,0.203587) node[fill=white] {ZC} edge[-Stealth] (0.321931,0.203587);

\end{axis}
\end{tikzpicture}%

%% file: FiguresTxt/DC_MSE_T_16_K_7_A_6_8.tex
% This file was created by matlab2tikz.
%
%The latest updates can be retrieved from
%  http://www.mathworks.com/matlabcentral/fileexchange/22022-matlab2tikz-matlab2tikz
%where you can also make suggestions and rate matlab2tikz.
%
\definecolor{mycolor1}{rgb}{0.00000,0.45000,0.74000}%
\definecolor{mycolor2}{rgb}{1.00000,0.00000,1.00000}%
\begin{tikzpicture}

\begin{axis}[%
width=4.739in,
height=4.33in,
at={(0.795in,0.584in)},
scale only axis,
xmin=0,
xmax=30,
xlabel style={font=\color{white!15!black}},
xlabel={Transmit Power [dBm]},
ymode=log,
ymin=1e-07,
ymax=0.001,
yminorticks=true,
ylabel style={font=\color{white!15!black}},
ylabel={Normalized MSE},
axis background/.style={fill=white},
xmajorgrids,
ymajorgrids,
yminorgrids,
legend style={legend cell align=left, align=left, draw=white!15!black}
]
\addplot [color=red, line width=1.5pt, mark size=3.0pt, mark=triangle, mark options={solid, red}]
  table[row sep=crcr]{%
0	0.000112539900530974\\
3.33333333333333	5.14696105485747e-05\\
6.66666666666667	2.45165822658279e-05\\
10	1.10350514745125e-05\\
13.3333333333333	5.25436319135824e-06\\
16.6666666666667	2.41463911307445e-06\\
20	1.13313921118927e-06\\
23.3333333333333	5.14704543861259e-07\\
26.6666666666667	2.37907918199943e-07\\
30	1.11406457201508e-07\\
};
\addlegendentry{LS}

\addplot [color=mycolor1, dashed, line width=1.5pt, mark size=6.0pt, mark=+, mark options={solid, mycolor1}]
  table[row sep=crcr]{%
0	0.000112775370649136\\
3.33333333333333	5.15350057947459e-05\\
6.66666666666667	2.4521829538617e-05\\
10	1.10362902519025e-05\\
13.3333333333333	5.25363339187941e-06\\
16.6666666666667	2.41439219413325e-06\\
20	1.13324996131611e-06\\
23.3333333333333	5.14719893388276e-07\\
26.6666666666667	2.37907295318926e-07\\
30	1.11403788669864e-07\\
};
\addlegendentry{Scaled LS}

\addplot [color=black!50!green, dotted, line width=1.5pt]
  table[row sep=crcr]{%
0	0.000112530659402144\\
3.33333333333333	5.14692394287237e-05\\
6.66666666666667	2.45163613587155e-05\\
10	1.10349349184059e-05\\
13.3333333333333	5.25429227892386e-06\\
16.6666666666667	2.41465475272344e-06\\
20	1.1331369137667e-06\\
23.3333333333333	5.14703809805549e-07\\
26.6666666666667	2.37908303648667e-07\\
30	1.1140629661594e-07\\
};
\addlegendentry{LMMSE}

\addplot [color=red, line width=1.5pt, only marks, mark size=4.5pt, mark=o, mark options={solid, red}]
  table[row sep=crcr]{%
0	0.000111476659082979\\
3.33333333333333	5.13791582069619e-05\\
6.66666666666667	2.38955553047018e-05\\
10	1.13533091436289e-05\\
13.3333333333333	5.17741251154969e-06\\
16.6666666666667	2.39426555316252e-06\\
20	1.09472909194961e-06\\
23.3333333333333	5.10085737800769e-07\\
26.6666666666667	2.42884514134948e-07\\
30	1.10244548394786e-07\\
};
\addlegendentry{MMSE - Simulation}

\addplot [color=black, dashdotted, line width=1.5pt]
  table[row sep=crcr]{%
0	0.00011150781300684\\
3.33333333333333	5.13835260837748e-05\\
6.66666666666667	2.39076659201943e-05\\
10	1.13545810591942e-05\\
13.3333333333333	5.18091067918929e-06\\
16.6666666666667	2.39503086663968e-06\\
20	1.09511677490205e-06\\
23.3333333333333	5.1047459645786e-07\\
26.6666666666667	2.42933446604424e-07\\
30	1.10326080238435e-07\\
};
\addlegendentry{MMSE - Analysis}

\addplot [color=mycolor2, line width=1.5pt, mark size=3.2pt, mark=square, mark options={solid, mycolor2}]
  table[row sep=crcr]{%
0	0.000888550927478349\\
3.33333333333333	0.000414475501613409\\
6.66666666666667	0.000192361378360889\\
10	8.93055876863835e-05\\
13.3333333333333	4.10843811555038e-05\\
16.6666666666667	1.9095238168132e-05\\
20	8.90678464421544e-06\\
23.3333333333333	4.20180546360035e-06\\
26.6666666666667	1.92805551238612e-06\\
30	8.92567623251483e-07\\
};
\addlegendentry{\cite{Abdallah2021,  Zhao2019, Liu2021, Zhu2018,  Yerzhanova2021}}

\end{axis}

\begin{axis}[%
width=6.115in,
height=5.312in,
at={(0in,0in)},
scale only axis,
xmin=0,
xmax=1,
ymin=0,
ymax=1,
axis line style={draw=none},
ticks=none,
axis x line*=bottom,
axis y line*=left
]
\end{axis}
\end{tikzpicture}%

%% file: FiguresTxt/BC_MSE_T_16_K_7_A_6_8.tex
% This file was created by matlab2tikz.
%
%The latest updates can be retrieved from
%  http://www.mathworks.com/matlabcentral/fileexchange/22022-matlab2tikz-matlab2tikz
%where you can also make suggestions and rate matlab2tikz.
%
\definecolor{mycolor1}{rgb}{0.00000,0.45000,0.74000}%
\definecolor{mycolor2}{rgb}{1.00000,0.00000,1.00000}%
\begin{tikzpicture}

\begin{axis}[%
width=4.739in,
height=4.33in,
at={(0.795in,0.584in)},
scale only axis,
xmin=0,
xmax=30,
xlabel style={font=\color{white!15!black}},
xlabel={Transmit Power [dBm]},
ymode=log,
ymin=0.007,
ymax=250,
yminorticks=true,
ylabel style={font=\color{white!15!black}},
ylabel={Normalized MSE},
axis background/.style={fill=white},
xmajorgrids,
ymajorgrids,
yminorgrids,
legend style={legend cell align=left, align=left, draw=white!15!black}
]
\addplot [color=red, line width=1.5pt, mark size=3.0pt, mark=triangle, mark options={solid, red}]
  table[row sep=crcr]{%
0	9.25681852472692\\
3.33333333333333	4.44798280507361\\
6.66666666666667	2.07646133990266\\
10	0.917251037691554\\
13.3333333333333	0.439184972978772\\
16.6666666666667	0.198016730337921\\
20	0.097308324789676\\
23.3333333333333	0.0414174873940807\\
26.6666666666667	0.0191712910552496\\
30	0.00952589877858841\\
};
\addlegendentry{LS}

\addplot [color=mycolor1, dashed, line width=1.5pt, mark size=6.0pt, mark=+, mark options={solid, mycolor1}]
  table[row sep=crcr]{%
0	9.24535820840761\\
3.33333333333333	4.44542973078682\\
6.66666666666667	2.07591354593415\\
10	0.917138214873146\\
13.3333333333333	0.439159955494685\\
16.6666666666667	0.198011522850288\\
20	0.0973071207609921\\
23.3333333333333	0.0414172440279877\\
26.6666666666667	0.0191712416992562\\
30	0.00952588707598865\\
};
\addlegendentry{Scaled LS}

\addplot [color=black!50!green, dotted, line width=1.5pt]
  table[row sep=crcr]{%
0	2.00917361495188\\
3.33333333333333	1.85428054950234\\
6.66666666666667	1.29867531243491\\
10	0.722464921020236\\
13.3333333333333	0.393198806451489\\
16.6666666666667	0.187683747652851\\
20	0.0948495052753954\\
23.3333333333333	0.0409469063679284\\
26.6666666666667	0.019064126470145\\
30	0.00950259729894515\\
};
\addlegendentry{LMMSE}

\addplot [color=red, line width=1.5pt, only marks, mark size=4.5pt, mark=o, mark options={solid, red}]
  table[row sep=crcr]{%
0	0.992551076717188\\
3.33333333333333	0.952997642890332\\
6.66666666666667	0.816936912220703\\
10	0.543162068403543\\
13.3333333333333	0.331156809859738\\
16.6666666666667	0.171945632730146\\
20	0.0910184811920962\\
23.3333333333333	0.0394855903138625\\
26.6666666666667	0.0193138157759394\\
30	0.00966870850496705\\
};
\addlegendentry{MMSE - Simulation}

\addplot [color=black, dashdotted, line width=1.5pt]
  table[row sep=crcr]{%
0	0.996379195121838\\
3.33333333333333	0.965006454551637\\
6.66666666666667	0.816349774045961\\
10	0.544380237091571\\
13.3333333333333	0.330380321135205\\
16.6666666666667	0.171683606144662\\
20	0.0911402927760562\\
23.3333333333333	0.0394667063577933\\
26.6666666666667	0.0193077339019471\\
30	0.00966415731764565\\
};
\addlegendentry{MMSE - Analysis}

\addplot [color=mycolor2, line width=1.5pt, mark size=3.2pt, mark=square, mark options={solid, mycolor2}]
  table[row sep=crcr]{%
0	148.7812040548\\
3.33333333333333	70.2711099261553\\
6.66666666666667	33.6077347456779\\
10	14.5879621725704\\
13.3333333333333	6.87926714297261\\
16.6666666666667	3.10720836368964\\
20	1.54981472996509\\
23.3333333333333	0.675039446652807\\
26.6666666666667	0.320547303036241\\
30	0.153814641393605\\
};
\addlegendentry{\cite{Abdallah2021,  Zhao2019, Liu2021, Zhu2018,  Yerzhanova2021}}

\addplot [color=mycolor2, line width=1.5pt, mark size=3.2pt, mark=square, mark options={solid, mycolor2}, forget plot]
  table[row sep=crcr]{%
0	192.060935321087\\
3.33333333333333	91.6485068736334\\
6.66666666666667	43.3682526771007\\
10	20.1124133769014\\
13.3333333333333	9.11210346450977\\
16.6666666666667	4.26612603383399\\
20	1.90296031603817\\
23.3333333333333	0.948846398159574\\
26.6666666666667	0.423770511217203\\
30	0.196117078671952\\
};
\addplot [color=red, line width=1.5pt, mark size=3.0pt, mark=triangle, mark options={solid, red}, forget plot]
  table[row sep=crcr]{%
0	11.9742888523245\\
3.33333333333333	5.7169792248848\\
6.66666666666667	2.68898242075064\\
10	1.26199967375973\\
13.3333333333333	0.58299223841785\\
16.6666666666667	0.259995363796914\\
20	0.120767697396051\\
23.3333333333333	0.0586568526940063\\
26.6666666666667	0.0263292005135881\\
30	0.0124779325224187\\
};
\addplot [color=black!50!green, dotted, line width=1.5pt, forget plot]
  table[row sep=crcr]{%
0	2.09415456987191\\
3.33333333333333	1.97074420544226\\
6.66666666666667	1.45733773559809\\
10	0.921214979507606\\
13.3333333333333	0.504342527260882\\
16.6666666666667	0.242682604643603\\
20	0.116762238988778\\
23.3333333333333	0.0577036783754242\\
26.6666666666667	0.0261066713944097\\
30	0.0124314940916347\\
};
\addplot [color=mycolor1, dashed, line width=1.5pt, mark size=6.0pt, mark=+, mark options={solid, mycolor1}, forget plot]
  table[row sep=crcr]{%
0	11.959622893695\\
3.33333333333333	5.71372561440307\\
6.66666666666667	2.68826681865986\\
10	1.26184447986398\\
13.3333333333333	0.582958709732212\\
16.6666666666667	0.259988458722532\\
20	0.120766177811998\\
23.3333333333333	0.058656528581913\\
26.6666666666667	0.026329135863058\\
30	0.0124779172151499\\
};
\addplot [color=black, dashdotted, line width=1.5pt, forget plot]
  table[row sep=crcr]{%
0	0.992066667745256\\
3.33333333333333	0.961063608218894\\
6.66666666666667	0.864249144951201\\
10	0.65885406918128\\
13.3333333333333	0.403006968823227\\
16.6666666666667	0.224042448366074\\
20	0.109584159864225\\
23.3333333333333	0.0549720613222226\\
26.6666666666667	0.0260334523406049\\
30	0.0124189494802328\\
};
\addplot [color=red, line width=1.5pt, only marks, mark size=4.5pt, mark=o, mark options={solid, red}, forget plot]
  table[row sep=crcr]{%
0	1.00320034942954\\
3.33333333333333	0.97001540814394\\
6.66666666666667	0.86016783640914\\
10	0.656054694623349\\
13.3333333333333	0.404819527301477\\
16.6666666666667	0.22468139724811\\
20	0.109609087507679\\
23.3333333333333	0.0550306213233763\\
26.6666666666667	0.0260283744071194\\
30	0.0124186614925612\\
};
\end{axis}

\begin{axis}[%
width=1.013in,
height=0.939in,
at={(1.314in,0.859in)},
scale only axis,
xmin=22.1376144277581,
xmax=24.4830277323239,
ymode=log,
ymin=0.0319068776874427,
ymax=0.086155494100082,
yminorticks=true,
axis background/.style={fill=white},
xmajorgrids,
ymajorgrids,
yminorgrids
]
\addplot [color=red, line width=1.5pt, mark size=3.0pt, mark=triangle, mark options={solid, red}, forget plot]
  table[row sep=crcr]{%
0	11.9742888523245\\
3.33333333333333	5.7169792248848\\
6.66666666666667	2.68898242075064\\
10	1.26199967375973\\
13.3333333333333	0.58299223841785\\
16.6666666666667	0.259995363796914\\
20	0.120767697396051\\
23.3333333333333	0.0586568526940063\\
26.6666666666667	0.0263292005135881\\
30	0.0124779325224187\\
};
\addplot [color=mycolor1, dashed, line width=1.5pt, mark size=6.0pt, mark=+, mark options={solid, mycolor1}, forget plot]
  table[row sep=crcr]{%
0	11.959622893695\\
3.33333333333333	5.71372561440307\\
6.66666666666667	2.68826681865986\\
10	1.26184447986398\\
13.3333333333333	0.582958709732212\\
16.6666666666667	0.259988458722532\\
20	0.120766177811998\\
23.3333333333333	0.058656528581913\\
26.6666666666667	0.026329135863058\\
30	0.0124779172151499\\
};
\addplot [color=black!50!green, dotted, line width=1.5pt, forget plot]
  table[row sep=crcr]{%
0	2.09415456987191\\
3.33333333333333	1.97074420544226\\
6.66666666666667	1.45733773559809\\
10	0.921214979507606\\
13.3333333333333	0.504342527260882\\
16.6666666666667	0.242682604643603\\
20	0.116762238988778\\
23.3333333333333	0.0577036783754242\\
26.6666666666667	0.0261066713944097\\
30	0.0124314940916347\\
};
\addplot [color=red, line width=1.5pt, only marks, mark size=4.5pt, mark=o, mark options={solid, red}, forget plot]
  table[row sep=crcr]{%
0	1.00320034942954\\
3.33333333333333	0.97001540814394\\
6.66666666666667	0.86016783640914\\
10	0.656054694623349\\
13.3333333333333	0.404819527301477\\
16.6666666666667	0.22468139724811\\
20	0.109609087507679\\
23.3333333333333	0.0550306213233763\\
26.6666666666667	0.0260283744071194\\
30	0.0124186614925612\\
};

%=====================================
\addplot [color=red, line width=1.5pt, only marks, mark size=4.5pt, mark=o, mark options={solid, red}, forget plot]
  table[row sep=crcr]{%
0	0.992551076717188\\
3.33333333333333	0.952997642890332\\
6.66666666666667	0.816936912220703\\
10	0.543162068403543\\
13.3333333333333	0.331156809859738\\
16.6666666666667	0.171945632730146\\
20	0.0910184811920962\\
23.3333333333333	0.0394855903138625\\
26.6666666666667	0.0193138157759394\\
30	0.00966870850496705\\
};

%=================================

\addplot [color=black, dashdotted, line width=1.5pt, forget plot]
  table[row sep=crcr]{%
0	0.992066667745256\\
3.33333333333333	0.961063608218894\\
6.66666666666667	0.864249144951201\\
10	0.65885406918128\\
13.3333333333333	0.403006968823227\\
16.6666666666667	0.224042448366074\\
20	0.109584159864225\\
23.3333333333333	0.0549720613222226\\
26.6666666666667	0.0260334523406049\\
30	0.0124189494802328\\
};
\addplot [color=red, line width=1.5pt, mark size=3.0pt, mark=triangle, mark options={solid, red}, forget plot]
  table[row sep=crcr]{%
0	9.25681852472692\\
3.33333333333333	4.44798280507361\\
6.66666666666667	2.07646133990266\\
10	0.917251037691554\\
13.3333333333333	0.439184972978772\\
16.6666666666667	0.198016730337921\\
20	0.097308324789676\\
23.3333333333333	0.0414174873940807\\
26.6666666666667	0.0191712910552496\\
30	0.00952589877858841\\
};
\addplot [color=black!50!green, dotted, line width=1.5pt, forget plot]
  table[row sep=crcr]{%
0	2.00917361495188\\
3.33333333333333	1.85428054950234\\
6.66666666666667	1.29867531243491\\
10	0.722464921020236\\
13.3333333333333	0.393198806451489\\
16.6666666666667	0.187683747652851\\
20	0.0948495052753954\\
23.3333333333333	0.0409469063679284\\
26.6666666666667	0.019064126470145\\
30	0.00950259729894515\\
};
\addplot [color=mycolor1, dashed, line width=1.5pt, mark size=6.0pt, mark=+, mark options={solid, mycolor1}, forget plot]
  table[row sep=crcr]{%
0	9.24535820840761\\
3.33333333333333	4.44542973078682\\
6.66666666666667	2.07591354593415\\
10	0.917138214873146\\
13.3333333333333	0.439159955494685\\
16.6666666666667	0.198011522850288\\
20	0.0973071207609921\\
23.3333333333333	0.0414172440279877\\
26.6666666666667	0.0191712416992562\\
30	0.00952588707598865\\
};
\addplot [color=black, dashdotted, line width=1.5pt, forget plot]
  table[row sep=crcr]{%
0	0.996379195121838\\
3.33333333333333	0.965006454551637\\
6.66666666666667	0.816349774045961\\
10	0.544380237091571\\
13.3333333333333	0.330380321135205\\
16.6666666666667	0.171683606144662\\
20	0.0911402927760562\\
23.3333333333333	0.0394667063577933\\
26.6666666666667	0.0193077339019471\\
30	0.00966415731764565\\
};
\end{axis}

\begin{axis}[%
width=6.115in,
height=5.312in,
at={(0in,0in)},
scale only axis,
xmin=0,
xmax=1,
ymin=0,
ymax=1,
axis line style={draw=none},
ticks=none,
axis x line*=bottom,
axis y line*=left
]
\draw [black, line width=1.0pt] (axis cs:0.299298,0.264468) ellipse [x radius=0.0158636, y radius=0.027689];
\draw [black, line width=1.0pt] (axis cs:0.297595,0.203684) ellipse [x radius=0.0158636, y radius=0.027689];
\draw[-{Stealth}, color=black, line width=1.0pt] (axis cs:0.63,0.305) -- (axis cs:0.382,0.339);
\draw [black, line width=1.0pt] (axis cs:0.732538,0.275792) ellipse [x radius=0.011925, y radius=0.0183258];
\draw[line width=1.0pt, draw=black] (axis cs:0.630658436213992,0.288235294117647) rectangle (axis cs:0.662551440329219,0.348684954751131);
\draw [black, line width=1.0pt] (axis cs:0.817717,0.188235) ellipse [x radius=0.011925, y radius=0.0196078];

\draw (0.459298,0.274468) node[fill=white] {$\alpha=0.6$} edge[-Stealth] (0.318298,0.274468);

\draw (0.459298,0.213684) node[fill=white] {$\alpha=0.8$} edge[-Stealth] (0.318298,0.213684);

\draw (0.85,0.29) node[fill=white] {$\alpha=0.6$} edge[-Stealth] (0.745,0.29);

\draw (0.706,0.175) node[fill=white] {$\alpha=0.8$} edge[-Stealth] (0.806,0.175);

\draw (0.418,0.795) node[fill=white] {$\alpha=0.6$} edge[-Stealth] (0.318,0.795);

\draw (0.275,0.70) node[fill=white] {$\alpha=0.8$} edge[-Stealth] (0.375,0.70);

\end{axis}
\end{tikzpicture}%